\documentclass[rmp,aps,nofootinbib,twocolumn,10pt]{revtex4-1}
\usepackage[T1]{fontenc}
\usepackage[colorlinks=true,     linkcolor=blue,     citecolor=blue,
urlcolor=blue]{hyperref}

\usepackage{graphics}
\usepackage{epsfig}
\usepackage{amssymb}
\usepackage{graphicx}
\usepackage{graphics}
\usepackage[english]{babel}
\usepackage{amsmath}        
\usepackage{amsfonts}       
\usepackage{amsbsy,amssymb}     
\usepackage{xcolor}
\definecolor{red}{rgb}{0.7,0,0}
\definecolor{green}{rgb}{0.,0.35,0.}
\definecolor{blue}{rgb}{0.2,0.2,0.7}
\usepackage{times}
\usepackage{latexsym}
\usepackage{fancyhdr}
\usepackage{verbatim}
\usepackage{epsf}
\usepackage{subfigure}
\usepackage{bm}
\usepackage{epsfig}
\usepackage{psfrag}
\usepackage{dcolumn}
\usepackage{makeidx}
\usepackage[T1]{fontenc}

\def\inbar{\,\vrule height1.5ex width.4pt depth0pt}
\def\IR{\relax{\rm I\kern-.18em R}}
\def\IC{\relax\hbox{$\inbar\kern-.3em{\rm C}$}}

\newcommand{\bs}{\bf}

\newcommand{\mymod}[1]{\left|{#1}\right|}              
\newcommand{\ud}{\mathrm{d}}

\definecolor{red}{rgb}{0.7,0,0}
\definecolor{green}{rgb}{0.,0.35,0.}
\definecolor{blue}{rgb}{0.2,0.2,0.7}

\newcommand{\ensm}[1]{\ensuremath{#1}}

\newcommand{\bra}[1]{\ensm{\langle #1|}}

\newcommand{\J}{\ensm{t}}

\newcommand{\ket}[1]{\ensm{| #1 \rangle}}

\newcommand{\V}{\ensm{V}}

\newcommand{\beq}{\begin{equation}}
\newcommand{\eeq}{\end{equation}}

\newcommand{\be}{\begin{equation}}
\newcommand{\ee}{\end{equation}}
\newcommand{\bea}{\begin{eqnarray}}
\newcommand{\eea}{\end{eqnarray}}

\newcommand{\vect}[1]{\mathbf{#1}}

\newcommand{\gdD}[1]{g}
\newcommand{\UdD}[1]{U}

\newcommand{\as}{a_{\rm s}}

\newcommand{\ba}{{\bf a}}

\newcommand{\noe}{\hat{n}}

 \def\F{{\mathbb{F}}}
 
\def\R{{\mathbb{R}}}

\def\ee{\mathord{\rm e}}

\def\bs#1{\boldsymbol{#1}}

\def\be{\begin{equation}}
\def\ba{\begin{align}}
\def\enda{\end{align}}
\def\bi{\begin{itemize}}
\def\ei{\end{itemize}}

\def\Rb87{^{87}\rm{Rb}}                 
\def\Li6{^{6}\rm{Li}}                   
 \usepackage{color}
\usepackage[normalem]{ulem} 

\newcommand{\xe}{\color{black}}												

\definecolor{color_dl}{rgb}{1 0 0} 
\newcommand{\xdl}{\color{black}}
\newcommand{\cutdl}[1]{\color{color_dl}\sout{#1 }\color{black}} 

\newcommand{\ofr}{(\mathbf{r})}

\newcommand{\dr}{\int\! d^3{r}\ }
\newcommand{\BC}{\mathrm{DI}}
\newcommand{\JBC}{T}
\newcommand{\hJBC}{ \hat T}
\newcommand{\JBCBF}{T_\BF}
\newcommand{\eff}{{\mathrm{eff}}}
\newcommand{\ER}{E_\mathrm{R}}
\newcommand{\up}{\uparrow} 
\newcommand{\down}{\downarrow} 
\newcommand{\B}{\mathrm{B}}
\renewcommand{\F}{\mathrm{F}}
\newcommand{\BF}{\mathrm{BF}}
\newcommand{\BB}{\mathrm{BB}} 
\newcommand{\expect}[1]{\langle #1 \rangle}
\newcommand{\f}{c}
\renewcommand{\a}{{\alpha}}

\newcommand{\hb}[2]{\hat{b}_\text{#1}^{#2}}
\newcommand{\hbd}[2]{\hat{b}_\text{#1}^{#2\dagger}}
\newcommand{\abcd}{{\alpha\beta\gamma\delta}}
\renewcommand{\R}{\mathrm{R}}
\newcommand{\nR}{{n_\mathrm{R}}}
\newcommand{\nL}{{n_\mathrm{L}}}
\newcommand{\hatt}{\tilde}

\newcommand{\tot}{{\mathrm{tot}}}

\newcommand{\Fig}[1]{Fig.~\ref{#1}}
\newcommand{\Eq}[1]{Eq.~\eqref{#1}}
\newcommand{\Sec}[1]{Sec.~\ref{#1}}

\newcommand{\cites}[1]{{\color{blue}\cite{#1}\color{black}}}
\newcommand{\from}[1]{ ---  Figure from \protect\cites{#1}. --- }

\newcommand{\adaptedfrom}[1]{ ---  Figure adapted from \protect\cites{#1}. --- }

\newcommand{\eqa}[1]{\begin{eqnarray} #1 \end{eqnarray}}
\newcommand{\fo}{\hat{f}}
\newcommand{\Psio}{\hat{\Psi}}
\newcommand{\sgn}{\mathrm{sgn}}

\begin{document}

\title{Non-standard Hubbard models in optical lattices}
\author{Omjyoti Dutta$^1$, Mariusz Gajda$^{2,3}$, Philipp Hauke$^{4,5}$, Maciej Lewenstein$^{6,7}$, Dirk-S\"oren L\"uhmann$^8$, Boris A. Malomed$^{6,9}$, Tomasz Sowi\'nski$^{2,3}$, and Jakub Zakrzewski$^{1,10}$}
\affiliation{
\mbox{$^1$ Instytut Fizyki imienia Mariana Smoluchowskiego, Uniwersytet Jagiello\'nski, Reymonta 4, 30-059 Krak\'ow, Poland}
\mbox{$^2$ Institute of Physics of the Polish Academy of Sciences, Al. Lotnik\'ow 32/46, PL-02-668 Warsaw, Poland}
\mbox{$^3$ Center for Theoretical Physics of the Polish Academy of Sciences, Al. Lotnik\'ow 32/46, PL-02-668 Warsaw, Poland}
\mbox{$^4$ Institute for Quantum Optics and Quantum Information of the Austrian Academy of Sciences, A-6020 Innsbruck, Austria}
\mbox{$^5$ Institute for Theoretical Physics, Innsbruck University, A-6020 Innsbruck, Austria}
\mbox{$^6$ ICFO-Institut de Ci\`encies Fot\`oniques, Mediterranean Technology Park, Av. C.F. Gauss 3, E-08860 Castelldefels (Barcelona), Spain}
\mbox{$^7$ ICREA-Instituci\'o Catalana de Recerca i Estudis Avan\c{c}ats, Lluis Company 23, E-08011 Barcelona, Spain}
\mbox{$^8$ Institut f\"ur Laser-Physik, Universit\"at Hamburg, Luruper Chaussee 149, 22761 Hamburg, Germany }
\mbox{$^9$ Department of Physical Electronics, School of Electrical Engineering, Faculty of Engineering,}
\mbox{\ \ Tel Aviv University, Tel Aviv 69978, Israel}
\mbox{$^{10}$ Mark Kac Complex Systems Research Center, Jagiellonian University, ul. Reymonta 4, 30-059 Krak\'ow, Poland}
}

\date{\today}

\begin{abstract}
Originally, the Hubbard model has been derived for describing the behaviour of strongly-correlated electrons in solids. However, since over a decade now, variations of it are also routinely being implemented with ultracold atoms in optical lattices, allowing their study in a clean, essentially defect-free environment. 
Here, we review some of the rich literature on this subject, with a focus on more recent non-standard forms of the Hubbard model. 
After an introduction to standard (fermionic and bosonic) Hubbard models, we discuss briefly common models for mixtures, as well as the so called extended Bose-Hubbard models, that include interactions between neighboring sites, next-neighboring sites, and so on. 
The main part of the review discusses the importance of additional terms appearing when refining the tight-binding approximation on the original physical Hamiltonian. 
Even when restricting the models to the lowest Bloch band is justified, the standard approach neglects the density-induced tunneling (which has the same origin as the usual on-site interaction). 
The importance of these contributions is discussed for both contact and dipolar interactions. 
For sufficiently strong interactions, also the effects related to higher Bloch bands become important even for deep optical lattices. Different approaches that aim at incorporating these effects, mainly via dressing the basis Wannier functions with interactions, leading to effective, density-dependent Hubbard-type models, are reviewed.  
We discuss also examples of Hubbard-like models that explicitly involve higher $p$-orbitals, as well as models that couple dynamically spin and orbital degrees of freedom. 
Finally, we review mean-field nonlinear-Schr\"odinger models of the Salerno
type that share with the non-standard Hubbard models the nonlinear
coupling between the adjacent sites. In that part, discrete solitons are the
main subject of the consideration. We conclude by listing some future open problems.    
\end{abstract}
\maketitle
\tableofcontents


\section{Introduction}  
\subsection{Hubbard models}

Hubbard models are relatively simple, yet complex enough lattice models of theoretical physics,
 capable of describing strongly-correlated states of quantum many-body systems. Quoting Wikipedia \footnote{http://en.wikipedia.org/ as of May, 30, 2014}: {\sl The Hubbard model is an approximate model used, especially in solid state physics, to describe the transition between conducting and insulating systems. The Hubbard model, named after John Hubbard, is the simplest model of interacting particles in a lattice, with only two terms in the Hamiltonian \cites{Hubbard1963}: a kinetic term allowing for tunneling (`hopping') of particles between sites of the lattice and a potential term consisting of an on-site interaction. The particles can either be fermions, as in Hubbard's original work, or bosons, when the model is referred to as the ``Bose--Hubbard model'' or the boson Hubbard model}. Let us note that  the lattice  model for bosons was first derived by  Gersch and Knollman \cites{Gersch63},  prior to Hubbard's fermionic counterpart.

The Hubbard model is a good approximation for particles in a periodic potential at sufficiently low temperatures. All the particles are then  in the lowest Bloch band, as long as any long-range interactions between the particles can be ignored. If interactions between particles on different sites of the lattice are included, the model is often referred to as the ``extended Hubbard model''.

John Hubbard introduced the Fermi Hubbard models in 1963 to describe electrons, i.e.\ spin $1/2$ fermions in solids. The model has been intensively studied since,  especially since there are no efficient methods of simulating it numerically in dimensions greater than one. \xdl Because of this complexity, various calculational methods, such as for instance  exact diagonalization, various perturbative expansions, mean field/pairing theory, mean field/cluster expansions, slave boson theory, fermionic Quantum Monte Carlo \cites{Lee06,Lee08,Troyer2005}, or more recent tensor network approaches (cf.\ \cites{Chung2014,Corboz2010,Corboz2010a} and references therein),  lead to contradicting  quantitative, and even qualitative results. \xe Only the one-dimensional Fermi Hubbard model is analytically soluble with the help of \xdl the \xe Bethe ansatz \cites{Essler05}. The 2D Fermi Hubbard model, or better to say, a weakly coupled array of 2D Fermi Hubbard models is in the center of interest of contemporary condensed-matter physics, since it is believed to describe high-temperature superconductivity of cuprates. In the end of the last century, the studies of various types of Hubbard models have intensified enormously due to the developments of physics of ultracold atoms, ions, and molecules. 

\subsection{Ultracold atoms in optical lattices}

The studies of ultracold atoms constitute one of the hottest areas of atomic, molecular, and optical (AMO) physics and quantum optics. They have been awarded  with the 1997 Nobel Prize for S. Chu \cites{Chu98}, C. Cohen-Tannoudji \cites{Cohen-Tannoudji98}, and W.D.  Phillips \cites{ Phillips98} for laser cooling, and the 2001 Nobel Prize for E. Cornell, C. Wieman \cites{Cornell02}, and W. Ketterle \cites{Ketterle02} for the first observation of the Bose--Einstein condensation (BEC). All these developments, despite their indisputable importance and beauty, concern the physics of weakly interacting systems. Many AMO theoreticians working in this area  suffered an (unfortunately to some extend justified) critics of their condensed-matter colleagues, that ``all of that has been  known before''.  The recent progress in this area, however, is by no means less spectacular. Particularly impressive are  recent advances  in the studies of ultracold gases in optical lattices. Optical lattices are formed by several laser 
beams in standing wave configurations. They provide practically ideal, loss-free potentials, in which ultracold atoms may move and interact one with another \cites{Grimm00,Windpassinger13}. In 1998, a theoretical paper of Jaksch and coworkers \cites{Jaksch98}, following the seminal work by condensed-matter theorists \cites{Fisher89}, has shown  that ultracold atoms in optical lattices may enter the regime of strongly correlated systems, and exhibit a, so called, superfluid-Mott insulator quantum phase transition. The following experiment at the Ludwig-Maximilian Universit\"at in Munich \index{Greiner02} confirmed this prediction, and in this manner the physics of ultracold atoms got an invitation to the ``High Table'' -- the frontiers of the modern condensed-matter physics and quantum field theory. Nowadays it is routinely possible to create  systems of ultracold bosonic or fermionic atoms, or their mixtures,  in  one-, two-, or three-dimensional optical lattices in strongly correlated states \cites{Auerbach94}, 
i.e.\ states in which genuine quantum correlations, such as entanglement, extend over large distances (for recent reviews see \cites{Lewenstein12,Lewenstein07,Giorgini08,Bloch08}).  Generic examples of such states are found when the system in question undergoes a, so called, quantum phase transition \cites{Sachdev99}. The transition from the Bose superfluid (where all atoms form a macroscopic coherent wave packet that is spread over the entire lattice) to the Mott-insulator state (where a fixed number of atoms is  localized in every lattice site) is a  paradigmatic  example of such a quantum phase transition. While the systems observed in experiments such as \cites{Greiner02} are of finite size, and are typically confined in some trapping potential, so that they might not exhibit a critical behaviour in the rigorous sense, there is no doubt about their strongly correlated nature. 

\subsection{Ultracold matter and quantum technologies}

\xdl The unprecedented control and precision with which one can engineer ultracold gases inspired many researchers to consider such systems as possible candidates for implementing quantum technologies, in particular quantum-information processing and high-precision metrology. In the 1990's, the main effort of the community was directed towards the realization of a universal scalable quantum computer, stimulated by the seminal work of Cirac and Zoller \cites{CiracZoller}, who have proposed the first experimental realization of a universal two-qubit gate with trapped ions.   In order to follow a similar approach with atoms, one would first  choose specific states of  atoms, or groups of atoms, as states of qubits (two level systems), or qudits (elementary systems with more than two internal quantum states). The second step would then consist in implementing quantum logical gates on the single-qubit and two-qubit level. Finally, one would aim at implementing complete quantum protocols and quantum error correction in such systems by employing interatomic interactions and/or interactions with external (electric, magnetic, laser) fields. Perhaps the first paper presenting such a vision with atoms proposed, in fact, to realize quantum computing  using ultracold atoms in an optical lattice \cites{Jaksch99}. It is also worth stressing that the pioneering  paper of Jaksch et al. \cites{Jaksch98}] was motivated by the quest for quantum computing: The transition to a Mott-insulator state was supposed to be in this context an efficient way of preparing  a quantum register with fixed number of atoms per site.

In  recent years, however, it became clear that while the prospects of universal quantum computing are still elusive, another approach to quantum computing, suggested by Feynman \cites{Feynman86}, may already now be realized with ultracold atoms and ions in laboratories. This approach  employs these highly controllable systems as quantum computers of special purpose, or, in other words, as quantum simulators \cites{Jaksch05}. There has been a considerable interest recently in both of these approaches in theory and experiment. In particular, it has been widely discussed that ultracold atoms, ion, photons, or superconducting circuits could serve as quantum simulators of various types of the so called Bose-- or Fermi--Hubbard models and related lattice spin models \cites{Lewenstein07,Lewenstein12,Jaksch05,Cirac04,Cirac12,Bloch12,Blatt12,Aspuru-Guzik12,Houck12,Hauke2011d}.
The basic idea of quantum simulators can be condensed in four points (see, e.g., \cites{Hauke2011d}):
\begin{itemize} 
\item A quantum simulator is an experimental system that mimics a simple model, or a family of simple models of condensed-matter physics, high-energy physics, quantum chemistry etc.
\item The simulated models have to be of some relevance for applications and/or our understanding of challenges of contemporary physics.
\item The simulated models should be computationally very hard for classical computers. Exceptions from this rule are possible for quantum simulators that exhibit novel, so far only theoretically predicted phenomena.
\item A quantum simulator should allow for a broad control of the parameters of the simulated model, and for control of preparation, manipulation, and detection of states of the system. It should allow for validation (calibration)!
\end{itemize}
Practically all  Hubbard models can hardly be simulated by classical computers for very large systems; at  least some of them are hard to simulate even for moderate system sizes due to the lack of scalable classical algorithms, caused for instance by the infamous sign problem in Quantum Monte Carlo (QMC) codes, or complexity caused by disorder. 
These Hubbard models describe a variety of condensed-matter systems (but not only), and thus are directly related to challenging problems of modern condensed-matter physics, concerning for instance high-temperature superconductivity (cf.\ \cites{Lee08}), Fermi superfluids (cf.\ \cites{Giorgini08,Bloch08}), or 
lattice gauge theories and quark confinement \cites{Montvay97} 
(for recent works in the area of ultracold atoms and lattice gauge theories cf.\ \cites{Luca-annals, Luca-natcomm,reznikPRL,Zoller-DalmontePRL,Wiese-review}).  
The family of Hubbard models thus by far satisfies the relevance and hardness criteria mentioned above, moving them into the focal point of attempts at building a quantum simulator. For these reasons, a better understanding of the experimental feasibility of quantum simulating Hubbard models is of great practical and technological importance.  \xe

\subsection{Beyond standard Hubbard models}

\xdl As a natural first step, one would like to realize standard Bose-- and Fermi--Hubbard models, i.e., those models that have only a kinetic term and one type of interactions, as mentioned in the introduction. The static properties of the Bose--Hubbard model are accessible to QMC simulations, but only for not too large and not too cold systems, while the out-of-equilibrium dynamics of this model can be only computed efficiently for short times. The case of Fermi--Hubbard model is even more difficult: here neither static nor dynamical properties can be simulated efficiently, even for moderate system sizes. These models are thus paradigm examples of systems that can be studied with quantum  simulations with ultracold atoms in optical lattices \cites{Lewenstein07},  provided they can be realized with a sufficient precision and control in laboratories.  

Interestingly, however, many Hubbard models that are simulated with ultracold atoms do not have a standard form; the corresponding Hamiltonians frequently contain terms that include correlated and occupation-dependent tunnelings within the lowest band, as well as correlated tunnelings and occupation of higher bands. 
These effects have been observed in the past decade in many different experiments, concerning 
\begin{itemize}
\item Observations of density-induced tunneling \cites{Meinert2013,Jurgensen2014b};
\item Shift of the Mott transition in Fermi--Bose mixtures \cites{Heinze2011,Best2009,Gunter2006,Ospelkaus06};
\item Mott insulator in  the bosonic system \cites{Mark2011};
\item Modifications of on-site interactions \cites{Bakr2011,Campbell2006,Mark2011,Mark2012,Uehlinger2013,Will2010};
\item Effects of excited bands \cites{Anderlini07,Browaeys2005,Kohl2005,Mueller2007,Oelschlaeger2011,Oelschlaeger2012,Oelschlaeger2013,Wirth10};
\item Dynamical spin effects \cites{Pasquiou2010,Pasquiou2011,Paz2012,Paz2013}.
\end{itemize}
One can view these non-standard terms in two ways: as an obstacle, or as an opportunity.  On one hand, one has to be careful in attempts to quantum simulate standard Hubbard models. On the other hand,  non-standard Hubbard models are extremely interesting by themselves: they exhibit novel exotic quantum phases, quantum phase transitions, and other quantum properties. Quantum simulating these is itself a formidable task!  Since such models are now within experimental reach,  it is necessary to study and understand them in order to describe experimental findings and make new prediction for ultracold quantum gases.  For this reason, there has been quite a progress of such studies in recent years, and this is the main motivation for this review. \xe

Our paper is organized as follows. Before we explain what the considered non-standard Hubbard models are, we discuss in short the form and variants of standard and standard extended Hubbard models in Section II. In Section III we present the main {\it dramatis personae} of this review: non-standard single- and non-single-band  Hubbard models. The section starts with a short historical glimpse describing  models introduced already in the '80s by Hirsch and others. All the models discussed here have a form of single band models, in the sense that the effect of higher bands is included in an effective manner, for instance through many-body modifications of the Wannier functions describing single-particle states in a given lattice site. In contrast, the non-standard models considered in Section IV include explicit contributions of excited bands, which, however, at least in some situations, can be still cast within ``effective single band models'' (for instance via appropriate modifications of the Wannier 
functions).  

Section V deals with $p$-band Hubbard models, while Section VI with Hubbard models appearing in the theory of ultracold dipolar gases and the phenomenon of the Einstein-de Haas effect. Section VII is devoted to weakly interacting non-standard Hubbard models, and in particular to various types of exotic solitons that can be generated in such systems. 
We conclude our review in Section VIII
\xdl pointing out some of the open problems. \xe

\xdl Let us also mention some topics that {\it will not} be discussed in this review, primarily to keep it within reasonable bounds. We consider extended optical lattices and do not discuss double or triple well systems where also interaction induced effects are important (for a recent example see \cites{FisherUwe2013}). 
We also do not go into the rapidly developing subject of modifications to Hubbard models by externally induced couplings. These may lead to the creation of artificial gauge fields or spin-orbit interactions via e.g.\ additional laser (for recent reviews see \cites{Dalibard11,Goldman2013}) or microwave \cites{Struck2014} couplings. Fast periodic modulations
of different Hamiltonian parameters (lattice positions or depth, or interactions) may lead to effective, time-averaged Hamiltonians with additional terms modifying Hubbard models (see e.g.\ \cites{Eckardt05,Lignier07,Eckardt10,Struck11,Hauke2012c,Rapp2012,Liberto2014}). Even faster modulations may be used to resonantly couple the lowest Bloch band with the excited ones, opening additional experimental possibilities \cites{Sowinski2012a,Lacki2013b,Dutta2014,Straeter2014}.

\xe


\section{Standard Hubbard models in optical lattices}  
%
%

Before we turn to the discussion of non-standard Hubbard models, let us first establish a clear meaning of what we mean by standard ones. We start this section by discussing a weakly interacting Bose gas in an optical lattice, and derive the discrete Gross--Pitaevskii, i.e.\ discrete nonlinear Schr\"odinger equation describing such a situation.  Subsequently, we give a short description  of Bose-- and Fermi--Hubbard models and their basic properties. These models allow the treatment of particles in the strongly-correlated regime.  Finally, we discuss the \textit{extended} Hubbard models with nearest-neighbor, next nearest-neighbor  interactions, etc.,  which provide a standard basis for the treatment  of dipolar gases in optical lattices.  

\subsection{Weakly interacting particles: The non-linear Schr\"odinger equation}
\label{BEC}

We start by providing a description of a weakly interacting Bose-Einstein condensate  placed in an optical lattice. The many-body Hamiltonian in the second-quantization formalism describing a gas of $N$ interacting bosons in an external potential, $V_{\rm ext}$, reads:
\begin{eqnarray}
\hat{H}(t)=\int& d\bs{r}&\,\hat{\Psi}^{\dagger} (\bs{r},t)\left[-\frac{\hbar^2}{2m}\nabla^2+V_{\rm ext} \right] \hat{\Psi}(\bs{r},t) \nonumber\\
+\frac{1}{2}\int& d\bs{r}&\, d\bs{r'}  \hat{\Psi}^{\dagger}(\bs{r},t)\hat{\Psi}^{\dagger}(\bs{r'},t) V(\bs{r}-\bs{r'}) \hat{\Psi}(\bs{r},t)\hat{\Psi}(\bs{r'},t),\nonumber \\
\label{hamiltonian}
\end{eqnarray}
where $\hat{\Psi}$ and $\hat{\Psi}^{\dagger}$ are the bosonic anihilation and creation field operators, respectively. Interactions between atoms are given by an isotropic short-range pseudopotential modelling s-wave  interactions \cites{Bloch08}  
\beq \label{pseudo}
V(\bs{r}-\bs{r'})=\frac{4\pi \hbar^2 a_s}{m}\delta(\vec r-\vec r')\frac{\partial}{\partial |\vec r-\vec r'| }|\vec r-\vec r'| 
\eeq
Here, $m$ is the atomic mass and $a_s$ the $s$--wave scattering length that characterizes the interactions---attractive (repulsive) for negative (positive) $a_s$---through elastic binary collisions\index{collisions ! binary elastic} at low energies between neutral atoms, independently of the actual interparticle two-body potential. This is due to the fact that for ultracold atoms the de Broglie wavelength is much larger than the effective extension of the interaction potential, implying that the interatomic potential can be replaced by a pseudopotential.  For non-singular  $\hat{\Psi}(\bs{r},t)$ the pseudopotential  is equivalent to a contact potential of the form
 \beq \label{contact}
 V(\bs{r}-\bs{r'})=(4\pi\hbar^2a_s/m)\delta(\bs{r}-\bs{r'})=g\delta(\bs{r}-\bs{r'}).
\eeq 
Note that this approximation is valid provided no long-range contributions exist (later we shall consider modifications due to long-range dipolar interactions)---for more details about scattering theory see for instance \cites{Landau87,Gribakin93}.

 If the bosonic gas is dilute, $na_s^3\ll1$, where $n$ is the density, the mean-field description applies, the basic idea of which was formulated by Bogoliubov \citeyear{Bogoliubov47}. It consists in writing the field operator in the Heisenberg representation as a sum of its expectation value (condensate wave function) plus a fluctuating field operator, 
\begin{equation}
\hat{\Psi}(\bs{r},t)=\Psi(\bs{r},t)+\delta\hat{\Psi}(\bs{r},t).
\label{mean}
\end{equation}
When classical and quantum fluctuations are neglected, the time evolution of the condensate wave function at temperature $T=0$ is governed by the Gross-Pitaevskii equation (GPE) \cites{Gross61,Pitaevskii61,Pitaevskii03}, obtained by using the Heisenberg equations and Eq.~(\ref{mean})\index{Gross-Pitaevskii equation (GPE) ! time evolution}, 
\begin{equation}
i\hbar \frac{d}{dt}\Psi(\bs{r},t)=-\frac{\hbar^2}{2m}\nabla^2\Psi(\bs{r},t)+\left[ V_{\rm ext}+g|\Psi(\bs{r},t)|^2\right]\Psi(\bs{r},t).
\label{gpe}
\end{equation}
The wave function of the condensate is normalized to the total number of particles $N$. Here we will consider 
the situation in which the external potential corresponds to an optical lattice, combined with a weak  harmonic trapping potential.

A BEC placed in an optical lattice can be described in the so-called tight-binding approximation\index{tight binding approximation} if the lattice depth is sufficiently large, such that the barrier between the neighboring sites is much higher than the chemical potential and the energy of the system is confined within the lowest band. This approximation corresponds
to decomposing the condensate order parameter \index{order parameter ! BEC}$\Psi(\bs{r},t)$ as a
sum of wave functions $\Theta(\bs{r}-\bs{R}_i)$ localized in each
well of the periodic potential, 
\begin{equation}
\Psi(\bs{r},t)=\sqrt{N}\sum_i \varphi_i\Theta(\bs{r}-\bs{R}_i),
\label{ansatz}
\end{equation}
where $\varphi_i=\sqrt{n_i(t)}e^{i\phi_i(t)}$ is the amplitude of
the $i$--th lattice site with $n_i=N_i/N$ and $N_i$ is the number
of particles in the $i$--th site. Introducing the ansatz given by
Eq.~(\ref{ansatz}) into Eq.~(\ref{gpe}) \cites{Trombettoni01},
one obtains the discrete non-linear Schr\"odinger (DNLS) equation, \index{Schr\"odinger equation! discrete non linear} which in its standard form reads
\begin{equation}
i\hbar\frac{\partial\varphi_i}{\partial t}=-K(\varphi_{i-1}-\varphi_{i+1})+(\epsilon_i+U|\varphi_i|^2)\varphi_i,\,
\label{dnlse}
\end{equation}
 where $K$ denotes the next-neighbor tunneling rate, 
 \begin{equation}
K=-\int d\bs{r}\left[\frac{\hbar^2}{2m}\bs{\nabla}\Theta_i\cdot\bs{\nabla}\Theta_{i+1}+\Theta_i
V_{ext}\Theta_{i+1}\right]. \label{tunneling}
\end{equation}
The on-site energies are given by
\begin{equation}
\epsilon_i=\int d\bs{r}\left[\frac{\hbar^2}{2m}(\bs{\nabla}\Theta_i)^2+V_{ext}\Theta^2_{i}\right],
\label{onsite}
\end{equation}
and the nonlinear coefficient by 
\begin{equation}
U=g N\int d\bs{r}\,\Theta^4_{i}\,.
\label{nonlinear}
\end{equation}
Here we have only reviewed the lowest order DNLS equation. Nevertheless,
it has been shown \cites{Smerzi03} that the effective
dimensionality of the BECs trapped in each well can modify the
degree of nonlinearity and the tunneling rate in the DNLS equation.
We will come back to the DNLS equation and its non-standard forms in Section VII. 

\subsection{Bose--Hubbard model\label{cha:standardHub}}

 In the strongly interacting regime, bosonic atoms in a periodic lattice potential are
well described by a Bose--Hubbard Hamiltonian
\cites{Fisher89}. In this section, we explain how the Bose--Hubbard Hamiltonian can be derived
from the many-body Hamiltonian in second quantization (\ref{hamiltonian}) by expressing the fields 
through the single-particle Wannier modes.  To be specific, we shall assume from now on the separable 3D lattice potential of the form
\begin{equation}
V_\text{ext}=\sum_{l=x,y,z} V_{0l}\sin^2(\pi l/a),
\label{pot}
\end{equation}
for which  Wannier functions are the product of one-dimensional standard Wannier functions \cites{Kohn1959}. In Eq.~(\ref{pot}), 
$a$ plays the role of the lattice constant (and is equal to half the wavelength of the lasers forming the standing-wave pattern).
By appropriately arranging the directions and relative phases of the laser beams, much richer lattice structures 
may be achieved \cites{Windpassinger13},  such as the celebrated triangular or kagom\'e lattices. The corresponding Wannier functions may then be found following the approach developed by Marzari and Vanderbild \cites{Marzari97,Marzari12}.   We shall, however, not consider here different geometrical aspects of possible optical lattices, but rather concentrate on the interaction-induced phenomena. Similarly, we do not discuss phenomena that are induced by next-nearest neighbor tunnelings.

\xdl Let us start by reminding the reader of the handbook approach \cites{Ashcroft1976}. The field operators can always be expanded in the basis of Bloch functions, which are the eigenfunctions of the single-particle Hamiltonian consisting of the kinetic term and the periodic lattice potential,  
\begin{equation}
\hat{\Psi}(\bs{r})=\sum_{n,k} \hat{b}_{n,k} \phi_{n,k}(\bs{r}).
\label{Bloch_field}
\end{equation}
The Bloch functions have indices denoting the band number $n$ and the quasi-momentum $k$. 
For sufficiently deep optical potentials, and at low temperatures, the band gap between the lowest and the first excited band may be large enough, so that the second and higher bands will practically not be populated and can be disregarded. 
Within the lowest Bloch  band of the periodic potential (\ref{pot}) the field operators may be expanded into an orthonormal Wannier basis, consisting of functions
localized around the lattice sites. 
More precisely, the Wannier
functions have the form $w_i(\bs{r})=w(\bs{r}-\bs{R}_i)$, with $\bs{R}_i$ corresponding to the minima of the lattice potential \cites{Jaksch98}
\begin{equation}
\hat{\Psi}(\bs{r})=\sum_i \hat{b}_i w_i(\bs{r}).
\label{wannier_field}
\end{equation}
This expansion  (known as tight-binding appoximation)\index{tight binding approximation} makes sense because the temperature is sufficiently low,  and because the typical interaction energies are not strong enough to excite higher vibrational states. 
Here,  $\hat{b}_i$ ($\hat{b}^{\dag}_i$) denote the annihilation (creation) operators of a particle localized at the $i$-th lattice site, which obey canonical commutation relations
$\left[\hat{b}_i,\hat{b}^{\dag}_j\right]=\delta_{ij}$.  
The impact of higher bands in multi-orbital Hubbard models is discussed in \Sec{sec:MO}, whereas  
the situation where particles are confined in a single higher band of the lattice is addressed in \Sec{orbitalhub}.
Introducing the above expansion into
the Hamiltonian given in Eq.~(\ref{hamiltonian}),   one obtains 
\begin{equation}
\hat{H}= -\sum_{\langle i,j\rangle}  t_{ij} \hat{b}_i^\dag \hat{b}_j 
+\frac{U}{2} \sum_i \hat{n}_i(\hat{n}_i-1) - \mu  \sum_i \hat{n}_i,
\label{BoseHubbard}
\end{equation}
where $\langle ,\rangle$ indicates the sum over nearest
neighbors (note that each $i,j$ pair appears twice in the summation ensuring hermicity of the first term). 
Further, $\hat{n}_i=\hat{b}^{\dag}_i \hat{b}_i$ is the boson
number operator at site $i$. In the above expression, $\mu$ denotes the chemical potential, which is introduced to control the total number of atoms.
In the standard approach, among all terms arriving from the expansion in the Wannier basis, only  tunneling between nearest neighbors is considered and only interactions between particles on the same lattice site are kept.  Note that this  may not be a good approximation for shallow lattices \cites{Trotzky12}.  Another way of looking at this problem is to realize that for sufficiently shallow lattice potentials,  the lowest band will not have a cosinus-like dispersion, and hence the single band tight binding approximation (as introduced above) will not be valid.
The tunneling matrix element between
adjacent sites is given by 
\begin{equation}
t_{ij}=-\int d\bs{r}\;w_i^{\star} \,(\bs{r}) \left[\frac{-\hbar^2\bs{\nabla}^2}{2m}+ V_{\rm ext}\right]w_j(\bs{r}).
\label{t}
\end{equation}
The subindex $(ij)$ can be omitted in the homogeneous case, when
the external optical potential is isotropic and tunneling is the
same along any direction. For a contact potential, the strength of the two-body on-site interactions $U$ reduces to 
\begin{equation}
U=g\int d\bs{r}\;\; |w_i(\bs{r})|^4.
\label{u}
\end{equation}
 If an external potential $V_\mathrm{ext}$  accounts also for a trapping potential
$V_{T}$, an additional term in the Bose--Hubbard Hamiltonian appears, 
accounting for the potential energy, 
\begin{equation}
\hat H_\mathrm{ext}=\sum_i \epsilon_i n_i\,,
\label{H_epsilon}
\end{equation} 
 with $\epsilon_i$ given by 
\begin{equation}
\epsilon_i=\int d\bs{r} \,V_T \,|w_i(\bs{r})|^2\approx V_T(\bs{R}_i).
\label{epsilon}
\end{equation}
This term describes an energy offset for each lattice site; typically it is absorbed into a site-dependent chemical potential, $\mu_i=\mu + \epsilon_i$. \xe

Within the harmonic approximation (i.e.\ the approximation in which the on-site potential is harmonic and the Wannier functions are Gaussian), 
it is possible to obtain analytical expressions for the 
integrals above. 
 While this approximation may provide qualitative information, often, even for 
deep lattices, exact Wannier-functions expansion provides much better quantitative results, in the sense that the tight-binding model closer represents the real physics in continuous space. The harmonic approximation
underestimates tunneling amplitudes due to assuming Gaussian tails of the wavefunctions, as compared with the real exponential tails of Wannier functions.  As we shall see later in \Sec{orbitalhub} and \Sec{orbitaldipole}, the two approaches may lead to qualitatively different physics also for excited bands.  For the same reason, even in the mean-field
DNLS approach, discussed in \Sec{BEC}, it is desirable to use Wannier functions in place of $\Theta$ localized functions introduced there. 

The Bose--Hubbard Hamiltonian, Eq.~(\ref{BoseHubbard})\index{Bose-Hubbard model}\index{Hubbard model ! Bose systems}, exhibits
two different quantum phases depending on the ratio between the
tunneling energy and the on-site repulsion energy: (i) a
superfluid, compressible, gapless phase when tunneling dominates, and (ii) an incompressible,  Mott-insulator ground state when the on-site interaction
dominates\index{SF-MI phase transition}.  A detailed discussion of methods of analysis \xdl (various kinds of mean-field approaches, Quantum Monte Carlo methods, strong coupling expansions, DMRG,  exact diagonalizations, etc.) \xe as well as properties of this standard model have been often reviewed \cites{Zwerger03,Lewenstein07,Bloch08,Cazalilla11,Lewenstein12}. In particular for high order expansions see \cites{Elstner99,Damski06}, while for the most recent works on this model see \cites{Rigol13,Lacki14}.

\subsection{Fermi--Hubbard model}
\label{FermiHubbardModel}

This section, describing the Hubbard model for a trapped gas of interacting spin 1/2  fermions, follows to a great extent the recent reviews of 
\cites{Bloch08,Lee08,Giorgini08,Radzihovsky10}.
The starting point is again a quantum field 
theory model similar to (\ref{hamiltonian}), reading
\begin{equation}
\label{eq:BFM}
\hat{H}_{F}=\int d\bs{r} \Bigg[\sum_{\sigma}\hat{\psi}_{\sigma}^{\dagger}
\big(-\frac{\hbar^2}{2m}\nabla^2+V_{\rm ext}\big)\hat{\psi}_{\sigma}+
  {g} \Big( \hat{\psi}_{\downarrow}^\dagger
\hat{\psi}^\dagger_{\uparrow}\hat{\psi}_{\uparrow}
\hat{\psi}_{\downarrow}\Big)\Bigg],
\end{equation}
where $\sigma=\{\uparrow, \downarrow\}$ denotes the spin, the
field operators obey fermionic anticommutation relations $\{
\hat{\psi}(\bs{r})_{\sigma},\hat{\psi}^{\dagger}(\bs{r}')_{\sigma'}\}=\delta_{\sigma\sigma'}
\delta(\bs{r}-\bs{r}')$. As previously for bosons, applying a standard 
tight-binding approximation, the electronic (or
for us atomic spin 1/2) Fermi-Hubbard model is obtained with the
Hamiltonian 
\begin{equation}
 \hat{H}= -\sum_{\langle i,j\rangle,\sigma}t_{ij}\hat{f}_{i\sigma}^\dag \hat{f}_{j\sigma}+ \frac{U}{2}\sum_i
\hat{f}_{i\uparrow}^\dag \hat{f}_{i\downarrow}^\dag \hat{f}_{i\downarrow}
\hat{f}_{i\uparrow}-\mu\sum_{i,\sigma}\hat{f}_{i\sigma}^\dag
\hat{f}_{i\sigma},
\label{eq:FHa}
\end{equation}
where $\hat{f}_{i\sigma}^\dag$ ($\hat{f}_{i\sigma}$) is the creation (annihilation) operator of $\sigma$
fermions at site $i$ and $\mu$ is the chemical potential. 
This model has fundamental importance for the theory of conducting electrons (or fermions in general).  

The BCS theory of superconductivity is essentially a theory of pairing, or a theory of
Gaussian fermionic states.  For weak interactions, when $U\ll t$ (assuming $t_{ij}=t$ for simplicity), one can replace the
quartic interaction  term in the Hamiltonian, by a ``Wick-averaged''
bilinear term
\begin{align}
 U\sum_i
\hat{f}_{i\uparrow}^\dag \hat{f}_{i\downarrow}^\dag
\hat{f}_{i\downarrow} \hat{f}_{i\uparrow}\simeq
&\left(\Delta_i\hat{f}_{i\uparrow}^\dag \hat{f}_{i\downarrow}^\dag
+ \Delta_i^*\hat{f}_{i\downarrow} \hat{f}_{i\uparrow} +
W_{i\downarrow}\hat{f}_{i\uparrow}^\dag  \hat{f}_{i\uparrow}\right. +\nonumber\\
&\left.W_{i\uparrow}\hat{f}_{i\downarrow}^\dag
\hat{f}_{i\downarrow} - V_{i}\hat{f}_{i\uparrow}^\dag
\hat{f}_{i\downarrow} - V^*_{i}\hat{f}_{i\downarrow}^\dag
\hat{f}_{i\uparrow}\right), \label{pairing}
\end{align}
where $\Delta_i=U\langle \hat{f}_{i\downarrow}
\hat{f}_{i\uparrow}\rangle$, $W_{i\sigma}=U\langle
\hat{f}_{i\sigma}^\dag  \hat{f}_{i\sigma}\rangle$, and
$V^*_{i}=U\langle \hat{f}_{i\downarrow}^\dag
\hat{f}_{i\uparrow}\rangle$.
 The further steps are
straightforward. For $T=0$ the ground state of the bilinear
Hamiltonian (\ref{eq:FHa}) is easily obtained by diagonalization.
Next, we calculate 
the ground state averages of $\Delta_i$, $W_{i\sigma}$, and $V_i$,
and obtain in this way self-consistent, highly nonlinear equations
for these quantities. Typically, they have to be then treated
numerically. Similarly, for $T>0$ the averages have to be performed with respect to the
quantum Boltzmann-Gibbs state, i.e.\ thermal canonical state, or
even better grand canonical state.

\begin{figure}[t]
\centerline{
\includegraphics[width=80mm,natwidth=610,natheight=642]{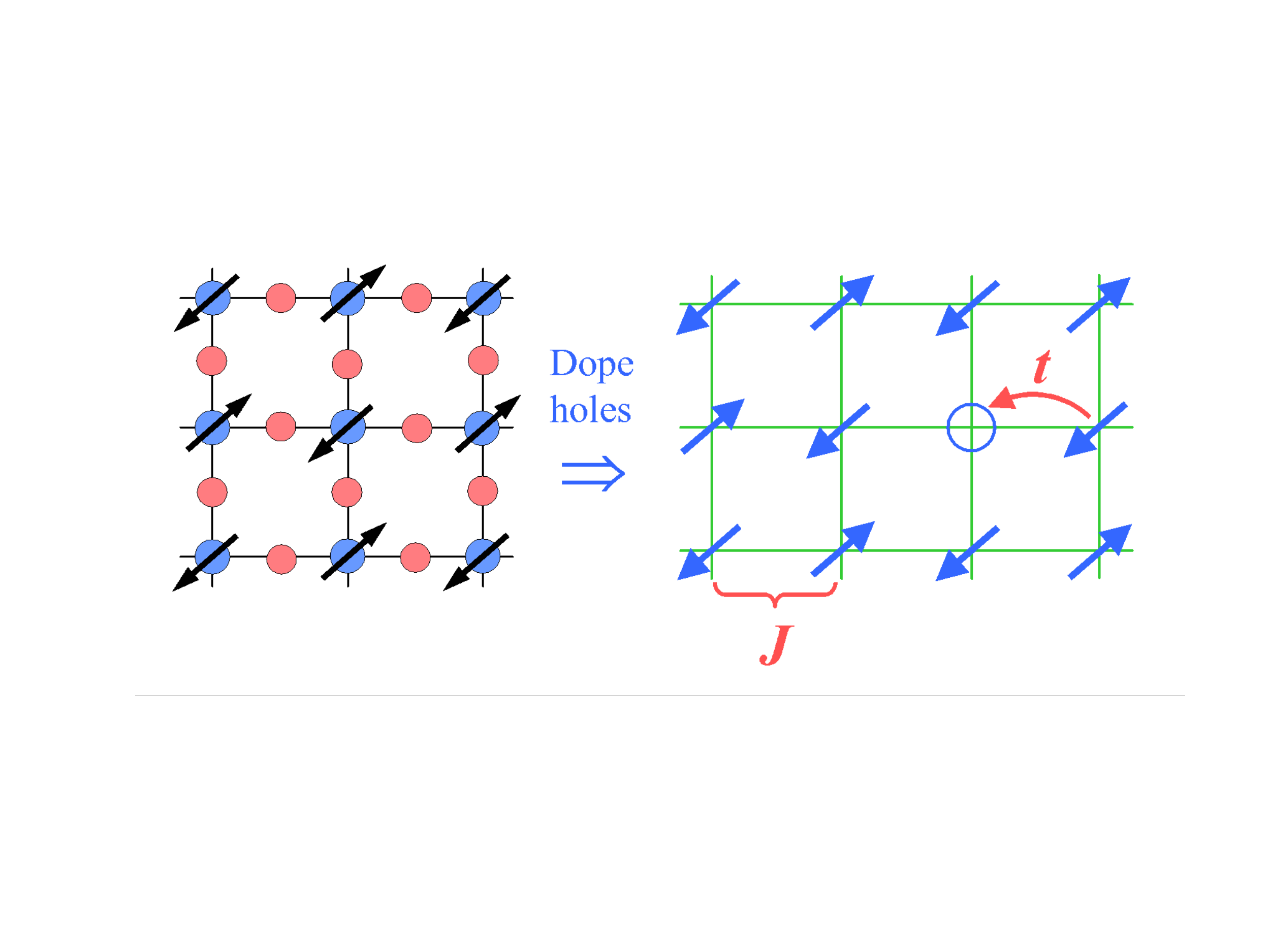}
} \caption{Schematics of a Cu-O layer (on the left) forming a typical cuprate.
Copper atoms sit on a square lattice with oxygen atoms in between. One band model 
with electron hopping rate $t$ (on the right) corresponding to the simplified electronic structure.
 $J$ denotes the antiferromagnetic super-exchange between spins on
neighboring sites.
Reprinted figure with permission from P.A.~Lee, N.~Nagaosa, and X.-G.~Wen, \href{http://link.aps.org/abstract/RMP/v78/p17}{Rev. Mod. Phys. {\bf 78}, 17 (2006)}. Copyright (2006) by the American Physical Society.}
\label{cuprates}
\end{figure}

Cuprates were the first high-temperature
superconductors discovered, and all of them have a layered structure, 
consisting typically of several oxygen-copper planes
(see Fig. \ref{cuprates}).
So far, there is
no consensus concerning mechanisms and nature of high-$T_c$
superconductivity\index{high $T_c$ superconductivity, cuprates}.
Nevertheless, many researchers believe that the Hubbard model can
provide important insights to help in understanding the high-$T_c$
superconductivity of \index{cuprates} cuprates.

 \begin{figure}[t]
\centerline{
\includegraphics[width=90mm,natwidth=610,natheight=642]{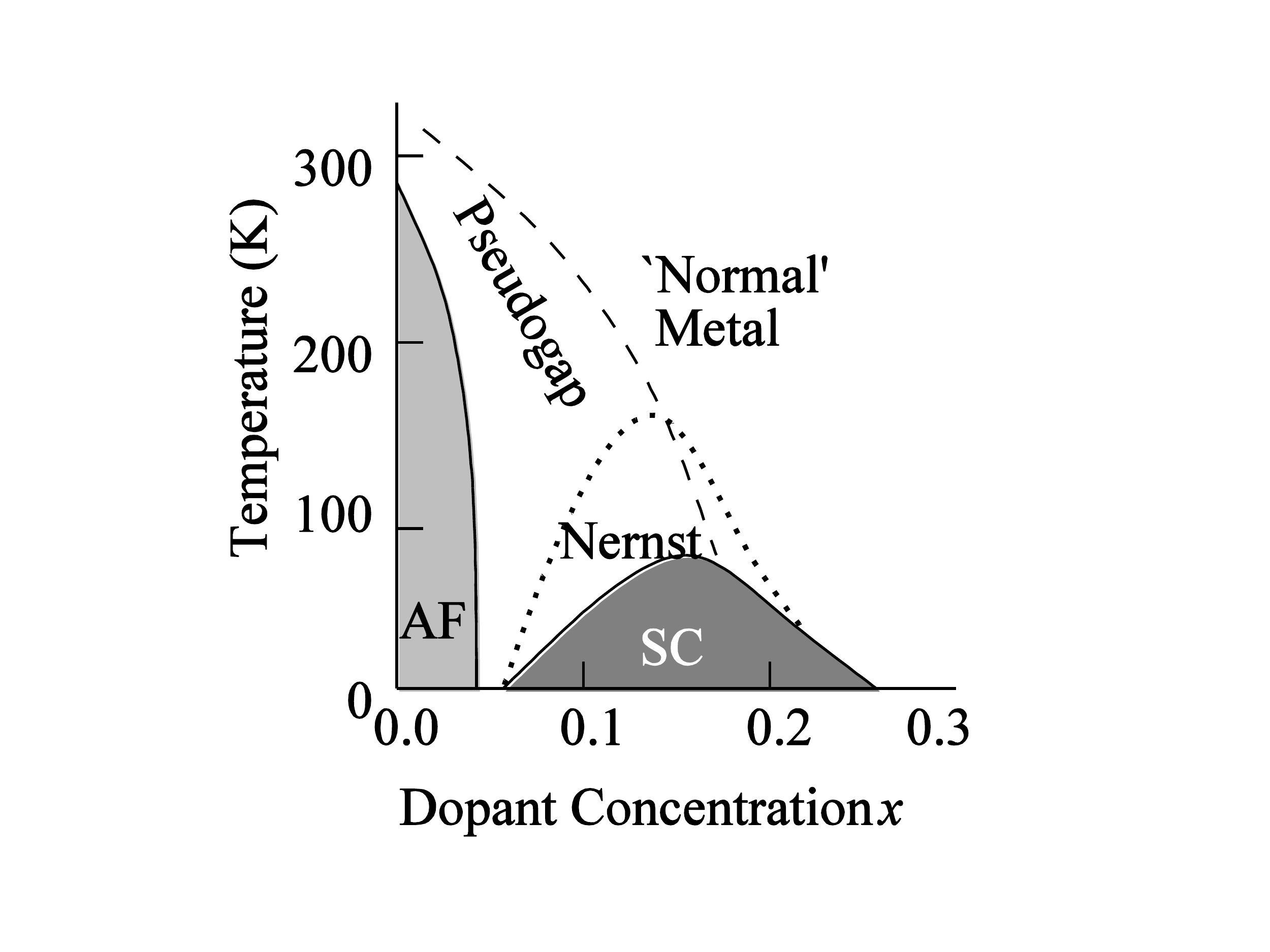}
}\caption{Schematic phase diagram of high-$T_c$ materials in the plane temperature versus dopant concentration $x$. 
AF and SC denote antiferromagnet and $d$-wave superconductor, respectively. Fluctuations of the SC appear below the dotted line corresponding to the Nernst effect. The pseudogap region extends below the dashed line. Reprinted figure with permission from P.A.~Lee, N.~Nagaosa, and X.-G.~Wen, \href{ http://link.aps.org/abstract/RMP/v78/p17 }{Rev. Mod. Phys. {\bf 78}, 17 (2006)}. Copyright (2006) by the American Physical Society.}
\label{fig_hightc}
\end{figure}

Consider again the Hubbard Hamiltonian (\ref{eq:FHa}). The hopping matrix element $t_{ij}$ between
sites $i$ and $j$ is in principle not restricted to the nearest neighbors. We
denote  nearest-neighbor hopping by $t$ and further-neighbor hoppings
by $t^\prime$, $t^{\prime\prime}$, and so on. At half-filling (one electron
per site) the system undergoes a metal-to-insulator transition \index{metal-to-insulator transition}
as the ratio $U/t$ is increased. The insulator is the Mott
insulator\index{fermionic Mott insulator} \cites{Mott49} that we
met already for bosons. There is exactly
one particle per site, and this effect is caused solely by strong
repulsion. This is in contrast to a band insulator, which has two
electrons of opposite spin per site, and cannot have more in the
lowest band due to the \index{Pauli exclusion principle} Pauli exclusion principle. For large enough
$U/t$, fermions remain localized at the lattice sites, because any
hopping leads to a double occupation of some site, with a large
energy cost $U$. The fermionic Mott insulator is additionally predicted to be
antiferromagnetic (AF), because the AF alignment permits virtual
hopping to gain a super-exchange energy $J=4t^2/U$, whereas for parallel
spins hopping is strictly forbidden by Pauli exclusion. The fermionic MI was realized in the
beautiful experiments \cites{Jordens08,Schneider08}, while the forming of an AF state seems to be very close to an experimental realization
\cites{Hulet-AF,Esslinger-pseudoAF,Imri2014}. Importantly, the first fermionic MI in 2D was also realized recently \cites{Uehlinger2013}.

Electron vacancies  (holes) can be introduced into the
copper-oxygen layers in a process called hole doping\index{hole
doping}---leading to even more complex and interesting physics.
In condensed matter, doping is typically realized by introducing a
charge reservoir away from the copper-oxygen planes, such that
it removes electrons from the plane. For ultracold
atoms the number of ``spin-up'' and ``spin-down'' atoms can be
controlled independently. Thus, in principle one can easily mimic
the effect of doping, although in the presence of the confining
harmonic potential it is difficult to achieve homogeneous doping
in a well controlled way. One can circumvent this problem in
repulsive Fermi-Bose mixtures. In such mixtures, composite
fermions consisting of a fermion (of spin up or down) and a
bosonic hole may form, and their number can
be controlled by adding bare bosons to the system \cites{Eckardt10a}.

Figure~\ref{fig_hightc} presents the schematic phase diagram \index{phase diagram ! high $T_c$ materials} that results from hole doping in
the plane spanned by temperature $T$ and hole concentration $x$. 
At low $x$ and low $T$, the AF
order is stable. With increasing $x$, the AF order is rapidly
destroyed by a few percent of holes. For even larger $x$,
a superconducting phase appears, which is believed to
be of $d$-wave\index{$d$-wave superconductivity} type. The
transition temperature reaches a maximum at the optimal doping of
about 15\%.  The high-$T_c$ SF region has a characteristic bell
shape for  all hole-doped cuprates, even though the maximum $T_c$
varies from  about 40~K to 93~K and higher. The region below the
dashed line in Fig.~\ref{fig_hightc}, above
$T_c$ in the underdoped region (where $x$ is smaller than optimal),
is an exotic metallic state, called \index{pseudogap phase} pseudogap phase. 
Below the dotted line, there is a region of strong fluctuations of the superconducting phase 
characterized by the, so called, \index{Nernst effect} Nernst effect \cites{Lee08}.

\subsection{Extended (dipolar) Hubbard models}
\label{sec:EBHM}

Let us now go beyond contact interactions and consider the tight-binding description for systems with longer than contact, or simply with long-range interactions. 
Instead of Coulomb interactions that appear between electrons in solids, in cold-atom physics a paradigmatic model can be realized with dipole-dipole interactions. This may have a magnetic origin, but strong interactions can occur for electric dipole interactions as, e.g.\ between polar molecules. Recent  reviews of ultracold dipolar gases \xdl in optical lattices  provide a detailed introduction and description of this subject for Fermi \cites{Baranov-dipoles} and Bose \cites{,Lahaye-dipoles,Trefzger11} systems (see also \cites{Lewenstein12}); here we present only the essentials. \xe 

Assuming a polarized sample where all dipoles point in the same direction, the total interaction potential consists of a contact term and a dipole-dipole part
\beq \label{fullpot}
V(\bs{r}-\bs{r'})=g\delta(\bs{r}-\bs{r'})+\frac{d^2}{4\pi\epsilon_0}\frac{1-3\cos^2\theta}{|\bs{r}-\bs{r'}|^3},
\eeq
where $\theta$ is the angle between the polarization direction of the dipoles and their relative position vector  $\bs{r}-\bs{r'}$, $d$ is the electric dipole moment, and $g$ is the amplitude of the contact interaction. Note that the classical interaction between two point dipoles contains also another $\delta$-type contribution, which is absent for effective atom-atom (molecule-molecule) interactions (or may be thought of as being incorporated into the contact term). For convenience, we denote the two parts of $V(\bf{r}-\bf{r'})$ as $U_{\rm c}$ and $U_{\rm dd}$, respectively. 

The interaction between the dipoles is highly anisotropic. We consider a stable two-dimensional geometry with a tight confinement in the direction of polarization of the dipoles. Applying an optical lattice in the perpendicular plane, the potential reads
\begin{equation}
\label{trap2d}
V_\mathbf{ext}({\bf r}) \!=\! V_0 \left[\cos^2(\pi x/a)+\cos^2(\pi y/a)\right] + \frac{1}{2}m \Omega^2_z z^2  \,.
\end{equation}
As previously, we use the expansion of the
field operators in the basis of  Wannier functions (stricly speaking a product of one-dimensional Wannier functions in $x$ and $y$ directions with the ground state of the harmonic trap  in $z$ with frequency $\Omega_z$), and restrict
ourselves to the lowest Bloch band. 

\xdl
\subsubsection{Dipolar Bose Hubbard models}

Within the above described  approximations, and for a one-component Bose system, the
Hamiltonian becomes the standard
Bose--Hubbard Hamiltonian \eqref{BoseHubbard} with addition of a dipolar contribution, which reads in the basis of Wannier functions \xe
\begin{equation}
\hat{H}_{\rm dd} = \sum_{ijkl} \frac{U_{ijkl}}{2}\; \hat{b}^\dag_i \hat{b}^\dag_j \hat{b}_k \hat{b}_l\,,
\end{equation} 
where the matrix elements $U_{ijkl}$ are given by the integral
\begin{equation}\begin{split}
\label{EQ:VIntegral}
 U_{ijkl} = \int & \ud^3r_1 \ud^3r_2 \ w_i^*({\bf r}_1)w_j^*({\bf r}_2)\\
 &\times U_{\rm dd}({\bf r}_1 - {\bf r}_2) w_k({\bf r}_1)w_l({\bf r}_2).
\end{split}\end{equation}
The Wannier functions are localized at the wells of the optical lattice with a spatial localization $\sigma$. For a deep enough lattice  $\sigma \ll a$, the Wannier functions $w_i({\bf r})$ are significantly non-vanishing for ${\bf r}$ close to the lattice centers ${\bf R_i}$, and thus the integral (\ref{EQ:VIntegral}) may be significantly non-zero for the
indices $i=k$ and $j=l$. Thus, there are two main
contributions to $U_{ijkl}$: the {\it off-site} term $U_{ijij}$, corresponding to $k=i\neq
j=l$, and the {\it on-site} term $U_{iiii}$, when all the indices are equal. 
\\\\*{\it Off-site contribution $-$} The dipolar potential $U_{\rm dd}({\bf r}_1 - {\bf r}_2)$ changes slowly on scales larger than $\sigma$.
Therefore, one may approximate
it with the constant $U_{\rm dd}({\bf R}_i - {\bf R}_j)$ and take it out of the integration. Then the integral reduces to
\begin{equation}
U_{ijij} \simeq U_{\rm dd}({\bf R}_i - {\bf R}_j) \int \! \ud^3r_1 \mymod{w_i({\bf r}_1)}^2
\int \! \ud^3r_2 \mymod{w_j({\bf r}_2)}^2,
\end{equation}
which leads to the off-site Hamiltonian
\begin{equation}
\hat{H}_{\rm dd}^\text{off-site} = \frac{1}{2}\sum_{i \neq j}\frac{V}{|i-j|^3}\, \hat{n}_i \hat{n}_j,
\label{vij}
\end{equation}
with $V=U_{ijij}$
 and  the sum running over all sites of the lattice. %
\begin{figure}[b]
\centering\includegraphics[width=0.9\linewidth]{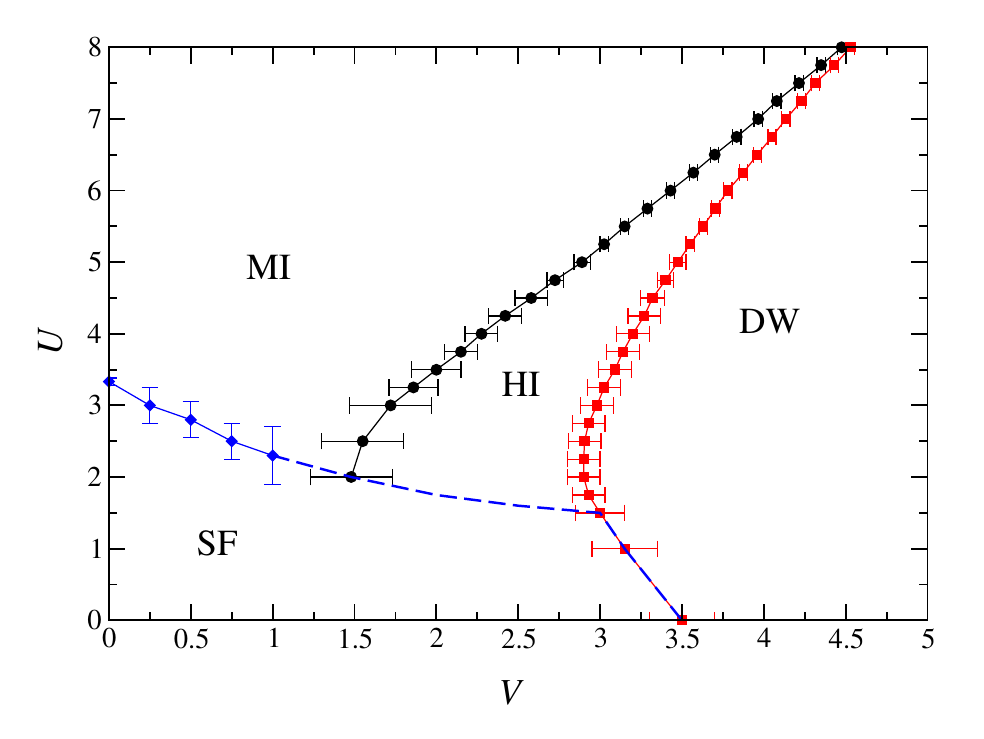}
\caption{Phase diagram of the extended 1D Bose--Hubbard model \eqref{EQ:eBHH} as a function of the on-site interaction $U$ and the nearest-neighbor interaction $V$ with $t=1$. It shows the superfluid phase (SF), the Mott insulator (MI), the density wave (DW) and the Haldane insulator (HI) for the filling per site $\rho=1$. \from{Rossini2012} } 
\label{Rossini2012Figure1}
\end{figure}
\\\\*{\it On-site contribution $-$} At the same lattice site $i$, where $|{\bf r}_1 - {\bf r}_2| \sim \sigma$, the dipolar potential
 changes very rapidly and
diverges for $|{\bf r}_1 - {\bf r}_2| \rightarrow 0$. Therefore, the integral
\begin{equation}
\label{EQ:VInt}
U_{iiii} = \int \ud^3r_1 \ud^3r_2 n({\bf r}_1) U_{\rm dd}({\bf r}_1 - {\bf r}_2) n({\bf r}_2),
\end{equation}
with $n({\bf r}) = \mymod{w({\bf r})}^2$ being the single-particle density, has to be calculated taking into account the
atomic spatial distribution at the lattice site. 
The solution can be found by Fourier transforming, i.e.\ 
\begin{eqnarray}
U_d= U_{iiii}  = \frac{1}{(2\pi)^3} \int \ud^3k \, \widetilde{U}_{\rm dd}({\bf k}) \;\widetilde{n}^{\;2}({\bf k}),
\end{eqnarray}
which leads to an on-site dipolar contribution to the Hamiltonian of the type
\begin{equation}
\label{EQ:HddContact}
\hat{H}_{\rm dd}^\text{on-site} =\frac{U_d}{2}  \sum_{i} \hat{n}_i(\hat{n}_i-1).
\end{equation}
Thus,  for dipolar gases the effective on-site interaction $U$ is given by
\begin{equation}
\label{EQ:OnSiteU}
U = g\int \ud^3r \mymod{w({\bf r})}^4  + \frac{1}{(2\pi)^3} \int \ud^3k \, \widetilde{U}_{\rm dd}({\bf k}) \;\widetilde{\rho}^{\;2}({\bf k}),
\end{equation}
which contains the contribution of the contact potential and the
dipolar contribution (\ref{fullpot}).

Let us note that the dipolar part of the on-site interaction $U_{iiii}=U_d$ (\ref{EQ:VInt})
is directly dependent on the atomic density in a lattice well, and thus  can be
increased or decreased by changing the anisotropy and strength of
the lattice confinement (see \cites{Lahaye-dipoles} for details). 

We may now write the simplest tight-binding Hamiltonian of the system. Often
one limits the off-site interaction term to nearest-neighbors, thus only obtaining the Hamiltonian
\begin{equation}\begin{split}
\label{EQ:eBHH}
\hat{H}_{\rm eBH} = &- t \sum_{\langle i,j \rangle} \; \hat{b}^\dag_i \hat{b}_j + \frac{U}{2}\sum_i \hat{n}_i(\hat{n}_i-1)  \\&+ \frac{V}{2} \sum_{\langle i,j \rangle}  \hat{n}_i\,\hat{n}_j - \sum_i \mu_i \hat{n}_i ,
\end{split}\end{equation}
 which is commonly referred to as \textit{extended} Bose-Hubbard model. Note that the sum over nearest-neighbors $\langle i,j \rangle$  leads to two identical terms
in off-site interaction $V$ for pairs $i,j$ and $j,i$. This is accounted for by the factor $1/2$ in the Hamiltonian. The dipolar Bose-Hubbard model with interactions not truncated to nearest neighbors is discussed at the end of this section. \xdl The particle number is fixed by the chemical potential $\mu_i$, which can be site-dependent, for instance due the presence of a trapping potential. For homogeneous systems, as discussed here, the chemical potential is constant, i.e.\ $\mu_i=\mu$. Slowly varying trapping potentials can be treated in the same framework by using the local density approximation. \xe 

For bosons, the phase diagram was intensely investigated in one dimension, 
where the superfluid to Mott-insulator transition is of  Berezinskii-Kosterlitz-Thouless (BKT) type \cites{Kuhner98,Kuhner2000}. The inclusion of nearest-neighbor interaction leads to a density-modulated insulating phase with crystalline, staggered diagonal order. 
Depending on the context, the phase is referred to as density-wave or charge-density wave (borrowed from electronic systems, where it is also used for metals with density fluctuations),  Mott crystal or Mott solid. In one dimension, the phase is denoted also as alternating or staggered Mott insulator, whereas in two dimensions it is often referred to as checkerboard phase (see Fig.~\ref{fig_lobes}b).
It was shown that there is a direct transition between the superfluid and the charge-density wave without an intermediate supersolid phase, showing superfluid \textit{and} crystalline order. Later it was realized that a bosonic Haldane insulator phase exists with non-local string correlations \cites{Rossini2012,Torre2006,Dalmonte2011,Deng2011}. While this gapped phase does not break the translational symmetry, particle-hole fluctuations appear in an alternating order. These fluctuations are separated by strings of equally populated sites. The corresponding phase diagram in one dimension and at filling (The density per site) $\rho=1$ is plotted in Fig.~\ref{Rossini2012Figure1}.

\begin{figure}
\centering
\includegraphics[width=0.8\linewidth]{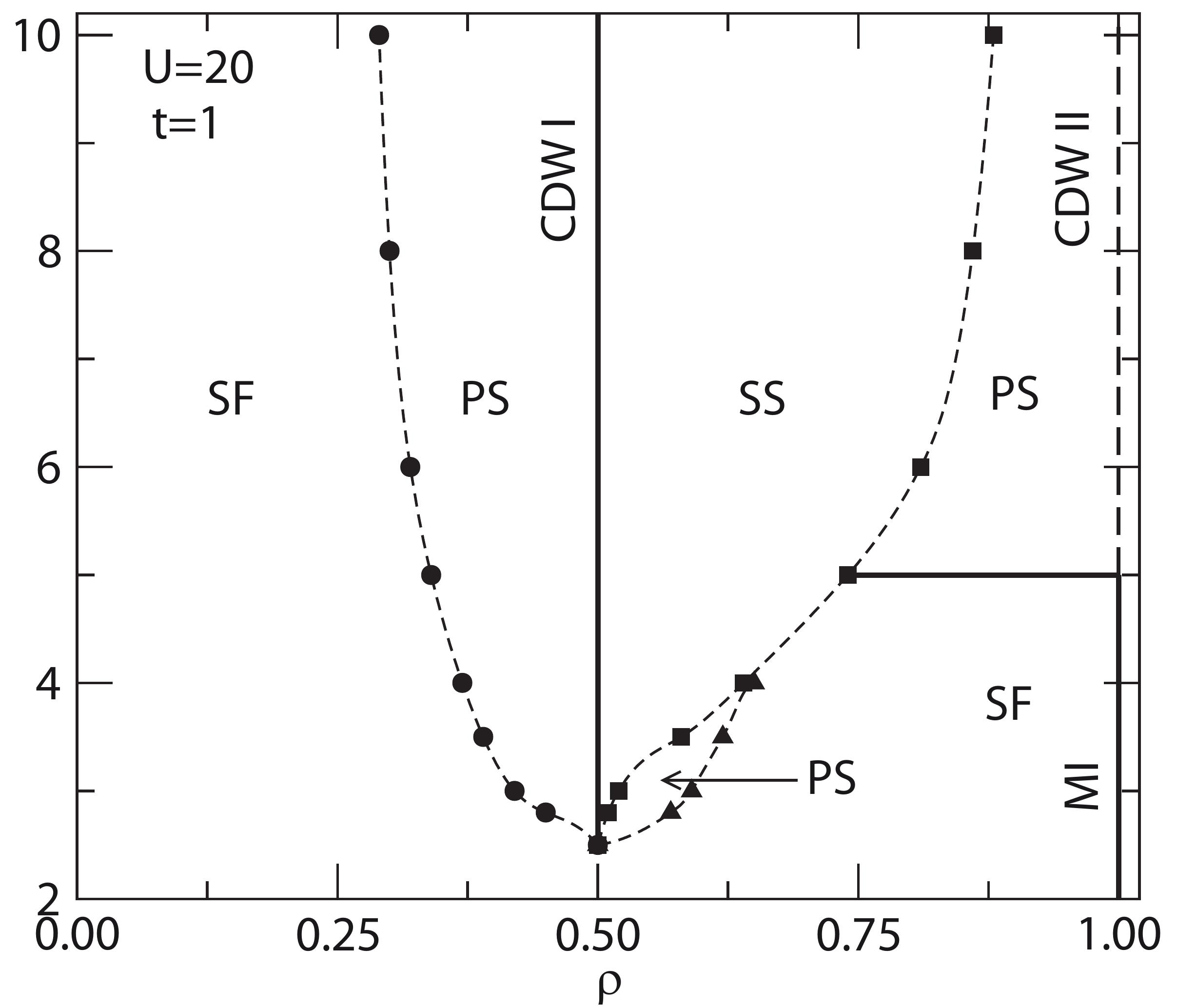}
\caption{The phase diagram in the filling $\rho-V$ parameter space  of the extended two-dimensional Bose-Hubbard model  $U = 20$. The energy unit is $t=1$. The thick solid vertical line indicates the charge-density wave (CDW I) at half filling; other phases present are the superfluid (SF), supersolid (SS), and at unit filling either Mott insulator (MI) or another charge-density wave (CDW II); PS denotes phase separated regions. Reprinted figure with permission from P.L.~Sengupta, P. Pryadko, F. Alet, M. Troyer, and G. Schmid, \href{http://link.aps.org/abstract/PRL/v94/p207202}{Phys. Rev. Lett. {\bf 94}, 207202 (2005)}. Copyright (2005) by the American Physical Society.
}
\label{maik:fig1-noT}
\end{figure}

For non-commensurate fillings the model is also quite rich. It has been studied using a quantum Monte Carlo
approach in two dimensions \cites{Sengupta05c} for fillings below unity. The phase diagram of the system for strong interactions $U$  is
reproduced in Fig.~\ref{maik:fig1-noT}. Two interesting novel phases appear. The elusive supersolid (SS) phase
shows a diagonal long-range order as revealed by a non-zero structure factor and simultaneously a non-zero superfluid density. As shown in Fig.~\ref{maik:fig1-noT}, additionally regions of phase separation (PS) appear, which are revealed as discontinuities (jumps) of the filling $\rho$ as a function of the chemical potential $\mu$ \cites{Sengupta05c}. When the on-site interaction becomes weaker, the SS phase becomes larger and PS regions disappear at filling larger than $1/2$ \cites{Maik13}. 
For half-integer and integer fillings an insulating charge-density wave (CDW) appears, which is also often referred to as checkerboard phase \cites{Sengupta05c,Batrouni06,Sowinski2012}. 
These findings were confirmed and further studied in one-dimensional Monte-Carlo \cites{Batrouni06} and DMRG analyses \cites{Mishra2009}. 
\begin{figure}
\centering
\includegraphics[width=0.8\linewidth]{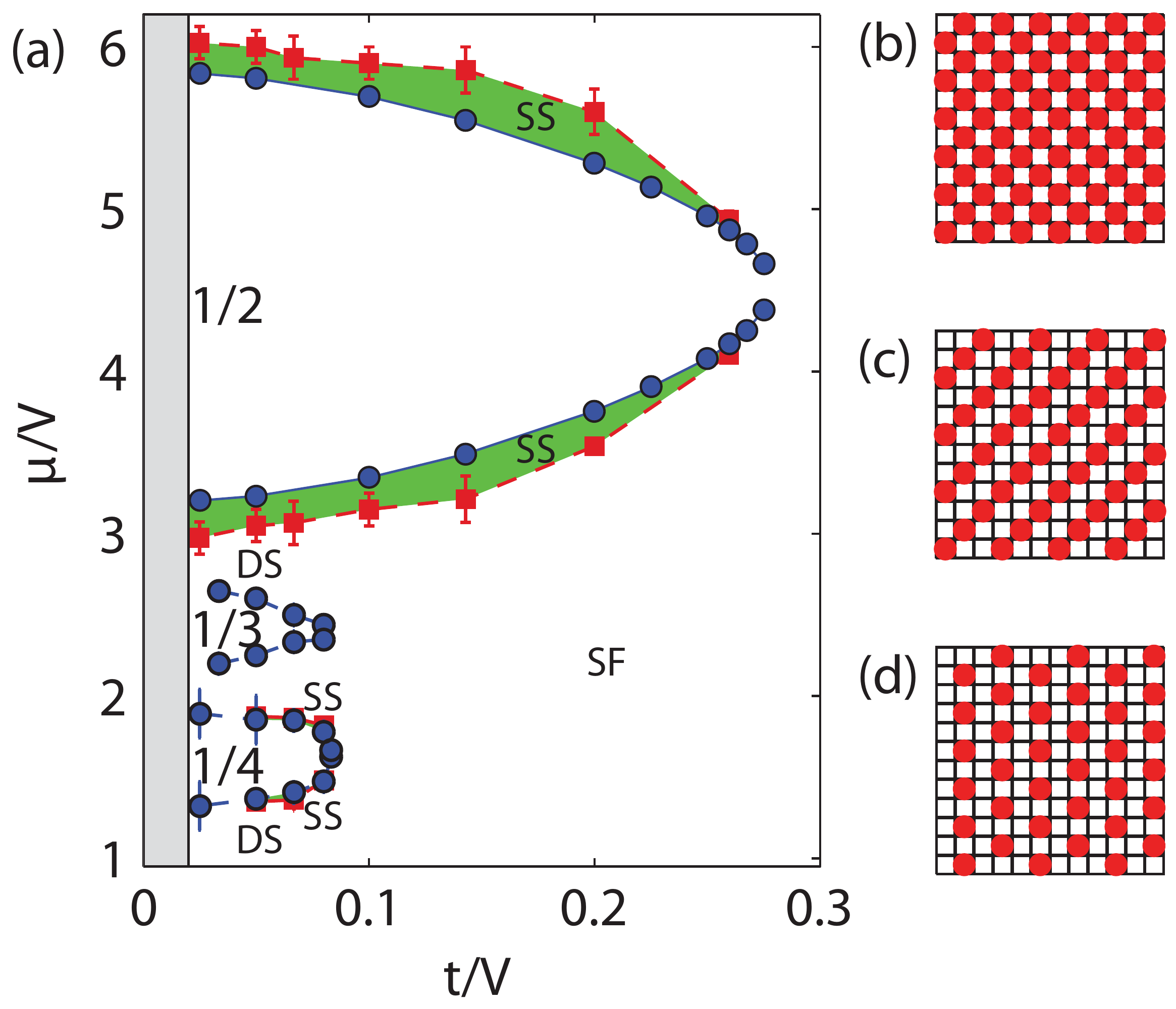}
\caption{The phase diagram of the dipolar two-dimensional Bose-Hubbard model in the hard-core limit. 
The lobes represent insulating density waves (also denoted as Mott solids) with densities indicated, SS denotes the supersolid phase, SF the superfluid phase and DS the devil's staircase.
The panels (b-d) are sketches of the ground-state configurations for the Mott solids with density (b) $\rho=1/2$ (checkerboard), (c) $\rho=1/3$ (stripe solid) and (d) $\rho=1/4$ (star solid). \from{Capogrosso10}}\label{fig_lobes}
\end{figure}
	
The phase diagram becomes even richer when the true long-range interactions for dipoles, Eq.~(\ref{vij}), are taken into account beyond nearest-neighbor interactions. \xdl The Hamiltonian reads then
\begin{equation}\begin{split}
\label{EQ:tdBHH}
\hat{H}_{\rm eBH} = &- t \sum_{\langle i,j \rangle} \; \hat{b}^\dag_i \hat{b}_j + \frac{U}{2}\sum_i \hat{n}_i(\hat{n}_i-1)  \\&+ \frac{1}{2} \sum_{ i\ne j }  \frac{V}{|i-j|^3}\hat{n}_i\,\hat{n}_j - \sum_i \mu_i \hat{n}_i .
\end{split}\end{equation}
Consider the case of low filling in the hard-core limit (with large on-site interaction $U$, excluding double occupancy).
 Such a case  was discussed in \cites{Capogrosso10} using large-scale Quantum Monte Carlo  (QMC) simulations. The Hamiltonian considered included the effects of a trap of frequency $\omega$, and was given by 
\begin{equation}
\label{EQ:htdBHH}
\hat{H}_{\rm eBH} = - t \sum_{\langle i,j \rangle} \; \hat{b}^\dag_i \hat{b}_j +  \frac{1}{2} \sum_{ i\ne j }  \frac{V}{|i-j|^3}\hat{n}_i\,\hat{n}_j - \sum_i  (\mu-\Omega i^2) \hat{n}_i ,
\end{equation}
with the requirement that the initial system
has no doubly occupied sites. The results are summarized in Fig.~\ref{fig_lobes}. \xe
For small-enough hopping $t/V \ll 0.1$, it is found that the low-energy phase
is incompressible ($\partial \rho / \partial \mu = 0$ with the filling factor $\rho$)
for most values of $\mu$.
This parameter region is denoted as DS in  Fig.~\ref{fig_lobes} and corresponds to the
classical {\it devil's staircase}. This is a succession of incompressible ground
states, dense in the interval $0 < \rho <1$, with a spatial structure
commensurate with the lattice for all rational fillings~\cites{Hubbard78,Fisher80} and no
analogue for shorter-range interactions. For finite $t$, three main Mott
lobes emerge with $\rho=1/2$, $1/3$, and $1/4$, named checkerboard, stripe, and star solids, respectively. Their ground-state configurations are visualized in Fig.~\ref{fig_lobes}(b-d). Interestingly, as found in \cites{Capogrosso10}  these phases
survive in the presence of a confining potential and at finite temperature.
Note that the shape of the Mott solids with $\rho=1/2$ and $1/4$ away from the tip of the lobe can be shown to be qualitatively captured by mean-field calculations, while this is not the case for the stripe solid at
filling 1/3 which has a pointy-like structure characteristic of
fluctuation-dominated 1D configurations. Mott lobes at other rational filling
factors, e.g. $\rho=1$ and $7/24$, have also been observed \cites{Capogrosso10},  but are not shown in the figure.
It is worth mentioning  that in the strongly correlated regime (at low $t/V$) the physics of the system is dominated by the presence of numerous metastable states resembling glassy systems and QMC calculations in this case become practically impossible. These metastable states were in fact correctly predicted by the generalized mean-field theory  \cites{Trefzger-Menotti}.

For large enough $t/V$, the low-energy phase is superfluid  for all values of the chemical potential $\mu$.
At intermediate values of $t/V$, however, doping the Mott solids
(either removing particles creating vacancies or adding extra particles) stabilizes
a supersolid phase, with coexisting superfluid and crystalline orders
(no evidence of this phase has been found in the absence of doping).
The solid/superfluid transition consists of two steps, with both transitions of second-order type and a supersolid as an intermediate phase. Remarkably, the long-range interactions stabilize
the supersolid in a wide range of parameters. For example, a vacancy supersolid is present
 for fillings $0.5> \rho \gtrsim 0.43$, roughly independent of the interaction strength.
 This is in contrast with typical extended Bose-Hubbard model results (compare Fig.~\ref{maik:fig1-noT}) where the supersolid phase appears only for $\rho>0.5$, i.e. no vacancy supersolid is observed. Similarly, the phase separation is not found when long-range interactions are taken into account \cites{Capogrosso10}.  Note, however, that in the former case soft bosons were considered while hard core bosons are studied in \cites{Capogrosso10}.

\xdl Let us note that it is still not a full story. As discussed above, the Hamiltonian \eqref{vij} is obtained assuming that
the dipolar potential changes slowly on scales of the width of the Wannier functions, $\sigma$. Corrections due to finite $\sigma$ have been discussed recently by Wall and Carr \cites{Wall2013}. These corrections lead to deviations from the inverse-cube power law at short and medium distances on the lattice scale---the dependence here is rather exponential with the power law recovered only for large distances. The resulting correction may be significant at moderate lattice depths and leads to quantitative differences in the phase diagram, as discussed in the one-dimensional case at unit filling \cites{Wall2013}.  The extent to which the full diagram is modified in 2D by these corrections is not yet known and is the subject of on-going studies. 
\xe

\xdl 

\subsubsection{Dipolar Fermi--Hubbard models}

The fermionic version of the extended Hubbard model (\ref{EQ:eBHH}) with nearest-neighbor interactions is also widely discussed in solid-state physics for both polarized (spinless) and spin 1/2 fermions (cf. 
\cites{Georges1996, Si2001,Gu2004,Raghu2008,Robaszkiewicz1981,Kivelson1987,Hirsch1984,Nasu1983}). There are much less papers on the 
model including the true long-range interactions for dipoles, described for spinless fermions by the Hamiltonian
\begin{equation}
\label{EQ:tdFHH}
\hat{H}_{\rm eFH} = - t \sum_{\langle i,j \rangle} \; \hat{f}^\dag_i \hat{f}_j +  \frac{1}{2} \sum_{ i\ne j }  \frac{V}{|i-j|^3}\hat{n}_i\,\hat{n}_j - \sum_i \mu_i \hat{n}_i .
\end{equation}
This model has been studied by Mikelsons and Freericks using a mean-field Ansatz \cite{Freericks}; in this way, a fermionic version of the phase diagram of Fig.~\ref{fig_lobes}(b-d) was derived for the homogenous case $\mu_i=\mu$. 

\begin{figure}
\centering
\includegraphics[width=0.7\linewidth]{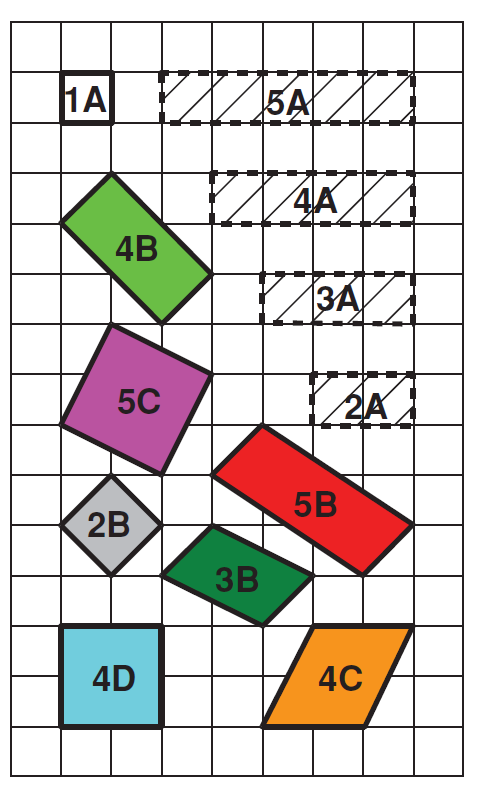}
\caption{ Unit cells corresponding to different  density wave phases. Vertexes indicate sites with higher density.
Only density wave orders corresponding to unit cells with the solid outline were found to be stabilized.  
Reprinted figure with permission from K. Mikelsons and J. K. Freericks, \href{http://link.aps.org/abstract/PRA/v83/p043609}{Phys. Rev. A{\bf 83}, 043609 (2011)}. Copyright (2011) by the American Physical Society.
}\label{fig_cdw_fermi}
\end{figure}

\begin{figure}
\centering
\includegraphics[width=0.9\linewidth]{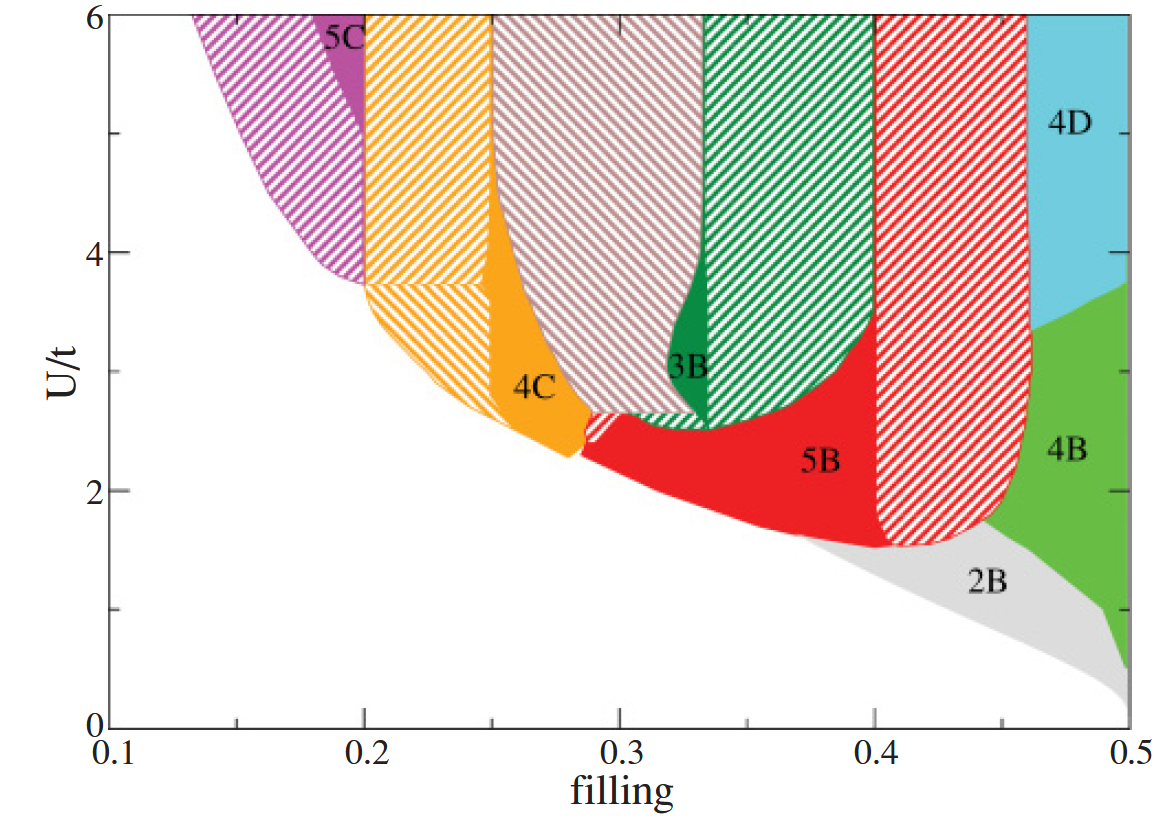}
\caption{ Phase diagram for $T=0$
including the phase separation (those regions are dashed).
 Decreasing the interaction
 contracts the range of filling of the ordered phases and progressively elliminates 
 phases commensurate with low values of filling. The only phase surviving down to $U/t=0$ is
 the checkerboard phase (2B).
 Phase separation replaces the 4D phase 
 near the filling $\rho=0.28$ and $\rho=0.36$ for larger $U/t$.
 In parts of the phase diagram, 4C and 5C phases show phase separation
 with the homogeneous state. Reprinted figure with permission from K. Mikelsons and J. K. Freericks, \href{http://link.aps.org/abstract/PRA/v83/p043609}{Phys. Rev. A {\bf 83}, 043609 (2011)}. Copyright (2011) by the American Physical Society.
 }\label{fig_lobes-Fermi}
\end{figure}
Mikelsons and Freericks solve the model using mean-field theory (MFT).
As they stress: ``This can be justified, since the the interaction is 
long range and consequently each site is effectively coupled to any other site.
In fact, due to the absence of a local interaction, 
the MFT is equivalent to the dynamical mean-field theory (DMFT) approach,
 which becomes exact in the infinite-dimensional limit. 
The absence of a spin degree of freedom also implies that 
the model is in the Ising universality class,
with a finite transition temperature in 2D.''
Within MFT one approximates the interaction part of the Hamiltonian by writing
\begin{equation}
\hat n_i \hat n_j \approx \hat n_i \langle n_j \rangle + \langle n_i \rangle \hat n_j 
 - \langle n_i \rangle \langle n_j \rangle \,,
\label{eq:MFT_approx}
\end{equation}
i.e.\ one neglects the density fluctuations, as it is done in the first-order (Hartree--Fock) self-consistent 
perturbation theory---it should be very accurate for small $U/t$.
In the MFT approximation, the mean density $\langle n_i \rangle$ is 
a fixed parameter in the Hamiltonian and acts as a site-dependent potential. 
The resulting MFT Hamiltonian is quadratic in the ($\hat f, \hat f^{\dag}$) operators and can 
be easily diagonalized for large, but finite lattices, especially assuming translational invariance at some level. 
MFT can  be regarded as a variational method and its results can be compared with another variational ansatz corresponding to phase separation. The results are presented in Figs.\ \ref{fig_cdw_fermi} and \ref{fig_lobes-Fermi}, where we present schematically unit cells, corresponding to different ``charge-desity wave'' orderings, and the phase diagram at zero temperature $T=0$.  

\xe


%
%
%
  
\section{Non-standard lowest-band Hubbard models}

\begin{figure*}
\centering\includegraphics[width=0.97\linewidth]{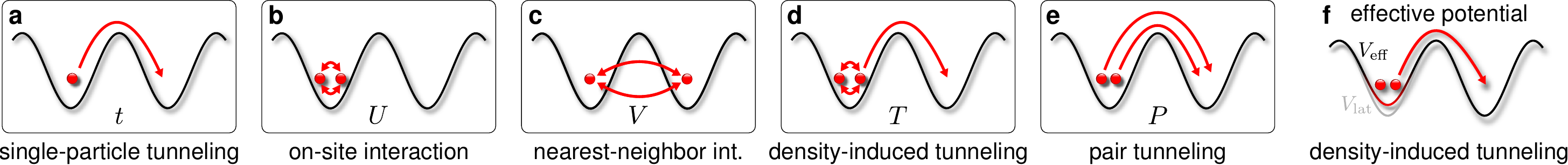}
\caption{On-site and nearest-neighbor off-site processes. In the Hubbard model, only (\textbf{a}) the tunneling $\J$ and (\textbf{b}) the on-site interaction $U$ are 
accounted for. Generalized Hubbard models can include (\textbf{c}) the nearest-neighbor interaction $V$,  
(\textbf{d})  the density-induced tunneling $\JBC$, and (\textbf{e})  the pair tunneling $P$. 
 The relative amplitudes of these processes depend on the interaction potential. They are plotted for contact interaction in \Fig{SingleBandAmplitudes} 
and for dipolar interaction in \Fig{DipoleAmp}.  
 (\textbf{f}) In optical lattices, the density-induced tunneling $\JBC$ has a relatively large amplitude and can therefore affect the tunneling
in the system. Effectively, it gives rise to a modified tunneling potential, which is shallower (shown here) for repulsive and
deeper for attractive interactions.\adaptedfrom{Jurgensen2012}}
\label{SingleBandProcesses}
\end{figure*}

\label{sec:sb_bose}

The original article on the Hubbard model has been published by J.~Hubbard in 1963 as a description of electrons in narrow bands \cites{Hubbard1963}. 
As discussed in \Sec{FermiHubbardModel}, in this framework the many-particle Hamiltonian is restricted to a tunneling matrix element $t$ and 
the on-site interaction $U$. Other two-particle interaction processes are considerably  smaller than the on-site term and are therefore neglected. Hubbard's article
gives also an estimation on the validity of the approximation (for common $d$-wave electron systems), where the (density-density) nearest-neighbor interaction $V$ is identified
as the first order correction (see \Sec{sec:EBHM}). \xdl  However, it was pointed out by F. Guinea, J.~E.~Hirsch and others 
\cites{Guinea88, Guinea88a,Guinea88b, Hirsch1989,Strack1993,Hirsch1994,Amadon1996} \xe that one of the neglected terms in
the two-body nearest-neighbor interaction describes the density mediated tunneling of an electron along a bond to a neighboring site. It therefore contributes to the tunneling  and was referred to as
\textit{bond-charge interaction} or \textit{density-induced tunneling}.  The main difference from the single-particle tunneling stems from the fact that the
operator depends on the density on the two neighboring sites. Strictly speaking, the simple Hubbard model is justified only if the bond-charge interaction is small
compared with the tunneling matrix element. It is worth noticing that bond-charge terms were already considered, although neglected, in the original paper of Hubbard of 1963. where he presented a non-perturbative approach based on decoupling of Green's  functions of   strongly interacting electron problem. Recently
Grzybowski and Chhajlany \cite{Ravi} applied the Hubbard method to a model
with strong bond-charge interaction term: these authors  divided the
tunneling terms into double-occupancy preserving and non-preserving ones, and treated the
latter as a perturbation.

For optical lattices, this density-induced tunneling \cites{Mazzarella2006,Mering2011,Luhmann2012,Jurgensen2012,Lacki2013} is of particular interest due to two points. 
First, unlike in solids, its amplitude can be rather large in optical lattices due to the characteristic shape of the Wannier functions for sinusoidal potentials. Second, the density-induced tunneling scales directly with
the filling factor, which enhances its impact for bosonic or multi-component systems.  In addition, ultracold atoms offer tunable interactions and differently
ranged interactions such as contact (\Sec{BEC}) or dipolar interaction potentials (\Sec{sec:EBHM}).

\xdl Before focusing on bosons, we start the discussion by reminding one of the classic papers on non-standard Fermi--Hubbard models. \xe 
In the following, different off-site interaction processes are discussed for bosons in optical lattices. We derive a generalized Hubbard model within the lowest band. 
Subsequently, the amplitudes of these off-site processes are calculated for both contact ($\delta$-function-shaped) interaction potentials and dipolar interactions. In the next sections, we focus on fermionic atoms and mixtures of different atomic species. 

\xdl \subsection{Non-standard Fermi--Hubbard models}

In order to give the reader an idea of what has been studied in the past in condensed-matter physics, we follow the 1996 paper by Amadon and Hirsch on  metallic ferromagnetism in a single-band model and effects of band filling and Coulomb interactions \cites{Amadon1996}. In this paper, the authors derive a single-band tight-binding model with on-site repulsion and nearest-neighbor exchange interactions as a simple model to describe metallic ferromagnetism. The main point  is to include
the effect of various other Coulomb matrix elements in the Hamiltonian that are expected to be of
appreciable magnitude in real materials. They compare results of exact diagonalization and mean-field theory in 1D. 
Quoting the authors: ``As the band filling decreases from 1/2, the tendency to ferromagnetism is
found to decrease in exact diagonalization, while mean-field theory predicts the opposite behavior. A nearest neighbor
Coulomb repulsion is found to suppress the tendency to ferromagnetism; however, the effect
becomes small for large on-site repulsion. A pair hopping interaction enhances the tendency to ferromagnetism.
A nearest-neighbor hybrid Coulomb matrix element breaks electron-hole symmetry and causes metallic
ferromagnetism to occur preferentially for more than half-filled rather than less-than-half-filled bands in this
model. Mean-field theory is found to yield qualitatively incorrect results for the effect of these interactions on
the tendency to ferromagnetism''. 

The starting point for the theory  is the single-band tight-binding Fermi Hamiltonian with all Coulomb
matrix elements included, 
\begin{equation}\begin{split}
 \hat{H}=& -\sum_{\langle i,j\rangle,\sigma}t_{ij}\left(\hat{f}_{i\sigma}^\dag \hat{f}_{j\sigma}+ {\rm h.c.}\right) \\
&+ \sum_{i,j,k,l, \sigma, \sigma'} \langle ij|1/r|kl\rangle \hat{f}_{i\sigma}^\dag \hat{f}_{j\sigma'}^\dag  \hat{f}_{l\sigma'}\hat{f}_{k\sigma},
\label{eq:FHaC}\end{split}
\end{equation}
where $\hat{f}_{i\sigma}^\dag$ creates  an electron of spin $\sigma$ in a Wannier orbital
at site i, which we denote $w_i(\bf r)$ . The Coulomb matrix elements
are given by the integrals
\begin{equation}
\label{VIntegralC}
 \langle ij|1/r|kl\rangle = \int  \ud{\bf r} \ud{\bf r'} w_i^*({\bf r})w_j^*({\bf r'})\frac{e^2}{|r-r'|} w_k({\bf r})w_l({\bf r'}).
\end{equation}
Restricting ourselves to only one- and two-center integrals
between nearest-neighbors, the following matrix elements
result:
\begin{eqnarray}
U&=&  \langle ii|1/r|ii\rangle, \\
V&=& \langle ij|1/r|ij\rangle, \\
J&=&  \langle ij|1/r|ji\rangle, \\
J'&=&  \langle ii|1/r|jj\rangle, \\
\Delta t&=&  \langle ii|1/r|ij\rangle.
\end{eqnarray}
As argued by the authors: ``Matrix elements involving three and four centers are likely to
be substantially smaller than these, as they involve additional
overlap factors. Even though the repulsion term  $V$ could
be of appreciable magnitude for sites further than nearest
neighbors, we assume that such terms will not change the
physics qualitatively''.

The resulting non-standard Fermi Hamiltonian reads 
\begin{equation}\begin{split}
 \hat{H}=& -\sum_{\langle i,j\rangle,\sigma}t^{\sigma}_{ij}\left(\hat{f}_{i\sigma}^\dag \hat{f}_{j\sigma}+ {\rm h.c.}\right) 
+U \sum_i
\hat{n}_{i\uparrow} \hat{n}_{i\downarrow} + V\sum_{\langle i,j\rangle}n_i n_j \\
& + J \sum_{\langle i,j\rangle,\sigma,\sigma'}  \hat{f}_{i\sigma}^\dag \hat{f}_{j\sigma'}^\dag  \hat{f}_{i\sigma'}\hat{f}_{j\sigma}
+ J' \hat{f}_{i\sigma}^\dag \hat{f}_{i\sigma'}^\dag  \hat{f}_{j\sigma'}\hat{f}_{j\sigma},
\label{eq:FHaCos}\end{split}
\end{equation}
with density dependent tunnelling
\begin{equation}
t^{\sigma}_{ij}=t_{ij}- \Delta t(\hat n_{i,-\sigma} + \hat n_{j,-\sigma}).
\end{equation}
In the situation considered in Ref. \cites{Amadon1996} ``all matrix elements in the above expressions are expected to be always
positive, except possibly for the hybrid matrix element $\Delta t$''.
However, with the convention that the single-particle hopping matrix
element $t$ is positive and that the operators describe electrons
rather than holes, the sign of $\Delta t$  is also
expected to be positive. This should be contrasted with the situations that we can approach with boson, see below. \xe

\subsection{Non-standard Bose--Hubbard models with density-induced tunneling}


\label{GHB}

We consider the same system as before with Hamiltonian (\ref{hamiltonian}) and optical lattice potential
(\ref{pot}), and restrict the Wannier functions expansion to the lowest Bloch band (\ref{wannier_field}) 
\xdl using the same procedure as in \Sec{cha:standardHub}. \xe
While previously we provided some heuristic arguments to drop various contributions of the interaction potential, we shall keep presently all the terms (restricting ourselves, however, to nearest neighbors only). 
For a general potential $V({\bf r}-{\bf r'})$ define
\begin{equation}
\label{VIntegral}
 V_{ijkl} = \int  \ud{\bf r} \ud{\bf r'} w_i^*({\bf r})w_j^*({\bf r'}) V({\bf r} - {\bf r'}) w_k({\bf r})w_l({\bf r'}).
\end{equation}

 The generalized lowest-band Hubbard Hamiltonian reads then
\begin{eqnarray}\label{modHamlow}
\!\!\!
\hat H_\text{GBH}&=	&- \J \sum_{\langle i,j \rangle} \hat b_i^\dagger \hat b_j 
			+ \frac{U}{2} \sum_i \hat n_i (\hat n_i -1) 
			+ \frac{V}{2} \sum_{\langle i,j \rangle} \hat n_i \hat n_j, \nonumber \\
			&&	- \JBC \sum_{\langle i,j \rangle} \hat b_i^\dagger ( \hat n_i + \hat n_j ) \hat b_j 
				+ \frac{P}{2} \sum_{\langle i,j \rangle}  \hat b_i^{\dagger 2} \hat b_j^{2}.
\end{eqnarray}
All contributing processes in this model are sketched in \Fig{SingleBandProcesses}.
The third term represents the nearest-neighbor interaction $V=V_{ijij}+{V}_{ijji}$, which was already introduced in \Sec{sec:EBHM}. 
Recall that the sum over nearest-neighbors $\langle i,j \rangle$  leads to two identical terms  $\hat n_i \hat n_j$ and  $\hat n_j \hat n_i$.
The forth term $\JBC=-({V}_{iiij}+{V}_{iiji})/2$ also origins from the interaction. 
  As illustrated in \Fig{SingleBandProcesses}, it constitutes a hopping process between neighboring sites and therefore directly affects the tunneling $\J$ in the lattice system. 
 This process is known as density- or
interaction-induced tunneling, density-dependent tunneling, and correlated tunneling, depending on the context. 
In the condensed-matter literature, this tunneling is also known as bond-charge interaction. The
last term $P={V}_{iijj}$ denotes pair-tunneling amplitude of the process when a pair of bosons hops from one site to the neighboring site. To get a general idea about the
relative importance of these terms we look into systems with i) contact interactions and ii) with contact and dipolar interactions. 

\begin{figure}
\centering\includegraphics[width=0.75\linewidth]{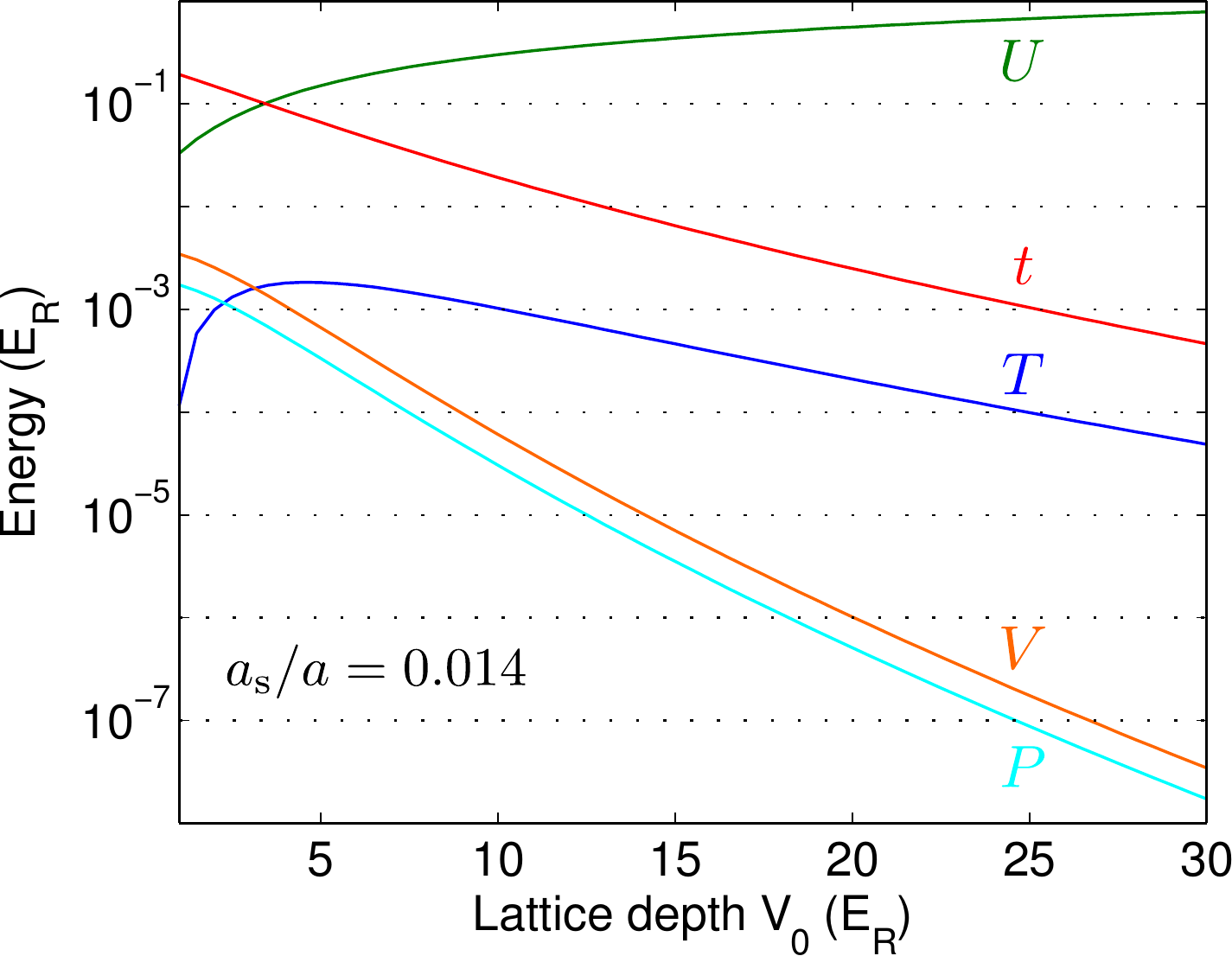}
\caption{(a) Lowest-band parameters  for the on-site interaction $U$, the tunneling $\J$, the interaction-induced tunneling $\JBC$, nearest-neighbor interaction $V$, and pair tunneling $P$. All interaction processes scale linearly with the scattering length $a_s/a$, whereas the tunneling $\J$ is unaffected. The amplitudes are plotted for an isotropic 3D optical lattice with lattice depth $V_0$ and scattering length $a_s/a=0.014$. 
\adaptedfrom{Luhmann2012} } 
\label{SingleBandAmplitudes}
\end{figure}


\subsubsection{Bosons with contact interaction} 
\label{DIT_contact}

Let us start with  correlated processes for ultracold bosonic atoms interacting via a contact interaction
 ${V}(\mathbf{r-r'})=g\delta(\mathbf{r-r'})$. Here, we assume an isotropic three-dimensional
optical lattice with lattice depths $V_x=V_y=V_z=V_0$. 
In the units of the recoil energy $\ER=h^2/(8ma^2)$, where $a$ is the lattice constant, and for Wannier functions $w_{\mathbf{i}}(\mathbf{r})$ in lattice coordinates $\mathbf{r}  \to \mathbf{r}/a $, the interaction integral can be expressed as
\begin{equation}
  U_{ijkl}=\frac{8}{\pi} \frac{a_s}{a} \int w^{0 *}_{\mathbf{i}}(\mathbf{r})w^{0 *}_{\mathbf{i}}(\mathbf{r})
w^{0}_{\mathbf{i}}(\mathbf{r})w^{0}_{\mathbf{j}}(\mathbf{r}) d\mathbf{r}.  \label{contactUijkl}
\end{equation}
This integral gives rise to various contributions, the on-site interaction $U_c$, next-neighbor interaction $V_c$, density-induced tunneling $\JBC_c$, 
and pair tunneling $P_c$ given by (with the subscript $c$ denoting, as before, contact interactions):
\begin{equation}\begin{split}
U_c / \ER &= U_{iiii}, \\
V_c / \ER &= U_{ijij}+U_{ijji}=2U_{ijij},\\
\JBC_c / \ER &= - (U_{iiij}+U_{iiji}) / 2 = - U_{iiij},\\
P_c / \ER &= U_{iijj}.
\label{contactT}
\end{split}\end{equation}
\xdl Since in this part we shall consider contact interactions only, we drop the subscript $c$ in the following for convenience. We shall reintroduce it later, when also dipolar interactions will be discussed. \xe 
From the integral expression, we see that the 
 amplitudes are proportional to the effective scattering length $a_s/a$ and depend solely on properties of the Wannier functions.  All amplitudes are plotted in \Fig{SingleBandAmplitudes}, where one sees that the on-site interaction $U$ is the dominating energy.  For neutral atoms, the nearest-neighbor interaction $V$ and the pair tunneling amplitude $P$ are much smaller than both $U$ and the (single-particle) tunneling amplitude $t$ (for $V_0\gtrsim 10 \ER$). However,  the amplitude, $T$,  of the density-induced channeling
\begin{equation}
   \label{eq:JBC}
    \hJBC = -\JBC \, \hat{b}_i^\dagger (\hat{n}_i + \hat n_j) \hat{b}_j 
	\end{equation}
is considerably larger than  $V$ and  $P$.
Due to the structure of this operator, we can combine it with the conventional single-particle tunneling $\J$ to an effective hopping 
\begin{equation}
   \label{eq:JBCtot}
	\hat\J_\eff=   	
	- \left[ \J + \JBC(\hat n_i + \hat n_j-1) \right] \hat{b}^{\dagger}_i \hat{b}_j.  
\end{equation}
Although this density-dependent hopping is small in comparison with the on-site interaction $U$, it can constitute a substantial contribution to the  tunneling process.
For repulsive interactions, as depicted in \Fig{SingleBandAmplitudes}, the value of $\JBC$ is positive and thus increases the magnitude of the overall tunneling, whereas attractive interactions decrease the overall magnitude.

The process of the density-induced tunneling \eqref{eq:JBCtot} can also be illustrated within an effective potential picture \cites{Luhmann2012} by inserting
the explicit expressions for the integral $\JBC_c$ \eqref{contactUijkl} and the tunneling amplitude $\J$ \eqref{t}. The term $\hat n_i + \hat n_j-1$ corresponds to the density
$n_\BC(\mathbf r)= n_i |w_i\ofr|^2 + {(n_j-1)}|w_j\ofr|^2$ on sites $i$ and $j$ excluding the hopping particle. The effective hopping operator~\eqref{eq:JBCtot} can then be written as  
 \begin{equation}
	\label{eq:effectivePotential}
	\hat\J_\eff=\!\int\! d^3 r\ w^{*}_i \!\left(\frac{\mathbf{p}^2}{2m}+V\ofr +g n_\BC \ofr \right)\! w_j\,  
	\hat{b}^{\dagger}_i \hat{b}_j.
\end{equation} 
Here, $V\ofr +g n_\BC\ofr$ can be identified as an effective tunneling potential, which is illustrated in \Fig{SingleBandProcesses}f. Since the density $n_\BC\ofr$ is maximal at the lattice site centers, the effective tunneling potential corresponds to a shallower lattice for repulsive interactions and therefore causes an increased tunneling. In this effective potential, the band structure and the Wannier functions are altered. Such a modified band structure was experimentally observed in optical lattices for an atomic Bose-Fermi mixture \cites{Heinze2011} (see \Sec{sec:sb_bose_fermi}). 

\begin{figure}
\centering\includegraphics[width=0.97\linewidth]{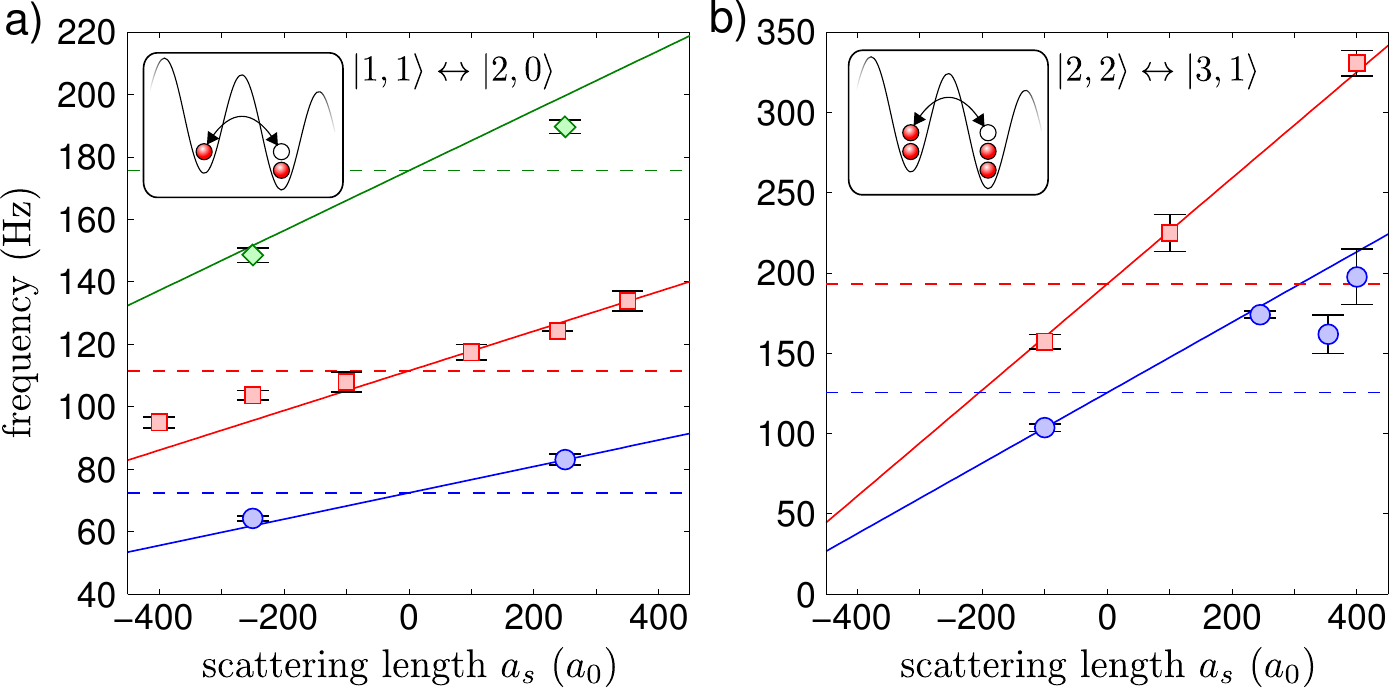}
\caption{The oscillation frequency of the doublon number in a tilted optical lattice in dependence on the scattering length $a_s$ for a filling (a) $n=1$ and (b) $n=2$.  The data sets are obtained for optical lattices with $V_x=V_y=20\ER$ and  $V_z=8\ER$ (green), $10\ER$ (red), and $12\ER$ (blue).  The scattering length is tuned via a Feshbach resonance, where $a_0$ is the Bohr radius. The solid lines represent the theoretical prediction $\nu_n  (\J+(2n-1)\,\JBC  )/h$, where $\JBC$ is proportional to the scattering length $a_s$ and $\nu_n$ is a prefactor for the resonant oscillation. The constant dashed line corresponds to the single-particle tunneling, i.e.\ $\nu_n  \J/h$.  \from{Jurgensen2014b}  }
\label{DensityInducedTunnelingExperiment}
\end{figure}

For standard $^{87}\mathrm{Rb}$ parameters\footnote{A scattering length of $\as=100\,a_0$ at a lattice spacing of $a=377\,\mathrm{nm}$ corresponds to $\as/a=0.014$.}, 
the bare amplitude $\JBC$ reaches roughly $10\%$ of the tunneling amplitude $\J$  for deep lattices (see \Fig{SingleBandAmplitudes}).  In addition, the density-induced tunneling scales with the particle number on neighboring
sites as  $n_i + n_j - 1$. At a filling factor of $n=3$, the correction is about $30\%$ at the superfluid to Mott-insulator transition point.  Note that all amplitudes except the tunneling $\J$ scale linearly with the interaction strength. 
By using Feshbach resonances to change the interaction strength, the amplitudes of  $\JBC$ and $U$ can be tuned independently from the lattice depth and thus the tunneling $\J$. In contrast, for contact interactions the ratio $\JBC/ U$ is only a function of the lattice depth.

The direct detection of density-induced tunneling was performed in an optical lattice experiment with Cs atoms and tunable interactions \cites{Jurgensen2014b,Meinert2013}. Here, a Mott insulator, prepared in a quasi one-dimensional lattice, is tilted by an offset energy $\epsilon$ per lattice site. By quenching the lattice into tunneling resonance, where the additional on-site energy $U$ of a hopping particle equals the tilt $\epsilon$, resonant oscillation can be observed (see insets of \Fig{DensityInducedTunnelingExperiment}). 
Due to the compensation of the dominating on-site interaction, the oscillation frequency is a direct measure for the (total) tunneling $\hat\J_\eff$ \eqref{eq:JBCtot} thereby revealing interaction effects on its amplitude. 
Figure~\ref{DensityInducedTunnelingExperiment} shows the observed oscillation frequency as a function of the interaction strength for filling factors $n=1$ and $n=2$. It shows the linear dependency of the density-dependent tunneling on both the scattering length $\JBC\propto a_s/a$ and the density  $\JBC\propto 2n - 1$. The solid lines depict the theoretical prediction for $\hat\J_\eff$, whereas the constant dashed line corresponds to single-particle tunneling in the standard Hubbard model.

 As a direct consequence of the density-induced tunneling, the critical point of the superfluid to Mott-insulator transition is affected depending on both scattering length and filling factor, and the transition is shifted towards deeper lattices for repulsive interactions. 
Since the nearest-neighbor interaction $V$ and the pair tunneling $P$ have very small amplitudes (\Fig{SingleBandAmplitudes}),  for neutral atoms we can neglect their contributions in the following.
 Mean-field theory allows to demonstrate how the interaction-induced tunneling affects the ground-state phase diagram of the generalized Bose-Hubbard Hamiltonian $\hat H_\text{GBH} -\mu \sum_i \hat n_i$ (Eq.~\eqref{modHamlow}) with $V=P=0$ , where $\mu$ is the chemical potential. In mean-field theory, a superfluid order parameter
	$\psi=\expect{\hatt b_i}=\expect{\hatt b_i^\dagger}$ is introduced, where $\psi\neq0$ corresponds to the superfluid phase (SF) and $\psi=0$ defines the Mott insulator (MI) with a fixed particle number per lattice site \cites{Oosten2001,Fisher1989}.
The decoupling of the lattice sites is achieved by neglecting the fluctuations between $\hatt b_i^\dagger$ and $\hatt b_j$ of quadratic order, i.e.,  
\begin{equation}
\hatt{b}_i^\dagger \hatt{b}_j \approx \hatt{b}_i^\dagger \hatt{b}_j - (\hatt{b}_i^\dagger - \expect{\hatt{b}_i^\dagger})(\hatt{b}_j - \expect{\hatt{b}_j}) =  \psi (\hatt{b}_i^\dagger + \hatt{b}_j) - \psi^2.
\end{equation}
 Analogously, the density-induced tunneling can be decoupled via 
 \begin{equation}
	\hatt{b}_i^\dagger \hatt{n}_i \hatt{b}_j + \hatt{b}_i^\dagger \hatt{n}_j \hatt{b}_j 
	\approx \psi(\hatt{b}_i^\dagger \hatt{n}_i + \hatt{n}_j \hatt{b}_j). \label{eq:DITDecoupling}
\end{equation}
disregarding terms of the order $\psi^3$ \cites{Luhmann2012}. 
With the decoupling above, on can perform second-order perturbation theory in $\psi$ of a Mott lobe with $n$ particles per site (see \cites{Oosten2001}).

\begin{figure}
\centering\includegraphics[width=0.8\linewidth]{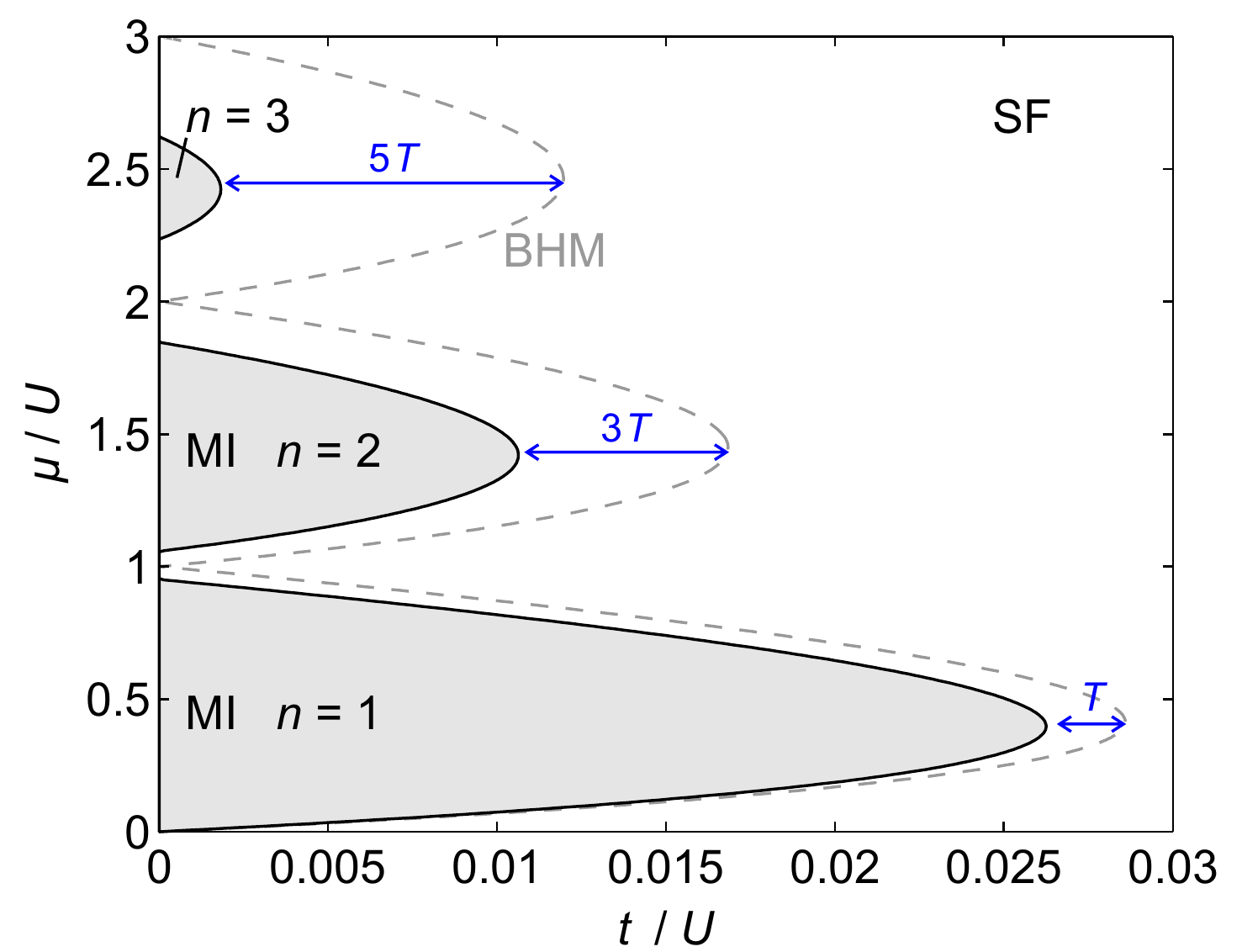}
\caption{Phase diagram of the generalized Bose--Hubbard model with density-induced tunneling $\JBC=0.002 U$ in an isotropic three-dimensional lattice with $z=6$ nearest-neighbors. It shows the transition from the superfluid phase (SF) to Mott-insulator phases (MI) with filling $n$. Between the Mott lobes at $J/U\to0$ the density-induced tunneling prohibits the Mott-insulating state. The dashed line depicts the results of the standard Bose--Hubbard model (BHM). The blue arrows indicate the occupation-dependent change of the overall tunneling by $(2n-1)\JBC$ for filling $n$ at the tips of the Mott lobes, which captures the main impact of the density-induced tunneling. The phase diagram is calculated by means of a Gutzwiller mean-field theory. Note that the second-order perturbation \eqref{eq:DITDecoupling} \xdl with a single mean-field parameter \xe fails at large values of $\JBC$.  
}
\label{BoseSingleBandPhases}
\end{figure}

 The results are plotted in \Fig{BoseSingleBandPhases} for the Bose-Hubbard model (dashed line) and the generalized Hubbard model 
with $\JBC/U=0.002$ (solid line). Although $\JBC$ is much smaller than $U$, the transition from the superfluid to the Mott-insulator phase is significantly shifted towards lower values of $\J/U$. The occupation-dependent nature of $\hJBC= -\JBC_c \hat{b}_i^\dagger (\hat{n}_i + \hat n_j) \hat{b}_j $ is reflected by the fact that the Mott lobes with higher filling factors $n$ are more strongly affected. In fact, for the given example lobes with $n\geq 4$ do not exist. The effect of interaction-induced tunneling can be mainly captured by the change of the overall tunneling as indicated by Eq.~\eqref{eq:JBCtot}.
For a filling factor $\hat n_i \to n$, the generalized and standard Bose-Hubbard model differ approximately by $(2n-1)\JBC$ at the tips of the Mott lobes. Below and above the tips hole and particle excitations, respectively, become more probable at the phase boundary. Thus, the shift of the Bose--Hubbard Mott lobes by the density-induced tunneling interpolates between the tips and can be approximated well by $2\JBC\,\mu/U$. Note that this type of phase diagram can be achieved experimentally by keeping the lattice depth $V_0$ (and therefore $\J$) fixed and tuning $\J/U$ by a Feshbach resonance.   

The modified phase diagram with a fixed interaction strength but variable lattice depth is shown in \Fig{BoseFermiSingleBandPhases} for $^{87}\mathrm{Rb}$ parameters in a three-dimensional lattice. The curve for bosons only corresponds to vanishing boson-fermion scattering length,  $a_\BF= 0 a_0$.  
The density-induced tunneling in combination with multi-orbital processes is discussed in \Sec{sec:MO}.    
In addition to ground-state properties, the density-induced tunneling influences also the dynamic behavior, which is discussed, e.g.\ in Refs.~\cites{Lacki2013,Lacki2013b}.


\subsubsection{Bosons with dipolar interaction} 
\label{DIT_dipolar}

For bosons with dipolar interaction, the situation can change drastically. For simplicity, we assume that the dipoles are polarized along the $z$-direction. 
We will consider two-dimensional lattice geometries with the potential given by \eqref{trap2d}, where the Wannier function along the $z$ direction is just  a harmonic oscillator ground-state eigenfunction, such that there is no aggregation of atoms or molecules along the $z$-direction due to attracting interactions. 
As for contact interactions, we work in dimensionless units by scaling the distance with respect to the lattice constant $a=\lambda/2$, i.e.\ $\pi x/a\rightarrow x$, and assume the recoil energy as a natural energy unit. Then, the interaction potential reads
\begin{equation}
{U}_{\rm dd}(\mathbf{r})= \mathcal{D}\frac{1-3\cos^2\theta}{r^3},
\end{equation}
[compare (\ref{fullpot})]
with the effective dipolar strength denoted by $\mathcal{D}=\frac{d^2 m}{2\pi^3\epsilon_0\hbar^2 a}$, where $d$ is the dipole moment of the polar molecules, 
and $\epsilon_0$ the vacuum permittivity. For atoms, the dipolar (magnetic) strength $\mathcal{D}=\frac{\mu_0\mu^2 m}{2\pi^3\hbar^2 a},$ where $\mu$ is the magnetic dipole moment and $\mu_0$ denotes the vacuum permeability. 

\begin{figure}[b]
\centering\includegraphics[width=1\linewidth]{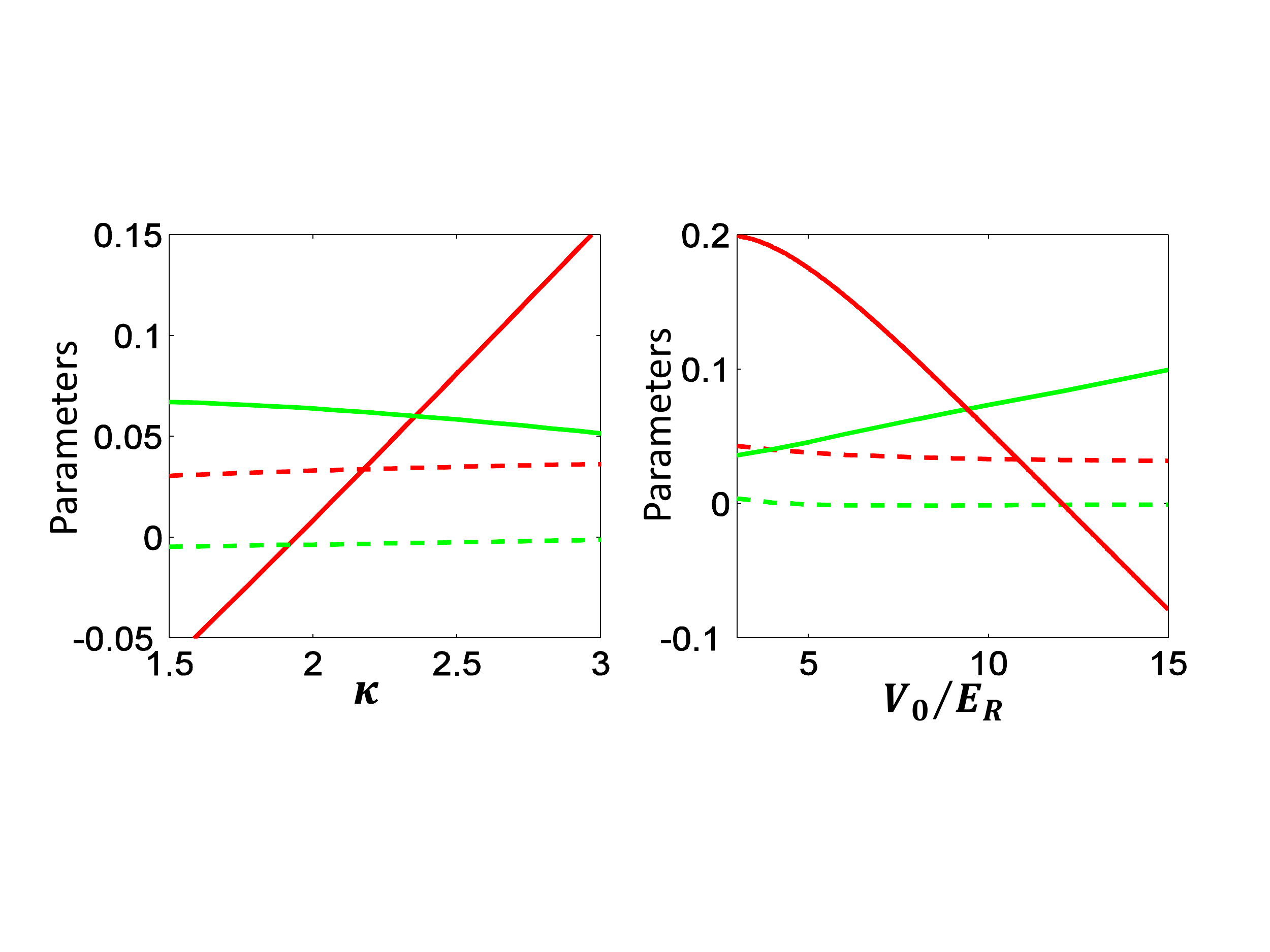}
\caption{The change of interaction parameters as (a) a function of the trap flattening $\kappa$ for lattice depth of $V_0=6E_R$, and (b) as a function of
lattice depth $V_0/E_R$ for a fixed trap flattening parameter $\kappa=3$. The red-solid line denotes on-site interaction $U_{\rm dd}/E_R$, the red-dashed 
line denotes nearest-neighbour interaction $V_{\rm dd}/E_R$, the green-solid line denotes interaction-induced tunneling $T_{\mathrm{dd}}/J$, and the green-dashed 
line denotes pair-tunneling amplitude $P_\mathrm{dd}/J$.  A sketch of the different processes can be found in \Fig{SingleBandProcesses}. 
}
\label{DipoleAmp}
\end{figure}

Dipolar interactions act together with contact interactions, affecting, e.g.\ the nearest-neighbour interactions and correlated 
tunneling amplitudes. 
The Hamiltonian now takes the form (\ref{modHamlow}) with the parameters $U=U_\text{c}+U_\text{dd}$,
$V=V_\text{c}+V_\text{dd}$, $P=P_\text{c}+ P_\text{dd}$ and $\JBC=\JBC_\text{c}-T_\text{dd}$ (we have reintroduced the subscript $c$ for contact interactions contribution). Note the minus sign between density-induced tunneling contributions due to
definitions in (\ref{contactT}) and (\ref{dipoleparameter}) below. 
Further on, we omit the terms with $V_\text{c}$ and $P_\text{c}$, as they are very small (see preceding section).
By means of the dipolar interaction integral 
\begin{equation}
D_{ijkl} = \mathcal{D}\int w^*_i(\mathbf{r})w^*_j(\mathbf{r'}) U_{\rm dd}(\mathbf{r-r'})
w_k(\mathbf{r'})w_l(\mathbf{r}) d\mathbf{r}d\mathbf{r'}\,,
\end{equation}
we can express the dipolar on-site ($U_{\rm dd}$), the nearest-neighbor interactions ($\V_{\rm dd}$), the density-induced tunneling ($T_\text{dd}$), and pair tunneling amplitude  ($P_\text{dd}$)  by 
\begin{equation}\begin{split}
U_\text{dd} / \ER &= D_{iiii}, \\
V_\text{dd} / \ER &= D_{ijij}+D_{ijji},\\
T_\text{dd} / \ER &=   D_{iiij},\\
P_\text{dd} / \ER &= D_{iijj},
\label{dipoleparameter}
\end{split}\end{equation}
where the amplitudes again vary linearly with the strength ${\cal D}$. But now the proportionality constant depends on the shape of the Wannier functions and the dipolar interaction. Thus, by tuning the lattice parameters and the trap frequency one only changes the proportionality factors in Eq.~\eqref{dipoleparameter}. We illustrate this dependency in Fig.~\ref{DipoleAmp} for a dipolar strength $\mathcal{D}=1$. To take into account the effect of the harmonic trap along the $z$-direction, we introduce a trap flattening parameter
$\kappa=\frac{\hbar\Omega}{2E_R}$. We see from Fig.~\ref{DipoleAmp}(a) that for a fixed lattice depth, the on-site interaction $U_{\rm dd}$ decreases with decreasing  $\kappa$ and 
\begin{figure}
\centering\includegraphics[width=0.9\linewidth]{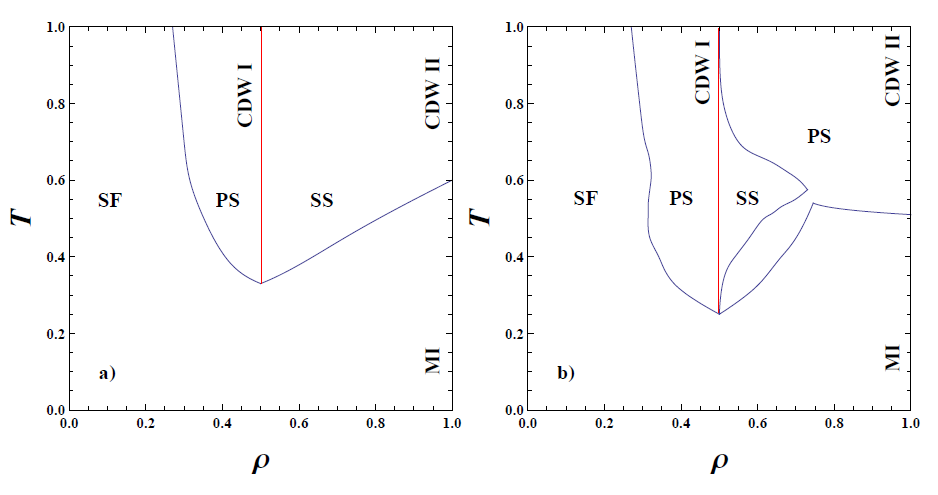}
\caption{The phase diagrams for $U = 20$ at finite $\JBC$ (with $V=10|\JBC|$ and $\J=1$ the unit of energy).
(a) If $\J$ and $\JBC$ are of the same sign, the relative importance of interactions decreases, leading to the disappearance of PS phases at greater than half filling. Compared to the $\JBC=0$ cases (\Fig{maik:fig1-noT})  this phase diagram resembles more the case for low $U=5$ . 
(b) If $\JBC$ and $\J$ compete due to opposite signs, the relative importance of interactions is enhanced, increasing the PS regions while supersolid phase region shrinks. \from{Maik13} }
\label{fig:maikt}
\end{figure}
when $\kappa$ becomes lower than a critical value, the interaction strength $U_{\rm dd}$ becomes negative. This critical points corresponds to the situation when the 
width of Wannier function is roughly equal to the trap length along $z$ direction. This shows that the on-site interaction strength can be tuned by changing the trap flattening ratio $\kappa$. On the other hand, the nearest-neighbor interaction $V_{\rm dd}$ does not show a strong dependence on the trap flattening $\kappa$, since it is mainly controlled by  the distance between the lattice sites. The single-particle correlated tunneling amplitude $T_{\rm dd}$, however, varies strongly with the lattice flattening, and it is positive even if the interactions are effectively repulsive. The plotted parameter in Fig.~\ref{DipoleAmp} for $\mathcal{D}=1$ corresponds to weak polar molecules, whereas for a dipole strength $\mathcal{D} \sim 10$ one can reach the limit where the correlated tunneling is of the same order as the single-particle tunneling  $T_{\rm dd} \sim t$. The total density-induced tunneling amplitude (defined above as $\JBC=\JBC_\text{c}-T_\text{dd}$) may thus change sign depending on $V_0$ or $\
kappa$.  Notice also that the pair-tunneling amplitude is much smaller than other parameters present in the Hamiltonian. In Fig.~\ref{DipoleAmp}(b), we plot the 
parameters 
as a function of the lattice depth $V_0/E_R$ for a fixed trap flattening ratio $\kappa=3$. Here also the parameters follow the same trend, as $T_{\rm dd}$ is positive and the on-site interaction decreases as the lattice depth gets stronger. As soon as the width of the Wannier functions along the $x-y$ plane becomes similar to the oscillator width along the $z$ direction, the on-site strength vanishes as before and an increase in the lattice depth leads to an attractive on-site interaction.  \xdl The sensitivity of the interaction parameters to the geometry of lattice sites described above was originally discussed in \cites{Sowinski2012}. It was also noticed and discussed in detail in \cites{Wall2013}, where it is shown that the effects become even more dramatic in a reduced quasi-1D geometry.\xe
\begin{figure}[b]
\centering\includegraphics[width=1\linewidth]{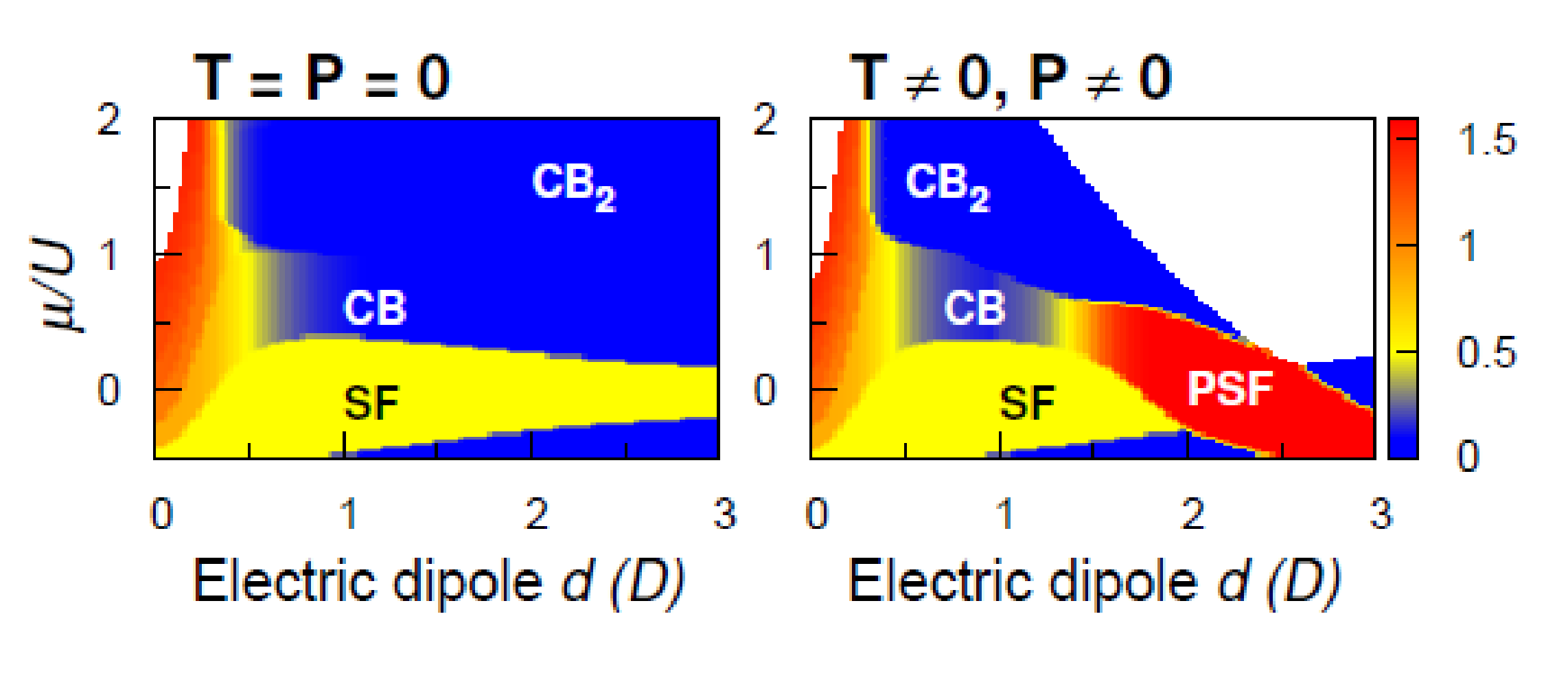}
\caption{ \label{polarphase} Phase diagram without (left) and with (right)
taking into account $T$ and $P$ for a model of polar molecules with parameters defined in the text. The color quantifies the superfluidity indicators $\phi_i$ and $\Phi_i$ (see text). Neglecting $T$ and $P$, for large
enough $d$ and $\mu$ the system is always in an insulating phase and
the average number of particles is a multiple of $1/2$. CB (CB$_2$)
denotes a checkerboard phase where sites with 0 and 1 (2) particles
alternate. Including the new terms, the insulating phases vanish for
large enough $d$, and pair superfluidity (PSF) appears. \from{Sowinski2012}
}
\label{Fig4}
\end{figure}

To appreciate the effect of the density-induced tunneling $\JBC$ for the physics of the extended model, we first consider the noncommensurate case for a two-dimensional system, with exemplary results presented in Fig.~\ref{fig:maikt}. The complete analysis obtained using Quantum Monte Carlo simulations can be found in \cites{Maik13}.
Following the discussion comparing the order of magnitude of different terms above, it is assumed that $|\JBC|\approx V/10$ (with either positive or negative sign). Since pair hopping $P$ is usually much smaller, it is omitted ($P=0$). Thus, Fig.~\ref{fig:maikt} presents the effects due to density-dependent tunneling, as compared with Fig.~\ref{maik:fig1-noT}. Observe that, while for the contact interactions the density-induced tunneling shifts the borders between different phases, for dipolar interactions these additional contributions may lead to a disappearance of the phase separation (PS) phase. A similar behavior appears for $T=0$ for smaller values of $U$  (compare \cites{Maik13}). 

 While the above analysis was carried out for chosen parameter values, one may also assume specific atomic parameters. Consider \cite{Sowinski2012} an ultracold gas of
dipolar molecules confined in an optical lattice with lattice depth
$V_0 = 6 E_R$, mass $m=127$a.m.u.\, and $\lambda = 790\,\mathrm{nm}$. We assume the $s$-wave scattering length
of the molecules,  to be $a_s \approx 100a_0$. For these parameters, $g \approx 1.06$ is approximately constant. We consider dipole moments
$d$ up to $\sim 3\,\mathrm{D}$ (Debye), which can be achievable for molecules like bosonic $\mathrm{RbCs}, \mathrm{KLi}$ \cites{Voigt2009}. We also
choose the lattice parameter $\kappa\approx 1.95$, making
(additionally to $\J$) the on-site interaction $U$ almost independent of the
dipole moment ($U_{\mathrm{dd}} \approx 0$). In this case, for a large
enough dipolar strength $\mathcal{D}$, we expect that with increasing $d$ the parameters
$V$, $T$, and $P$ determine the system properties. 
For clarity, we restrict ourselves to a 1D chain of $N$ lattice sites with periodic
boundary conditions. We analyze the influence of the additional terms $T$ and $P$ on the grand-canonical phase diagram, where the particle number is not conserved.
For this, we add a chemical potential term $-\mu\sum_i \hat{n}_i$ to
Hamiltonian (\ref{modHamlow}).
In Fig.\ \ref{polarphase}, we present the phase diagram as well as the average number of
particles per site for exact-diagonalization  calculations of 4 sites with occupation truncated at 4 particles per site. 
Without the modified terms, with increasing dipolar strength the system becomes insulating with checkerboard order (above also referred to as charge-density wave) due to the increased 
nearest-neighbour repulsion. The right hand plot reveals that  the inclusion of the density-induced tunneling changes 
the phase diagram. A novel pair-superfluid phase  arises (characterized by a non-zero pair-superfluid order parameter $\sum \langle \hat b_i^2\rangle$) as one increases
the dipolar strength.  Since in the exact diagonalization the particle number is conserved, superfluid and pair-superfluid phases are not identified with the typical order parameters but rather with large first, $\phi_i=\sum_j\langle\hat b^\dagger_j \hat b_i\rangle$, and second,  $\Phi_i=\sum_j\langle\hat b^\dagger_j\hat b^\dagger_j \hat b_i\rangle \hat b_i\rangle$, correlation functions, respectively. Apparently, a sufficiently large tunneling  $T$ destroys the insulating checkerboard  phase, making place for a pair superfluid.

\subsection{Density-induced tunneling in fermionic systems}

\label{sec:sb_fermi}

For neutral fermionic atoms in optical lattices with spin $\sigma=\{-\frac12, \frac12\}$, the Fermi-Hubbard model (\Sec{FermiHubbardModel}) is altered in a similar way as described in the previous section. 
When omitting the pair tunneling $P$, the generalized Fermi-Hubbard Hamiltonian reads
\begin{equation}\begin{split}
\!\!\!
 \hat H&=	- \J \sum_{\langle i,j \rangle, \sigma} \hat \f_{i\sigma}^\dagger \hat \f_{j\sigma} 
			+ U_\text{c} \sum_i \hat n_{i\up} \hat n_{i\down}  
			+ \frac{V_\text{c}}{2} \sum_{\langle i,j \rangle} \hat n_i \hat n_j \\
			&\ - \JBC \sum_{\langle i,j \rangle, \sigma} \hat \f_{i\sigma}^\dagger  \hat \f_{j\sigma} ( \hat n_{i,-\sigma} + \hat n_{j,-\sigma} ) 
		     + \frac{P}{2} \sum_{\langle i,j \rangle}  \hat \f_i^{\dagger 2} \hat \f_j^{2}, 
\end{split}\end{equation}
with matrix elements $U_c=U_{iiii}$, $\JBC=-U_{iiij}$, $V_c =2 U_{ijij}$, and $P = U_{iijj}$ \ in units of $\ER$ as defined above and $\hat n_i=\hat n_{i\up} + \hat n_{i\down}$ \cites{Hirsch1989}. 
The density-induced tunneling 
$- \JBC \,\hat \f_{i\sigma}^\dagger  \hat \f_{j\sigma} ( \hat n_{i,-\sigma} + \hat n_{j,-\sigma} ) $
gives rise to three different situations for a hopping fermion from site $i$ to $j$ with spin $\sigma$. Depending on the filling of the other spin component, the energy gain for a hopping particle 
\begin{equation}
	\hat\J_\eff=   	
	-\sum_{\langle i,j \rangle, \sigma} \left( \J  + \JBC   ( \hat n_{i,-\sigma} + \hat n_{j,-\sigma} ) \right) \hat \f_{i\sigma}^\dagger  \hat \f_{j\sigma}   
\end{equation}
can either be $\J_0=\J$,  $\J_1=\J+\JBC$, or $\J_2=\J+2 \JBC$. 

\subsection{Non-standard Bose--Fermi--Hubbard models}

\label{sec:sb_bose_fermi}

The density-induced tunneling discussed above plays also an important role in multi-component systems, where either the atoms have a spin degree of freedom or represent different atomic species. In particular, the interspecies interaction directly induces tunneling within both components. 
 Here, the most interesting case of a mixture of bosonic and fermionic species is discussed. However, several aspects can be transferred to other multi-component systems. The Bose--Fermi--Hubbard model presented in the following describes the case where bosonic and fermionic species are spin-polarized and interact via contact interaction. The experimental realizations of atomic mixtures of bosonic and fermionic particles in optical lattices 
\cites{Gunter2006,Ospelkaus2006,Best2009,Heinze2011} have triggered a vivid discussion about the role of inter- and intra-species interactions
\cites{Luhmann2008b,Lutchyn2009,Mering2011,Cramer2008,Cramer2011,Jurgensen2012}.  

The standard Bose-Fermi-Hubbard Hamiltonian \cites{Albus2003} is given by
\begin{eqnarray}
\!\!\!
	 \hat H_\text{BFH}&=\! & 	-\!  \sum_{\expect{i,j}} (\J_\B \hat b_i^\dagger \hat b_j   
												\!+\! \J_\F  \hat \f_i^\dagger \hat \f_j )
												+\frac{U_\BB}{2}  \sum_i \hat n_i (\hat n_i-1) \nonumber  \\
										&&	+ \!\sum_i U_\BF \hat n_i \hat m_i  
												- \sum_i (\mu_\B \hat n_i + \mu_\F \hat m_i ),
\label{eq:BFHM}\end{eqnarray}
where $\J_\B$ is the tunneling matrix element for bosons and $\J_\F$ for fermions. The intra- and inter-species interaction is restricted to  the on-site interaction $U_\BB$ and $U_\BF$, respectively. Here, $\hat b_i$ ($\hat \f_i$) is the bosonic (fermionic) annihilation operator and $\hat n_i$ ($\hat m_i$) the respective particle number operator, where the total number of bosonic and fermionic atoms are fixed by the chemical potentials $\mu_\B$ and $\mu_\F$. Let us assume for simplicity that the fermions are in a perfect band-insulator phase where Pauli-blocking prohibits the fermionic tunneling. This freezes out the fermionic degrees of freedom and the resulting Hamiltonian captures the behavior of the bosonic component under the influence of exactly one fermion per lattice site ($\langle \hat m_i\rangle=1$). Consequently, the Bose-Fermi-Hubbard Hamiltonian simplifies to an effective bosonic Hamiltonian 
\begin{equation}\begin{split}
	 \hat H_\text{FBI}=& 	-  \sum_{\expect{i,j}} \J_\B \hat b_i^\dagger \hat b_j   
												+\frac{U_\BB}{2}  \sum_i \hat n_i (\hat n_i-1)  \\
										&		+\sum_i (U_\BF-\mu_\B) \hat n_ i  .
\end{split}\label{eq:FBI_BFHM}\end{equation}
In this case, the interaction energy $U_\BF$ between bosons and fermions can be fully absorbed into an effective chemical potential $\mu_\eff=\mu_\B-U_\BF$. Hence, the resulting effective Hamiltonian does not differ from the standard Bose-Hubbard model except for a modification of the chemical potential. As a consequence, the behavior of the bosons is not influenced by the homogeneously distributed fermions, which is in contradiction to the experimental observations \cites{Ospelkaus2006,Gunter2006,Best2009}. Omitting the band-insulator assumption above and taking into account the experimental confinement has also only little influence \cites{Pollet2008}. 
Therefore, extended interspecies processes must play a role that are not covered in the Bose-Fermi-Hubbard model \cites{Mering2011,Jurgensen2012}.

The off-site processes arising from the boson-boson interaction (see \Fig{SingleBandProcesses}) are elaborated in section \ref{GHB}. The Bose-Fermi interaction leads to additional distinct processes, since the interacting particles are distinguishable, such as the cross tunneling, where bosonic and fermionic particles interchange sites. For the density-induced tunneling, either a boson 
or a fermion can tunnel. 

\begin{figure}
{\centering
\includegraphics[width=0.8\linewidth]{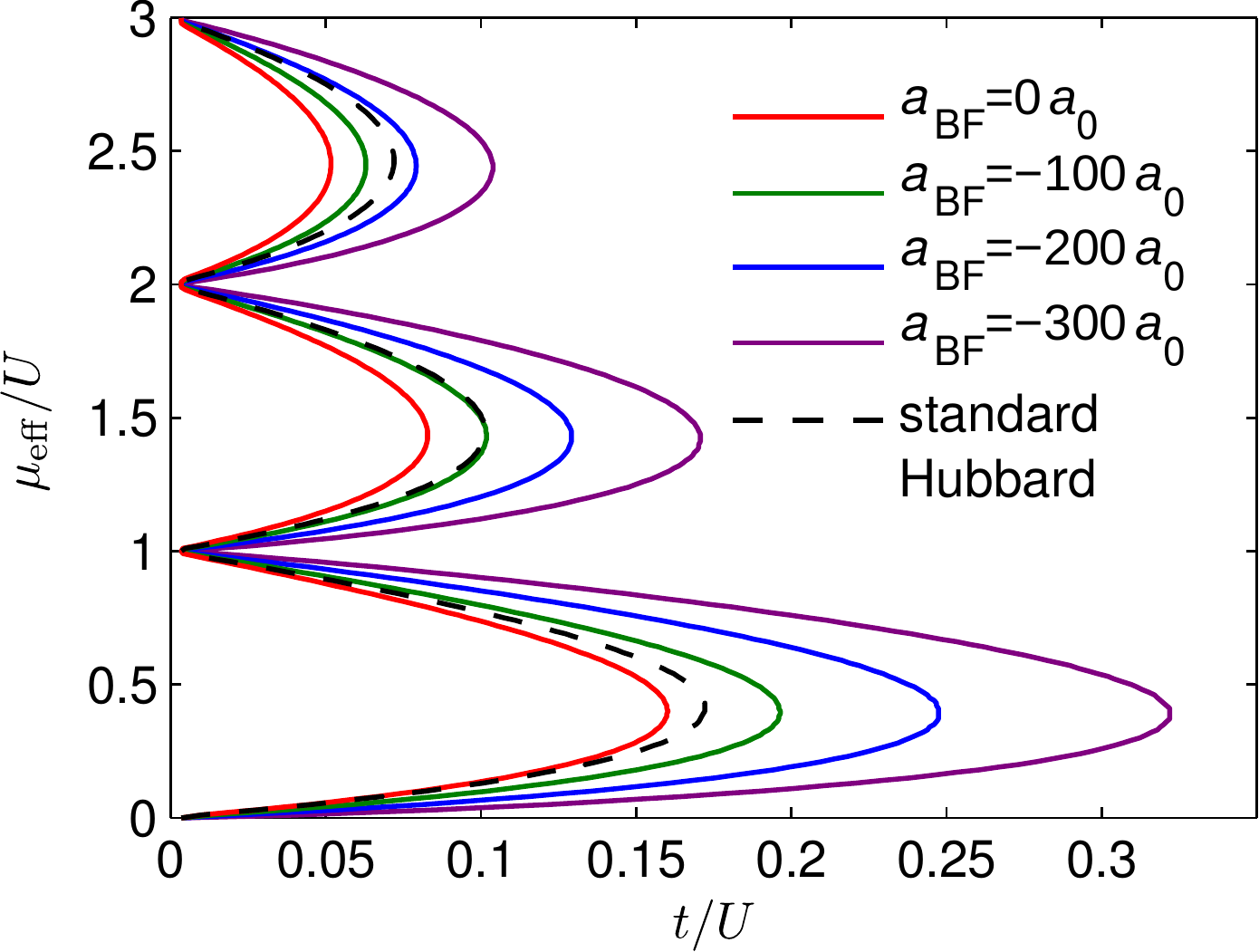}}
\caption{Phase diagram of the effective bosonic Hamiltonian \eqref{eq:extended_BFHM} with density-induced tunneling for different Bose-Fermi scattering lengths, $a_\text{BF}$,  within Gutzwiller mean-field theory. For comparison, the results of the standard Hubbard model are shown as a dashed black line. The calculation is performed for bosonic $^{87}\mathrm{Rb}$ and fermionic $^{40}\mathrm{K}$ in an optical lattice with a spacing of $a=377 \, \mathrm{nm}$ (experimental parameters of Ref.~\cites{Best2009}).
For the considered wavelength, the Wannier functions of both species are almost identical. The interaction between the bosonic atoms is fixed to a repulsive scattering length of $a_\BB=102\, a_0$ \cites{Will2010}, while the attractive interaction between the two species is tunable over a wide range using a Feshbach resonance \cites{Ferlaino2006, Best2009}.\from{Jurgensen2012}
\label{BoseFermiSingleBandPhases}
}
\end{figure}

However, for a fermionic band insulator all processes that involve the hopping of a fermion are forbidden. 
In this case, only on-site interactions and the bond-charge tunneling of bosons have to be taken into account, since other processes are prohibited or contribute only with small amplitudes (compare \Fig{SingleBandAmplitudes}).
The generalized, effective Hubbard model of the lowest band including these processes reads~\cites{Jurgensen2012}
\begin{equation}\begin{split}
 \hat H=& - \sum_{\expect{i,j}} \big[\J_\B\!+\! \JBC (\hat n_i + \hat n_j - 1) \!+\! 2\, \JBCBF \big]\hat b_i^\dagger \hat b_j\\
&+\frac{U_\BB}{2}  \sum_i \hat n_i (\hat n_i-1) - \mu_\eff \sum_i\hat n_i,
\end{split}\label{eq:extended_BFHM}\end{equation}
with the density-induced tunneling $\JBC$ mediated by boson-boson interaction (defined in \Eq{contactT}) and 
$\JBCBF= g_{\BF} \dr  w^{\B*}_i\ofr w^{\F*}_i\ofr w_i^\F\ofr w_j^\B\ofr $ mediated by the interspecies interaction.
The interaction parameter is  $g_\BF = \frac{2 \pi \hbar^2}{m_\mathrm{r}} a_\BF$, where  $m_\mathrm{r}$ is the reduced mass  and $a_\BF$ the interspecies scattering length. 
While the repulsive interaction between the bosons increases the total tunneling, the fermions reduce or enhance the bosonic mobility
depending on the sign of the boson-fermion interaction. 
As a consequence, and in strong contrast to the predictions of the standard Hubbard model, the superfluid to Mott-insulator transition is affected and the phase boundaries are shifted depending on the interspecies interaction strength. The phase diagram is shown in \Fig{BoseFermiSingleBandPhases} for different attractive Bose-Fermi interaction strengths. For strong Bose-Fermi attraction and low bosonic filling, the transition occurs at much shallower lattices, since the total tunneling is reduced. In the picture of an effective potential (see section \ref{DIT_contact}), this corresponds to a deeper tunneling potential. The effect is reversed when the repulsion between the bosons becomes stronger than the attraction to the fermions, which is the case for weaker Bose-Fermi interaction and higher bosonic filling. In this case, the effective tunneling potential is shallower and tunneling is enhanced. \xdl The Mott-insulator transition in Bose-Fermi systems is discussed further in \Sec{MOHamiltonians}.  \xe

%
%

\section{Multi-orbital Hubbard models}

\label{sec:MO}

Along with the off-site interactions discussed in the last section, taking into account higher bands is an important extension of standard Bose--Hubbard models. In the Hubbard model, only the lowest single-particle band is assumed to be occupied, since higher bands are energetically separated. 
In strongly correlated systems, the interaction-induced coupling between the orbital bands is, however, strong enough that higher bands are mixed with the lowest band. Due to their dominating contribution to the total energy, the orbital occupation is determined by on-site interaction processes. 
Within a mean-field treatment (\Sec{MOMFT} ), the occupation of higher orbitals corresponds to a modified on-site wave function of the particles in order to minimize the on-site interaction energy. Due to the population of higher orbitals, also the effective wave-function overlap on neighboring lattice sites changes. As a consequence, the tunneling amplitude is modified and becomes occupation-dependent.

First, we formulate a multi-orbital Hubbard model to define appropriate notation in \Sec{MOH}.
After a mean-field description for the orbital degrees of freedom (\Sec{MOMFT}), the correlated many-particle on-site problem is discussed (\Sec{MOOnSite}). The results can directly be used to compute the orbital dressing of off-site processes (\Sec{MODressing}). This leads intrinsically to occupation-dependent Hamiltonians (\Sec{MOHamiltonians}). Hubbard models where particles are only confined to higher bands of the lattice are discussed in \Sec{orbitalhub}. \xdl The analysis presented below is restricted to interacting bosons only that were studied in detail. Effects of higher bands for impurities embedded in a one-dimensional sea of fermions in a periodic potential  were considered in \cites{Doggen2014}. \xe

\subsection{Multi-orbital Hubbard models} \label{MOH}

Again the basic Hamiltonian in second quantization is given by \Eq{hamiltonian}. Now, however, we
expand the atom field taking explicitly excited bands into account 
\beq \label{wanmult}
\hat{\Psi}(\mathbf{r})=\sum_{i,\alpha} \hat{b}^{\alpha}_{i} w^{\alpha}_{i}(\mathbf{r}),
\eeq
where $w^{\alpha}_i(\mathbf{r})$ is a Wannier function of the band $\alpha$ localized at site $i$ while 
$\hat{b}^{\alpha\dagger}_{i}, \hat{b}^{\alpha}_{i}$ are the creation and annihilation operator for a boson at site $i$ and energy band $\alpha$. The single particle part of the Hamiltonian (\ref{hamiltonian}) yields tunnelings and energies in different orbitals
\begin{eqnarray} \label{eq:HubbardOrbitalTunnelingAndEnergy}
\J^{\alpha}_{ij} &=& -\int w^{\alpha *}_{i}(\mathbf{r}) \left [ -\frac{\hbar^2}{2m}\nabla^2 + V_{\rm ext}(\mathbf{r}) \right ] w^{\alpha}_{j}(\mathbf{r}) d\mathbf{r} ,\\
\epsilon^{\alpha}_{i} &=& \int w^{\alpha *}_{i}(\mathbf{r}) \left [ -\frac{\hbar^2}{2m}\nabla^2 + V_{\rm ext}(\mathbf{r}) \right ] w^{\alpha}_{i}(\mathbf{r}) d\mathbf{r}.
\end{eqnarray}
Similarly the interaction part of the Hamiltonian may be expressed as 
\begin{equation}\label{genHam}
H_{\rm int}= \frac{1}{2}\sum_{\alpha\beta\gamma\delta}\sum_{ijkl} U^{\alpha\beta\gamma\delta}_{ijkl} 
\hat{b}^{\alpha\dagger}_{i}\hat{b}^{\beta\dagger}_{j}\hat{b}^{\gamma}_{k}\hat{b}^{\delta}_{l}
\end{equation}
with the interaction integrals
\beq \label{intint}
U^{\alpha\beta\gamma\delta}_{ijkl}=\int w^{\alpha *}_{i}(\mathbf{r})w^{\beta *}_{j}(\mathbf{r'})V(\mathbf{r-r'})
w^{\gamma}_{k}(\mathbf{r'})w^{\delta}_{l}(\mathbf{r})d\mathbf{r}d\mathbf{r'}.
\eeq
Combining different terms we obtain the multi-orbital Hubbard model in its full glory. The summations over site indices may be, as before, limited to nearest neighbors, but e.g.\ the tunneling between next-nearest-neighbor sites can also be included in the model depending on the specific problem or lattice geometry.
 
The full description of lattice and orbital degrees of freedom captured in a multi-orbital Hubbard model leads to an extremely complex many-particle problem. Also, for very strong interactions it may lead to convergence problems (see discussions in e.g.\ \cites{Lacki2013} and references therein). The goal of this section is therefore rather to define effective Hubbard models within a single band. This \textit{interaction-dressed} band includes the orbital degrees of freedom and can be treated by common single-band methods for lattice models.  The individual processes such as on-site interaction, tunneling, and density-induced tunneling are affected and renormalized by this treatment.  

\subsection{Mean-field description of higher orbitals}  \label{MOMFT}

As described in \Sec{sec:sb_bose}, we find that different types of extensions of the  Hubbard
model become relevant when the interaction between the particles is enhanced,
e.g.\ by means of a Feshbach resonance or by reducing the lattice constant. When the interaction is sufficiently
weak compared to the lattice potential, the bosonic system can be approximately
modeled using lowest-band single-particle Wannier states, which are localized at the minima of the lattice.
Under these conditions, the Hubbard interaction $U$ and
tunneling parameter $J$ are given by respective matrix elements with respect
to the single-particle Wannier states. This approximation breaks down for stronger interaction as the interaction-induced coupling to higher energy Wannier states starts playing a role. 

To describe such a system,
one can introduce modified Wannier-like orbitals with a dependence on the lattice-site occupation numbers $n_{j}$. Such
Wannier-like orbitals will have admixtures from higher bands, depending on
the occupation. The most significant effect of the repulsive interaction
will be a broadening of the Wannier-like orbitals with increasing
occupation, effectively enhancing $J$ and decreasing $U$. In terms of the
Hubbard description, we take this into account by replacing $J$ and $U$ by
functions $J_{\hat{n}_{i},\hat{n}_{j}}$ and $U_{\hat{n}_{i}}$ of the
number operators $\hat{n}_{i}$. Quantitative consequences of this kind of
modification to the plain bosonic Hubbard model have been studied by
several authors at a theoretical level \cites{Li2006, Johnson2009, Hazzard2010, Dutta2011}. Considering an
interaction-induced modification of the Wannier functions, also additional
Mott-insulator phases have been predicted \cites{Alon2005}. Variational time-dependent approach in which Wannier functions
adopt dynamically to lattice dynamics and interactions has been proposed \cites{Sakmann2011}. Unfortunately this original approach does not take efficiently into account the interaction induced multiparticle entanglement being, at the present stage, inferior to the multi-oribital expansion \cites{Major2014}. 
In Ref. \cites{Larson2009}, the effect of the interaction-induced coupling to the first
excited band on the Mott transition was considered. Re-entrant behaviour in the superfluid-Mott
transition has also been predicted due to the interaction-induced modification of Hubbard parameters \cites{Larson2009, Cetoli2010}.
The effect of interaction on the tunneling dynamics in one-dimensional double-well and triple-well potentials have been studied, e.g., in 
Refs. \cites{Zollner2008, Cao2011} where the authors found enhanced correlated pair tunneling near the fermionization limit.

For bosons with contact interaction, we rewrite the total Hamiltonian in terms of the field operators as
\begin{eqnarray}
\hat{H} &=& \int \hat{\Psi}^{\dagger}(\mathbf{r}) \left [ -\frac{\hbar^2}{2m}\nabla^2 + V_{\rm ext} \right ] \hat{\Psi}(\mathbf{r}) \nonumber\\
&+& \frac{g}{2}\int\hat{\Psi}^{\dagger}(\mathbf{r})\hat{\Psi}^{\dagger}(\mathbf{r}) 
\hat{\Psi}(\mathbf{r})\hat{\Psi}(\mathbf{r})d\mathbf{r}.
\end{eqnarray}
To derive a Hubbard-type description, the field operators
$\hat{\Psi}(\mathbf{r})$ are expanded in terms of Wannier-like orbitals
$\omega_i(\mathbf{r},\hat{n}_i)=\omega(\mathbf{r-R_i},\hat{n}_i)$
localized at the lattice minima
$\mathbf{R_i}$, namely $\hat{\Psi}(\mathbf{r})
=\sum_i \hat{b}_i \omega_i(\mathbf{r}, \hat{n}_i)$ with bosonic
annihilation and number operators $\hat{b}_i$ and
$\hat{n}_i=\hat{b}^\dagger_i\hat{b}_i$. Note that the ``wave function''
$\omega_i$ depends on the number operator $\hat{n}_i$ in order to take into
account interaction-induced occupation-dependent broadening. Keeping only the 
on-site interaction as well as the density-induced tunneling, we arrive at the effective single-band Hamiltonian
\begin{eqnarray}  \label{hub2}
\hat{H} & = & -\sum_{\langle i,j \rangle} \J_{\hat{n}_i,\hat{n}_j}\hat{b}^{\dagger}_ib_j - 
 \sum_{\langle i , j \rangle} \left [ T^1_{\hat{n}_i,\hat{n}_j} \hat b^\dagger_i \hat n_i \hat b_j + 
T^2_{\hat{n}_i,\hat{n}_j} \hat b^\dagger_i \hat n_j\hat b_j \right ] \nonumber\\
&+& \frac{1}{2}\sum_i U_{\hat{n}_i} \hat{n}_i(\hat{n}_i-1),
\end{eqnarray}
where
\begin{eqnarray}
\J_{\hat{n}_i,\hat{n}_j}&=&-\int d\mathbf{r}\, \omega_i(\mathbf{r}, \hat{n}_i)
\Big[-\frac{\hbar^2}{2m} \nabla^2   
+V_{\mathrm{ext}}(\mathbf{r}) \Big] \omega_j(\mathbf{r}, \hat{n}_j+1),  \nonumber \\
U_{\hat{n}_i}&=& g \int d\mathbf{r}\, \omega^2_i(\mathbf{r}, \hat{n}_i)\omega^2_i(\mathbf{r}, \hat{n}_i-1), \nonumber\\
T^1_{\hat{n}_i,\hat{n}_j} &=& -g \int d\mathbf{r}\, \omega_i(\mathbf{r}, \hat{n}_i+1) \omega^2_i(\mathbf{r}, \hat{n}_i)\omega_j(\mathbf{r}, \hat{n}_j), \nonumber\\
T^2_{\hat{n}_i,\hat{n}_j} &=& -g \int d\mathbf{r}\, \omega_i(\mathbf{r}, \hat{n}_i+1) 
\omega^2_j(\mathbf{r}, \hat{n}_j-1)\omega_j(\mathbf{r}, \hat{n}_j).
\end{eqnarray}

In order to estimate the occupation number dependence of the effective Wannier functions, we express them as
\begin{equation}
\omega_i(\mathbf{r}, \hat{n}_i)=\frac{1}{\sqrt{\mathcal{N}_{\hat{n}_i,\hat{n}_j}}}\left[ \phi_i(\mathbf{r}, \hat{n}_i) - 
\sum_{\langle j\rangle} \mathcal{A}_{\hat{n}_i,\hat{n}_j} \phi_j(\mathbf{r}, \hat{n}_j) \right ],
\end{equation}
where in a mean-field treatment we make a Gaussian ansatz for the localized wave functions at site $i$ with occupation number operator $\hat{n}_i$,
$\phi (\mathbf{r-R_i};\hat{n}_i)=\frac{1}{\pi^{3/4}d^{3/2}(n_i)}\exp (-(\mathbf{r-R})^{2}/d^{2}(\hat{n}_i))$,
and the width $d(\hat{n}_i)$ is a variational parameter depending on the particle
number $n_{i}$ \cites{Chiofalo2000,Vignolo2003,Schaff2010}. We introduced $\mathcal{A}_{\hat{n}_i,\hat{n}_j}$
to fulfill the requirement that the effective Wannier functions at neighbouring sites are orthogonal whereas
$\mathcal{N}_{\hat{n}_i,\hat{n}_j}$ takes care of the normalization of the Wannier functions. \xdl
A more rigorous, fully-correlated treatment can be found in \Sec{MODressing}. \xe

For deep enough lattice
depths, we can assume that the width is much smaller than the lattice constant, i.e. $a/d(n_i) \gg 1$. Consequently, one can define the overlap function
between Gaussians centered at neighboring sites as,
\begin{eqnarray}
\mathcal{S}_{\hat{n}_i,\hat{n}_j} &=& \int d\mathbf{r} \phi_i(\mathbf{r}, \hat{n}_i)\phi_j(\mathbf{r}, \hat{n}_j) \nonumber\\
& = & \left[\frac{2 d(\hat{n}_i)d(\hat{n}_j)}{d^2(\hat{n}_i)+d^2(\hat{n}_j)}\right]^{3/2}
\exp\left\{-\frac{a^2}{2[d^2(\hat{n}_i)+d^2(\hat{n}_j)]} \right\}. \nonumber\\
&&
\end{eqnarray}
In the limit of $\mathcal{S}_{\hat{n}_i,\hat{n}_j}\ll 1$, from the orthonormalization constraints, one gets 
\begin{eqnarray}
\mathcal{A}_{\hat{n}_i,\hat{n}_j} &=& 1-\sqrt{1-\mathcal{S}_{\hat{n}_i,\hat{n}_j}}, \\
\mathcal{N}^{-2}_{\hat{n}_i,\hat{n}_j} &=& 1-2\sum_{\langle j\rangle} \mathcal{A}_{\hat{n}_i,\hat{n}_j}
\mathcal{S}_{\hat{n}_i,\hat{n}_j}+\sum_{\langle j\rangle} \mathcal{A}^2_{\hat{n}_i,\hat{n}_j}.
\end{eqnarray}

To find the occupation-dependent width of the site-centered Gaussian $\phi (\mathbf{r-R_i};{n}_i)$, we minimize the Gross-Pitaevskii energy functional. 
Taking into account the full lattice potential (i.e.\ not employing a quadratic approximation for the
lattice minima), for a given $n_i$ this leads to
\begin{eqnarray}
&&\left[ \frac{d(n_{i})}{d_{0}}\right] ^{5}\exp \left[ -\pi ^{2}\frac{
d^{2}(n_{i})}{a^{2}}\right] =\nonumber\\
&&\frac{d(n_{i})}{d_{0}}+\sqrt{2\pi }\left[
\frac{V_{0}}{E_{R}}\right] ^{1/4}\frac{a_{s}}{a}(n_{i}-1).  
\label{sig} 
\end{eqnarray}
We have introduced $d_{0}/a=\left[ \frac{V_{0}}{E_{R}}\right] ^{-1/4}/\pi $
for the width of $\phi $ in the limit $V_{0}\gg E_{R}$. Note that Eq.~(\ref{sig})
has a solution only as long as $\sqrt{V_{0}/E_{R}}\gg d^{2}(n_{i})/d_{0}^{2}$.

%
%
%

\subsection{Multi-orbital on-site interaction}

\label{MOOnSite}

\begin{figure}
{\centering
\includegraphics[width=1\linewidth]{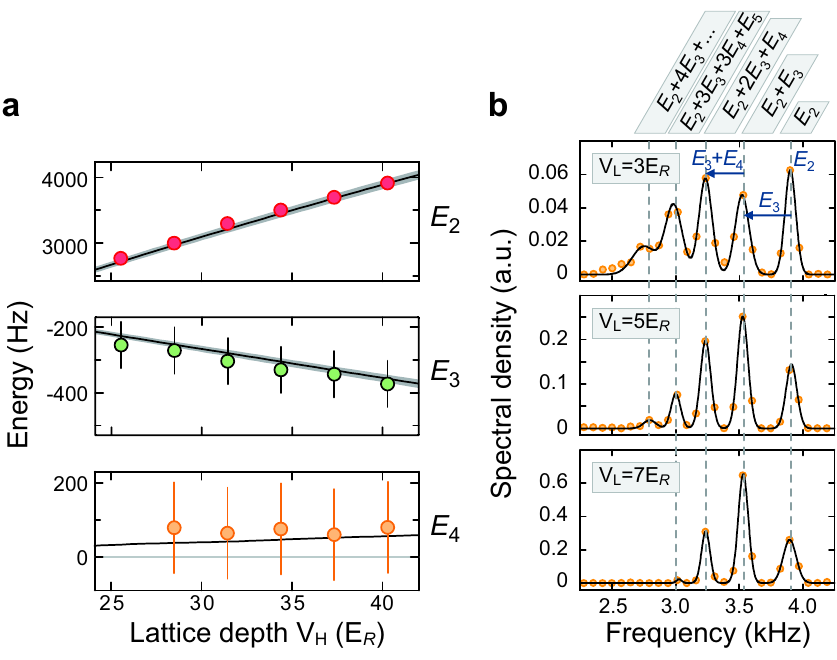}}
\caption{ 
The occupation-dependency of the on-site energy $U_n$ due to the population of higher orbitals was observed
in a collapse and revival experiment \cites{Will2010}.  
(a) Effective multi-body interaction energies $E_n$ using the expansion \eqref{eq:MultiBodyEnergies} as a function of the final lattice depth $V_\text{H}$ after the quench, where the circles correspond to the measured values ($V_\text{L}=8\ER$). The lines are theoretical predictions using exact diagonalization in a many-particle basis \xdl (\Eq{eq:ManyParticleGS}). \xe 
(b) The experimental energies are determined by Fourier transformation of time-resolved traces ($V_\text{H}=40\ER$), where the relative peak height depends on the number distribution of the superfluid state in the shallow lattice $V_\text{L}$. The dashed gray lines are the theoretical values.\from{Will2010} 
\label{Will2010}
}
\end{figure}

The on-site energy $U$ in the Hubbard model represents the interaction energy of particles on the same lattice site, calculated using the Wannier function 
of the lowest band. It is clear, however, that the
respective wave function is not an eigenfunction of the single-site problem with interactions, 
since repulsive interaction broadens and attractive interaction narrows the on-site density. In
the language of orbitals, this corresponds to the admixture of higher orbitals to the lowest orbital. The occupation of higher orbitals is particle-number dependent and is
a function of interaction strength and lattice depth. For bosonic atoms in optical lattices, the occupation-dependent population of higher orbitals could be observed
experimentally via spectroscopy measurements \cites{Campbell2006,Mark2011,Bakr2011,Mark2012} and via a quantum phase evolution measurement (\Fig{Will2010}) after a sudden 
quench of the lattice depth \cites{Will2010}. For two-component fermionic atoms, modulation spectroscopy was used to measure the on-site interaction in a honeycomb-lattice. For large interactions, the on-site interaction deviates from the theoretical single-particle on-site interaction computed from the honeycomb Wannier functions indicating the influence of higher-bands \cites{Uehlinger2013}.
Theoretically, the occupation-dependence was studied with mean-field approaches \cites{Li2006,Hazzard2010,Dutta2011}, field-theoretical
methods \cites{Johnson2009}, variational approaches \cites{Sakmann2011,Major2014}, direct-space quantum Monte-Carlo \cites{Pilati2012}, and different types of diagonalization approaches \cites{Busch1998,Buchler2010,Luhmann2012,Bissbort2012,Lacki2013}.

In the simplified case of two atoms with contact interaction in the harmonic confinement, the Schr\"odinger equation can be solved exactly \cites{Busch1998}. While the $\delta$-interaction
potential for neutral atoms is easily applied to the single-band problem, it must be taken with care when dealing with an (infinite) orbital degree of freedom, since the corresponding
Hamiltonian is not self-adjoint in higher than one dimension.  For two or three dimensions, a regularized $\delta$-potential can be used to circumvent this problem.
In Ref.~\cites{Busch1998}, analytical expressions for the energy and the wave functions are derived. The great advantage of the harmonic oscillator potential is the separability
in relative and center-of-mass coordinates. Transferring these results directly to optical lattices is problematic: While the Gaussian is a reasonable approximation for the lowest-band Wannier
function of a lattice site when dealing with on-site properties, higher-band Wannier functions differ strongly from their harmonic counterpart. As the regularization
in Ref.~\cites{Busch1998} explicitly accounts for the infinite series of higher-orbital wave functions, the results can not be transferred quantitatively.

To circumvent the subtleties of the $\delta$-potential, different types of interaction potentials can be applied. In Ref.~\cites{Buchler2010}, the two channels of
the Feshbach resonance are modeled to solve the problem using the Bloch functions of the optical lattice. 
By comparison with the standard Bose-Hubbard model, this allows to obtain the multi-orbital on-site energy. The great advantage of this treatment is that it models directly the experimental technique for tuning the interaction strength. 
A simpler approach is to use a finite-ranged model
interaction potential, where one has to assure that the results depend only weakly on the specific shape of the potential \cites{Pilati2012,Luhmann2012}. Note that the finite range
of the potentials leads to a high-energy cut-off, since fast oscillating wave functions of very high orbitals are averaged out within the interaction integrals. 
Note that for scattering resonances the assumption of a finite ranged interaction potential may break down. 
Numerically, a scaling with respect to the number of orbitals can be applied to predict the actual value of the problem with an infinite number of orbitals \cites{Buchler2010,Jurgensen2012,Lacki2013}.
  
For short-ranged interaction potentials $V(\mathbf{r}-\mathbf{r'})$, we can write the on-site problem for $n$ particles in a local many-particle Fock basis with states $\ket{N} = \ket{ n_0, n_1, ... }$, where $n_\alpha$ is the number of particles in orbital $\alpha$. Dropping the site index, the orbitals are the Wannier functions and  the Hamiltonian for a single lattice site reads, compare (\ref{genHam}),
\begin{equation}
	\label{eq:Hsite}
	\hat{H}_\text{site}=\sum_\alpha \epsilon^\a \hat{n}^\a + \frac{1}{2} \sum_{\alpha\beta\gamma\delta} U^\abcd\, \hat b^{\dagger\alpha}\hat b^{\dagger\beta}\hat b^{\gamma}\hat b^{\delta},
\end{equation}
where $\hat n^\a=\hbd{}{\alpha}\hb{}{\alpha}$, $\hbd{}{\alpha}$ creates and $\hb{}{\alpha}$ annihilates a particle in the Wannier orbital $\alpha$ with single-particle energies $\epsilon^\a$. 
The interaction integrals are given in \Eq{intint}. 
The many-particle ground state for $n$ particles
\begin{equation}
\ket{ \Psi(n) } = \sum_N c_N(n) \ket{ N(n) }
\label{eq:ManyParticleGS}
\end{equation}
is a superposition of local Fock states with $n$ particles, with real coefficients $c_N(n)$.
While for the non-interacting ground state $\ket{ \Psi_0 (n) }=\ket{n,0,0,...}$ all atoms occupy the single-particle ground state, the interaction promotes particles also to higher orbitals. 
On the mean-field level, the change of the single-site wave function is attributed to an interaction broadening of the density.
However, the significant change of the many-particle state lies also within modified higher-order correlations allowing the particles mutually to reduce their spatial overlap \cites{Bissbort2012}. Therefore, this effect can not be captured on an effective single-particle level. 
   
The eigenvalues $e_n$ of the Hamiltonian \Eq{eq:Hsite} for $n$ particles directly relates to the multi-orbital on-site energy $U_n$ per particle pair via
\begin{equation}
	U_n=\frac{2}{n (n-1)}\ e_n .
	\label{eq:Un}
\end{equation}
This is the occupation-dependent on-site energy for an effective Hubbard model. Note that the on-site energy decreases with the number of particles, i.e. $U_{n+1}<U_n$. From a different point of view, the occupation-dependent on-site energy can be understood as effective $n$-body collisions with energies $E_n$ \cites{Will2010,Johnson2009,Bissbort2012}. Expanded in terms of $n$-body collisions, the on-site energy for $n$ particles can be written as
\begin{equation}\begin{split}
\label{eq:MultiBodyEnergies}
	U_n=  &\frac{E_2}{2} {n(n-1)} + \frac{E_3}{6} {n(n-1)(n-2)}\\
& + \frac{E_4}{24}  {n(n-1)(n-2)(n-3)}+ ...
\end{split}\end{equation}
Using the occupation-dependent energies $U_n$, we can set $E_2=U_2$, $E_3=3U_3 - 3 E_2$, $E_4=6U_4-6E_2-4E_3$, $...$ 
The differences between the occupation-dependent energies $U_n$ have been observed in a collapse and revival experiment with bosonic atoms
after a quench from a shallow lattice ($V_\text{L}$) to a deep lattice ($V_\text{H}$) \cite{Will2010}. The local particle number distribution in the superfluid regime, which is Poissonian or number-squeezed depending on the lattice depth $V_\text{L}$, is preserved during the quench. The time-evolution of the matter wave field $\psi=\bra{\phi(t)}\,\hat b\,\ket{\phi(t)} $ in the deep lattice reflects the occupation-dependency of the on-site energy \eqref{eq:Un} via
\begin{equation}
|\psi|^2= \sum_{n,m=0}^{\infty} C_{n,m} \,\mathrm{e}^{-i(e_{n+1}-e_n-e_{m+1}+e_m)t/\hbar }, 
\end{equation}  
where the relative contribution $C_{n,m}$ depends on the particle number distribution in the superfluid state.
In \Fig{Will2010}, the results are shown as effective $n$-body collision energies $E_n$. 

\subsection{Bose-Hubbard models with local three-body interaction}

\label{Three-body}

Truncating the effective description of
(\ref{eq:MultiBodyEnergies}) to the first two terms, one may build a
 particular Bose-Hubbard model with local three-body interactions, with the Hamiltonian 
\begin{align} \label{threebodyHam}
\hat{H} &= -t\sum_{\langle i,j\rangle} \hat{b}^\dagger_i\hat{b}_j + \frac{U}{2}\sum_i \hat{n}_i(\hat{n}_i-1) + \nonumber \\ &+\frac{W}{6}\sum_i \hat{n}_i(\hat{n}_i-1)(\hat{n}_i-2) - \mu\sum_i\hat{n}_i,
\end{align}
\xdl where $\mu$ is the chemical potential fixing the particle number. \xe
While for contact interaction (compare Fig.~\ref{Will2010}a)  the strengths of two-body and three-body terms may be modified in a limited range by changing the lattice depth and geometry, one may assume
that the three-body term controlled by $W$ can be experimentally tuned independently of the two-body term $U$ (e.g.\ for dipolar or other type of interactions). It is in fact for polar molecules  in optical lattices that such a three-body potential term was introduced \cites{Buchler07b}. The following quantum Monte Carlo study \cites{Schmidt2008} revealed the existence of both solid and supersolid phases in the system. Another early discussion of the model (\ref{threebodyHam}) was done on the mean-field level in \cites{PhysRevA.78.043603}. It was shown that, depending on three-body term, the second insulating lobe ($\rho=2$)  changes its area. In contrast, the first insulating lobe ($\rho=1$) is insensitive to the three-body interactions. These results are intuitively straightforward. It is clear that for $\rho=2$, in contrast to the $\rho = 1$ case, tunneling  has to compete with not only two-body but also with three-body interactions to destroy the insulating phase. A pedagogical explanation of these facts 
and a comparison with the Gutzwiller mean-field approach was presented recently \cites{1402-4896-2014-T160-014038}.

\begin{figure}
\centering
\includegraphics{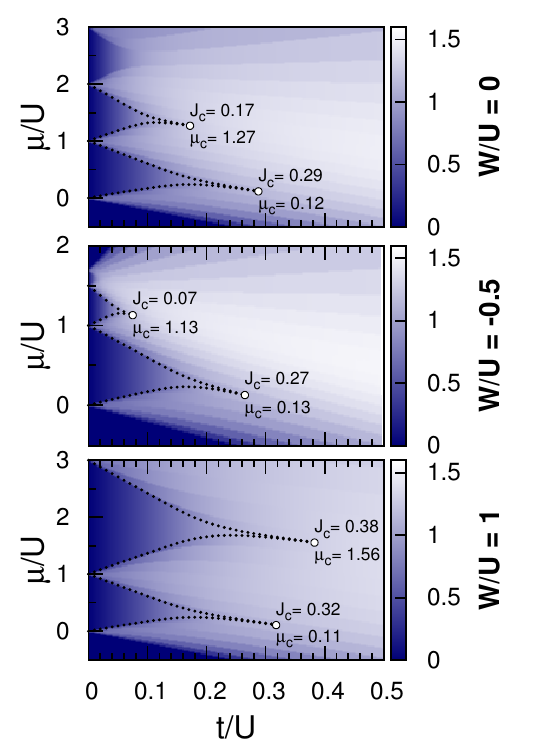}
\caption{ The phase diagrams of one-dimensional system described by the Hamiltonian \eqref{threebodyHam} for different values of the three-body interactions $W$. Open circles mark the transition from MI to SF, as estimated from the exact diagonalization of small systems and extrapolated to the thermodynamic limit. In the background of each phase diagram, the density plots of the correlation function $\langle \hat{a}^\dagger_i\hat{a}_j\rangle$ are visualized. For $\rho=1$, the insulating phase does not change significantly, but for $\rho = 2$ the size of the MI phase crucially depends on the three-body interaction parameter. Figure adopted from \cites{PhysRevA.85.065601}.
\label{fig:threebody}
}
\end{figure}

A more precise discussion of the model was done for the one-dimensional case. First, using the DMRG approach \cites{PhysReVA.84.065601}, it was shown that for strong enough three-body term the first insulting lobe, in contrast to what is predicted by mean-field results, changes its shape and the tip of the lob is shifted. However, the phase transition from MI to SF remains in the Berenzinskii-Kosterlitz-Thouless (BKT) universality class. In \cites{PhysRevA.85.065601}, these results were supported with exact diagonalization calculations, and the extension to an attractive three-body term was proposed (See Fig.\ref{fig:threebody}). Independently, a two-dimensional system with strong three-body attraction was studied with a Quantum Monte Carlo approach in \cites{PhysRevLett.109.135302}. In addition, some effects of finite temperatures were discussed in that article. Recently, a summary of  properties of the one-dimensional model \eqref{threebodyHam} on the basis of dynamical DMRG method was also presented \
cites{PhysRevA.88.063625}.

Further extensions of the model \eqref{threebodyHam} have also been studied. In particular,  {\it (i)} an extension adopting long-range dipole-dipole interactions was proposed and discussed in \cites{PhysRevA.82.013634}; {\it (ii)} a discussion of the influence of a magnetic field on the properties of the model  was provided in \cites{Huang20104364}; {\it (iii)} additional effects arising in a superlattice potential were studied with mean-field and DMRG approaches in \cites{PhysRevA.85.051604}.

Finally, it is worth noting that the seemingly exotic version of the model \eqref{threebodyHam} with a vanishing two-body term $U=0$ has also been discussed in detail in \cites{EurPhysJB.85.161,ArXiv.1307.6852}, where the one-dimensional case has been addressed using DMRG calculations. For that model,  the first insulating lobe for $\rho=1$ vanishes and the stability of higher Mott lobes increases with increasing filling $\rho$. As previously, in the vicinity of the phase transition the system remains in the BKT universality class.

Another proposition  \cites{PhysRevLett.102.040402} considers an attractive two-body term $U<0$ and strong three-body repulsion, a model which may mimic strong three-body losses. In \cites{PhysRevA.81.061604}, it was shown that for vanishing tunnelings and filling $0<\rho<2$, an additional pair-superfluid phase is present in the system. With increasing tunneling, the system undergoes a second-order phase transition to a normal SF phase. On this basis, in \cites{ArXiv.1304.4835} the model with large but finite repulsive $W$ was discussed, where it was shown that the critical exponents and the central charge governing the quantum phase transition have repulsion-dependent features. In consequence, the model \eqref{threebodyHam} with attractive 
two-body and repulsive three-body interactions extends the list of known systems violating the universality hypothesis. While some of these models seem unrealistic at first, it is also known how to control the relative strength of three-body interactions, as exemplified in \cites{Mazza10} for Raman induced couplings. A more recent work \cites{arXiv1311.1783} has shown how to engineer practically at will three-body interactions via photon-assisted tunneling.


\subsection{Multi-orbital dressing of off-site processes}
\label{MODressing}

\begin{figure}
\includegraphics[width=0.75\linewidth]{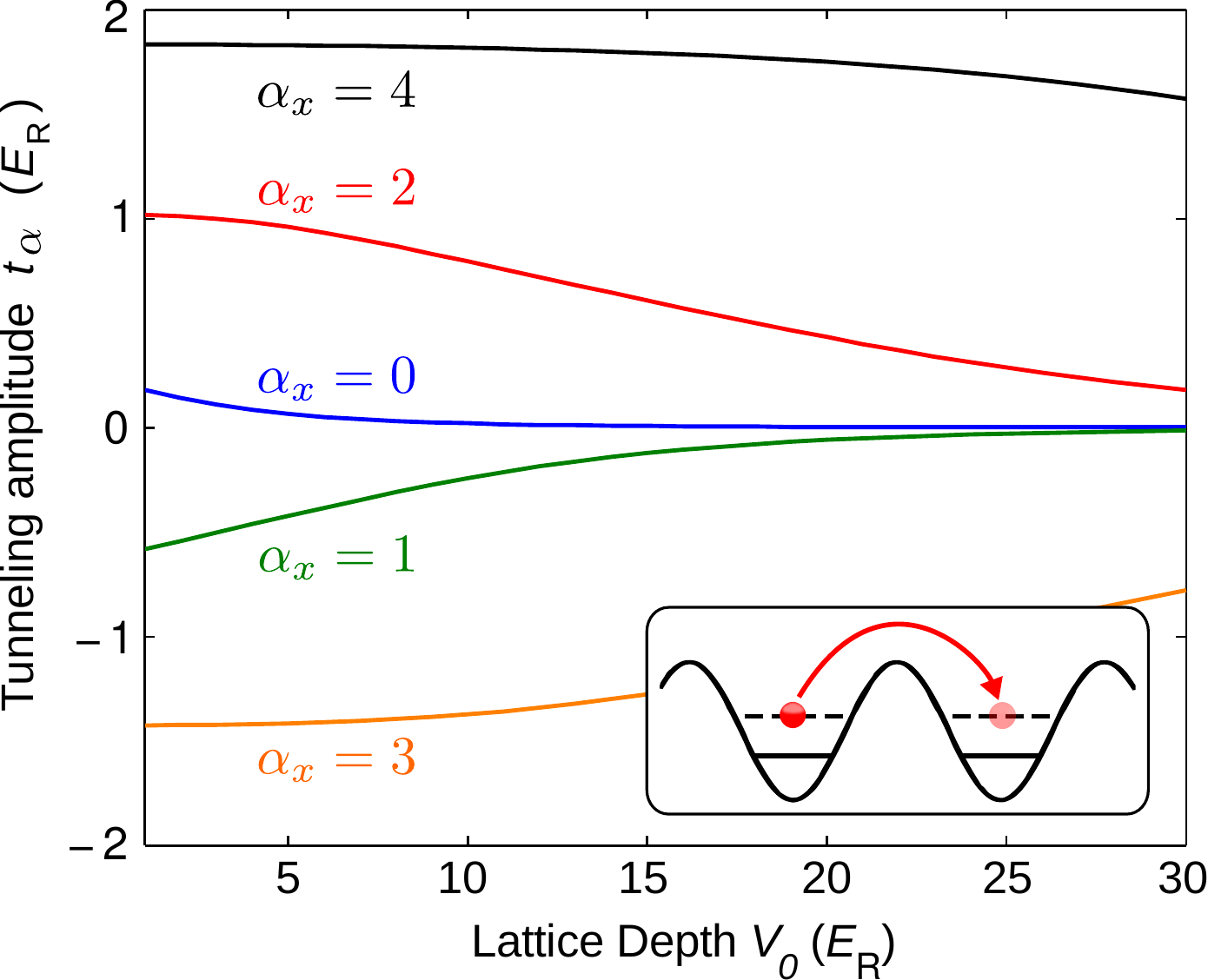}
\caption{Single-particle tunneling matrix elements $t^\a$ for the lowest five bands as a function of the lattice depth $V_0$.\from{Luhmann2012}
\label{MultiOrbitalTunnelingMatrixElements}}
\end{figure}

While the last section shows how on-site properties, i.e. the on-site wave function and the energy $U$, are influenced by orbital degrees of freedom, in the following their impact on off-site properties is discussed. As a result of the population of higher orbitals, the effective wave function overlap of particles on neighboring lattice sites changes. This leads to modified amplitudes of the tunneling and the off-site interactions (\Sec{GHB}).
Since the occupation of higher orbitals is typically few percents or lower, one would expect the effect on the hopping to be only marginal. However, as shown in \Fig{MultiOrbitalTunnelingMatrixElements} the tunneling matrix elements $J^\a$ in higher bands can be exponentially large compared with the lowest band. Therefore, the tunneling in higher orbitals can have a large net effect on the total tunneling amplitude. 
In optical lattices, the effect of bosonic tunneling in higher bands were discussed by variational mean-field methods \cites {Li2006,Larson2009,Hazzard2010,Dutta2011} (see \Sec{MOMFT}) and by numerical exact methods mainly restricted either to double- or triple-well systems, e.g. \cites{Sakmann2009,Sakmann2010,Cao2011}. The effect was also discussed for experiments with Bose-Fermi mixtures \cites{Luhmann2008b,Lutchyn2009,Mering2011,Jurgensen2012}.
 
When dealing with both lattice and orbital degrees of freedoms, one could be tempted to formulate the multi-orbital Hubbard model, i.e.
\begin{equation}\begin{split}
	\label{eq:HMOHM}
	\hat{H}= &-\sum_{\expect{i,j},\alpha}    \J^\a  \hbd{i}{\alpha}\hb{j}{\alpha}  +\sum_{i,\alpha} \epsilon^\a \hat{n}_i^\a  \\
   &+ \frac{1}{2} \sum_{i,\alpha\beta\gamma\delta} U^\abcd\, \hbd{i}{\alpha}\hbd{i}{\beta}\hb{i}{\gamma}\hb{i}{\delta},
\end{split}\end{equation}
(compare with \Eq{eq:Hsite}). Here, $\J^{\a}$ is the tunneling amplitude between neighboring sites $i$ and $j$ in band $\alpha$ \eqref{eq:HubbardOrbitalTunnelingAndEnergy}. Although this model is already a strong simplification of the full two-body Hamiltonian \eqref{genHam} as it disregards any off-site interactions, the complexity of this problem is enormous.  
The idea is therefore to switch from the non-interacting basis to a basis that is more adapted regarding the interactions.  
This basis is constructed from the solution of the multi-orbital on-site problem \eqref{eq:Hsite} as described in \cites{Luhmann2012,Jurgensen2012,Bissbort2012}. Since we restrict the single-site solutions only to the lowest energy state, we truncate the basis thereby to a single band, which is constructed from correlated single-site states \eqref{eq:ManyParticleGS}.  By construction, the second and the third term of \Eq{eq:HMOHM} are diagonal in this basis. In particular, the on-site interaction in the \textit{dressed band} is given by (the operators within the dressed band are denoted with a tilde) 
\begin{equation}
\sum_i U_{\tilde n_i} {\tilde n_i} ({\tilde n_i}-1). 
\label{eq:DressedU}
\end{equation}
The on-site interaction parameter, which is occupation dependent, can be expressed formally as a projection $U_{\tilde n_i} = \mathcal{P}_i^\dagger \sum_n  U_{n} \ket{\tilde n}_i \bra{\tilde n}_i \mathcal{P}_i $. Here, $U_n$ are the eigenenergies normalized per particle pair \eqref{eq:Un}, and $\mathcal{P}_i$ projects the many-site state to site $i$. 

\xdl It is important to note that other processes such as the \xe \cutdl{The} multi-orbital tunneling matrix element is also transformed \xdl in \xe \cutdl{into} the dressed basis.  \xdl Since the orbital dressing is a basis transformation that is block diagonal with respect to the particle subspaces, the usual commutation relations $\big[\tilde{b}_i,\tilde{b}^{\dag}_j\big]=\delta_{ij}$  are fulfilled. \xe  The appropriate procedure \xdl of the transformation to the interaction-dressed basis \xe  is described in \Sec{Appendix}. 
From a practical point of view, it is important that once the Hamiltonian is expressed in the dressed basis it remains a single-band  lattice problem (compare Eq.~\eqref{eq:HMOBC}). It is inherently occupation-dependent, but has otherwise the same complexity as the single-band Hubbard Hamiltonian. The dressed-band model allows to apply standard single-band methods to calculate the phase diagram such as e.g. the mean-field or quantum Monte-Carlo approaches.

\begin{figure}
\includegraphics[width=1\linewidth]{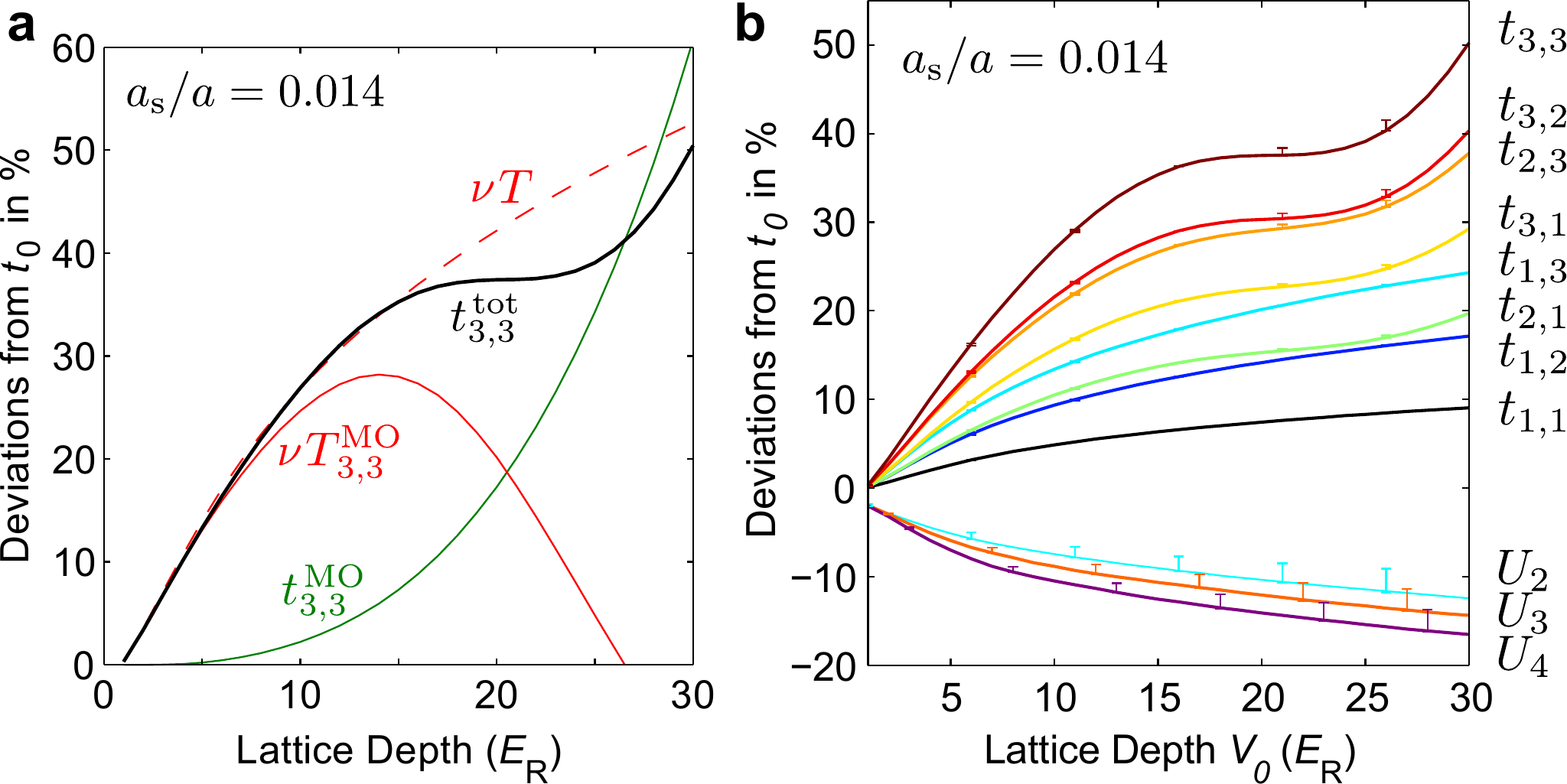}
\caption{ (\textbf{a}) Contributions to the effective tunneling $t^\mathrm{tot}_{n_j,n_i}$ with $n_i=n_j=3$ by multi-orbital tunneling $t_{n_j,n_i}$, density-induced tunneling $\JBC$, and multi-orbital density-induced tunneling $\JBC_{n_j,n_i}$, where the latter two scale with the prefactor $\nu=n_i+n_j-1$.  The total tunneling is the sum of the two multi-orbital dressed processes. 
The plot shows the deviations from the single-particle tunneling $t$ as a function of the lattice depth $V_0$. 
(\textbf{b}) Occupation-dependent total tunneling $t^\mathrm{tot}_{n_j, n_i}$ and on-site interactions $U_n$ for a box-shaped interaction potential (width of $W=5\, \mathrm{nm}$ and a lattice constant $a=377\, \mathrm{nm}$). The results are only weakly affected by changes in the scattering potential (error bars correspond to $W=25\, \mathrm{nm}$).\from{Luhmann2012} 
}\label{MultiOrbitalTunneling}
\end{figure}

\begin{figure}
\centering\includegraphics[width=0.75\linewidth]{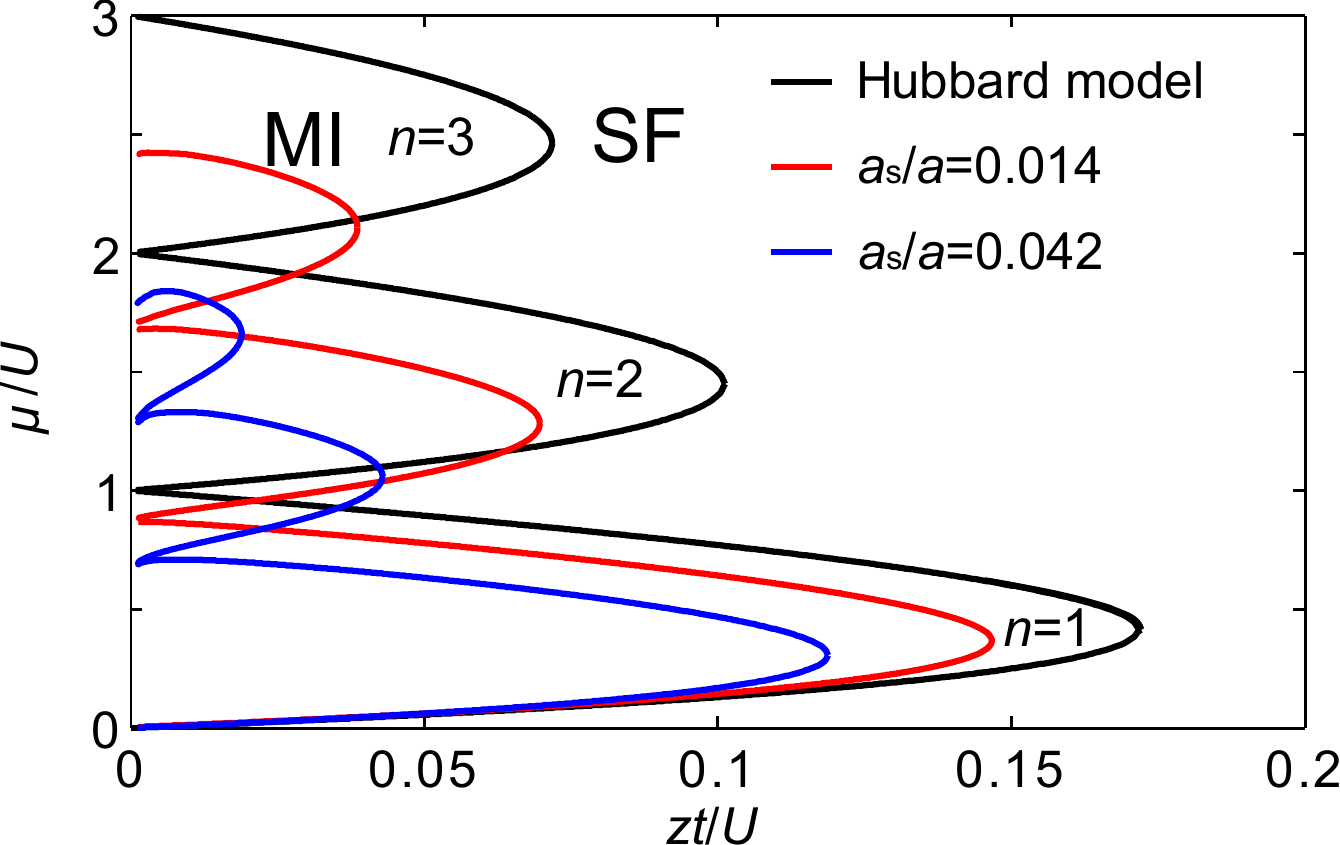}
\caption{Phase diagrams showing the superfluid (SF) to Mott-insulator (MI) transition for the generalized multi-orbital-dressed Hubbard Hamiltonian \eqref{eq:HMOBC} in the Gutzwiller approximation.  The phase boundaries are plotted for the interaction strengths $\as/a=0.014$ (red, $^{87}$Rb parameters) and $\as/a=0.042$ (blue) as well as for the Bose-Hubbard model (black).\from{Luhmann2012} 
\label{MultiOrbitalPhaseDiagramBosons}
}
\end{figure}

\subsection{Multi-orbital occupation-dependent Hamiltonians}
\label{MOHamiltonians}

The multi-orbital dressing of both interactions and tunneling leads to intrinsically occupation-dependent Hubbard models. As discussed in \Sec{MOOnSite} and \Sec{MODressing}, the multi-orbital renormalization of the on-site interaction, tunneling and other off-site processes causes the amplitudes to depend on the particle numbers on the participating sites. In optical lattices, the multi-orbital corrections can be on the same order of magnitude as the density-induced tunneling as discussed in \Sec{GHB}. Therefore, the combination of both effects is essential for a correct description. 

\begin{figure}
\centering\includegraphics[width=0.85\linewidth]{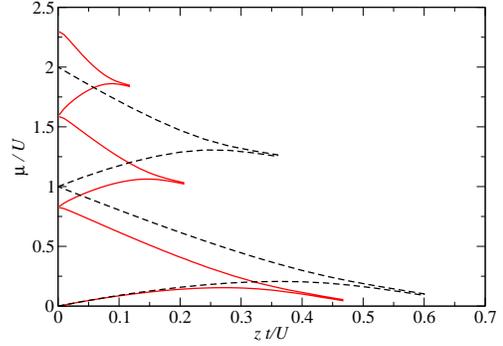}
\caption{Superfluid to Mott-insulator transition in one dimension for the generalized Hubbard Hamiltonian \eqref{eq:HMOBC} including density-induced tunneling and multi-orbital effects (solid red line). The phase diagram is obtained by means of a TEBD algorithm with 100 lattice sites for an interaction strength $\as/a=0.014$ and a vertical confinement of $34.8\ER$.  The dashed black line corresponds to the Bose-Hubbard model. \from{Lacki2013} 
\label{MultiOrbitalPhaseDiagramBosons1D}
}
\end{figure}

\begin{figure*}
{\centering
\includegraphics[width=0.8\linewidth]{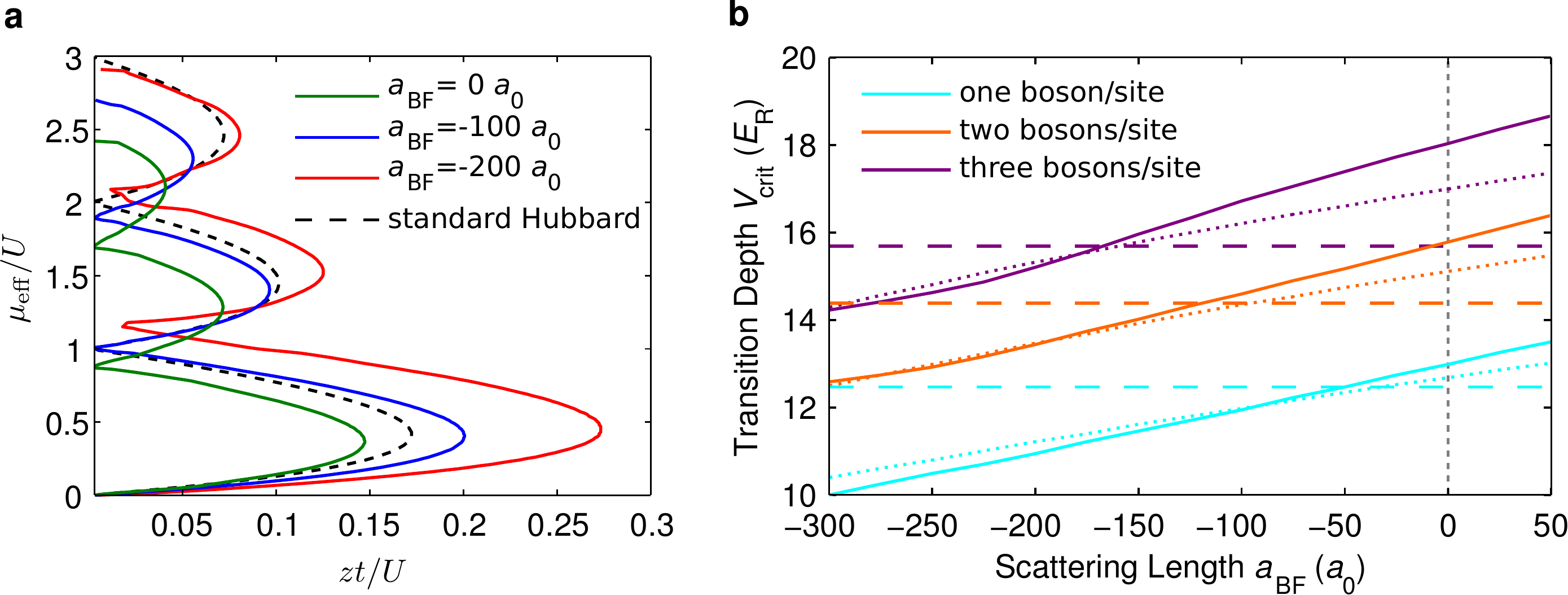}}
\caption{\textbf{a} Phase diagram for the superfluid to Mott-insulator transition of bosons interacting attractively with a fermionic band-insulator. The predictions of the standard Hubbard models are shown as a dashed black line. The attractive interaction effectively reduces the total tunneling resulting and extends Mott insulating phases in dependence on the interspecies scattering length $a_\BF$. \textbf{b} The critical lattice depth of the superfluid to Mott-insulator transition as a function of the interspecies scattering length $a_\BF$. The transition occurs at significantly shallower lattices than in the purely bosonic system $(a_\BF=0)$. The dashed lines correspond to the standard Bose-Fermi Hubbard model \eqref{eq:FBI_BFHM} and the dotted lines to the generalized lowest-band model \eqref{eq:extended_BFHM} with density-induced tunneling in \Sec{sec:sb_bose_fermi}.\from{Jurgensen2012}
\label{MultiOrbitalBoseFermi}
}
\end{figure*} 

In the multi-orbital dressed band (\Sec{MODressing}), the occupation-dependent Hamiltonian for bosons in an optical lattice is given by \cites{Luhmann2012}
\begin{equation}\begin{split}
	\hat H=& -\sum_{\langle i,j \rangle}  \hatt{b}_i^\dagger \hatt{b}_j \J_{\hatt n_j,\hatt n_i} 
	-  \sum_{\langle i,j \rangle} \hatt{b}_i^\dagger (\hatt{n}_i + \hatt{n}_j)\hatt{b}_j \JBC_{\hatt{n}_j,\hatt{n}_i} \\
	&+ \frac{1}{2} \sum_i U_{\hatt n_i} \hatt n_i (\hatt n_i -1),
	\label{eq:HMOBC}
\end{split}\end{equation}
where the second term represents the density-induced tunneling (see \Eq{modHamlow} with $V$, $P=0$).
The total tunneling consists of normal and density-induced tunneling, both of which effectively include higher orbital processes.  
For a given occupation of lattice site $i$ and $j$, the total tunneling can be evaluated as
\begin{equation} 
	\J^\tot_{n_j,n_i}= \J_{n_j,n_i} + (n_i\! +\! n_j \! - \! 1) \JBC_{n_j,n_i}, 
\end{equation}
where its individual contributions are shown in \Fig{MultiOrbitalTunneling}(a) as a function of the lattice depth. Note that both density-induced tunneling and the amount of multi-orbital corrections scale with the interaction strength. 
In shallow lattices, the multi-orbital renormalization of the tunneling and bond-charge tunneling is in general weak and gets substantial only at intermediate lattice depths ($V_0\gtrsim 15 \ER$). Interestingly, the higher-orbital contributions of tunneling and bond-charge interaction partly compensate each other at intermediate lattice depths.  Figure \ref{MultiOrbitalTunneling}({b}) demonstrates the occupation dependency of the total tunneling amplitude $J^\tot_{n_j,n_i}$ for different occupations $n_i$ and $n_J$.   
 
Using perturbative mean-field theory \cite{Oosten2001}, occupation-dependent amplitudes such as $\J^\tot_{n_i\pm1,n_i}$ must be approximated by $\J^\tot_{n_i,n_i}$. However, Gutzwiller calculations for the ground state without this restriction give very similar results.
The phase diagram is shown in \Fig{MultiOrbitalPhaseDiagramBosons} for two different interaction strengths $\as$ as a function of $z\J/U$, where $z=6$ is the number of nearest neighbors for three-dimensional cubic lattices. The superfluid phase is enlarged for repulsive interactions and the tips of the Mott lobes are shifted towards smaller values of $z\J/U$. This corresponds to a significant shift of the critical lattice depth of the superfluid to Mott insulator transition due to an effectively increased tunneling and reduced on-site interaction. The deformation along the $\mu/U$ axis is due to the occupation-dependent on-site interaction $U_n$. 
 Phase diagrams for 1D and 2D lattices are computed in Ref.~\cite{Lacki2013} using mean-field and  the time evolving block decimation algorithm (TEBD) in 1D. 
The corresponding phase diagram is shown in \Fig{MultiOrbitalPhaseDiagramBosons1D}, where the Mott lobes are affected in the same way as in the mean-field treatment. 
In addition to the band-dressing technique discussed here, also direct-space quantum Monte-Carlo methods have been applied using different interaction potentials \cites{Pilati2012}. 
Experimentally, the shift of this transition has been studied for a filling $n=1$ and tunable interactions \cites{Mark2011}. In general, however, the shift is considerably more pronounced for higher fillings since the density-induced tunneling and the multi-orbital renormalization scale with the particle number. 

\begin{figure}
{\centering
\includegraphics[width=0.85\linewidth]{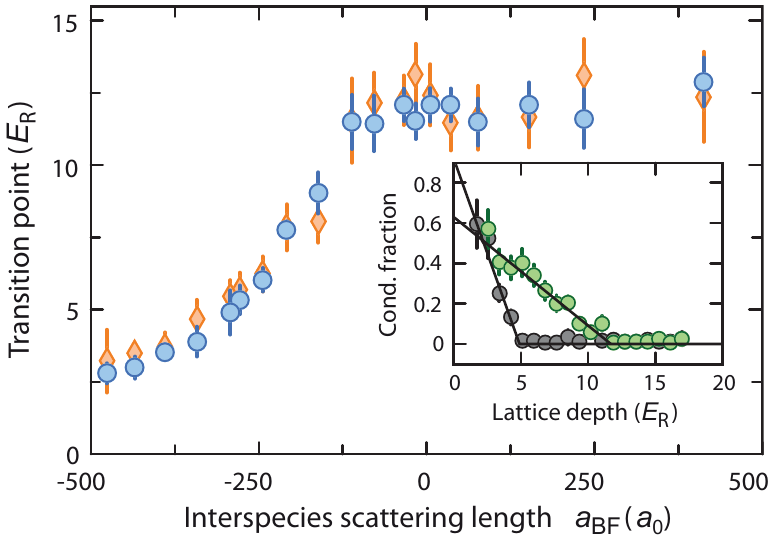}} \xdl
\caption{Superfluid to Mott insulator transition in a mixture of bosonic ${^{87}\mathrm{Rb}}$ and fermionic${^{40}\mathrm{K}}$ atoms, where the interspecies interaction $a_\mathrm{BF}$ was tuned by using a Feshbach resonance. The diamonds and circles represent experimental results for a ratio 0.5 and 0.75 of ${^{40}\mathrm{K}}$ to  ${^{87}\mathrm{Rb}}$ atoms, respectively. The transition point has been determined as the point of vanishing condensate fraction of ${^{87}\mathrm{Rb}}$ (inset). \from{Best2009}  \xe
\label{BoseFermiBest2009}
}
\end{figure} 

Another example for the realization of occupation-dependent models are Bose-Fermi mixtures, where effects of higher bands were discussed in Refs.~\cites{Luhmann2008b,Lutchyn2009,Mering2011,Jurgensen2012}. Here, the on-site energy (\Sec{MOOnSite}) and multi-orbital dressing of the tunneling (\Sec{MODressing}) must be treated in a many-particle product basis of $n$ bosons and $n_F$ fermions.
In addition, also density-induced tunneling $\JBC$ (boson-assisted) and $\JBCBF$ (fermion-assisted) crucially affect the phase diagram as discussed in \Sec{sec:sb_bose_fermi}. Therefore, it is important to treat both effects at the same time \cites{Mering2011,Jurgensen2012}.
Using the simplification of a fermionic band insulator ($n_F=1$), where all fermionic degrees of freedom are frozen out, the system can be described using an effectively bosonic Hamiltonian (cf. \Eq{eq:extended_BFHM})
\begin{equation}\begin{split}
	\tilde H=& -\sum_{\langle i,j \rangle}  \tilde{b}_i^\dagger \tilde{b}_j \left(\J_{\tilde n_j,\tilde n_i}+ 2 \JBC_{\BF,{\tilde n_j,\tilde n_i}} \right)\\
	&-\sum_{\langle i,j \rangle}  \tilde{b}_i^\dagger ( \tilde n_i+ \tilde n_j ) \tilde{b}_j\ \JBC_{\tilde n_j, \tilde n_i}  \\
	&+ \sum_i E_{\tilde{n}_i} - \mu \sum_i \tilde{n}_i.
\end{split}\end{equation}
In this case, the on-site energy for $n$ bosons is given by 
\begin{equation}
E_n = n \epsilon_{\mathrm{B},n} +  \epsilon_{\mathrm{F},n} + \frac12 n(n-1)  U_n + n  U_{\mathrm{BF},n},
\end{equation} 
containing the occupation-dependent (repulsive) interaction energies between the bosons $U_n$ and 
(attractive) interaction between bosons and fermions $U_{\mathrm{BF},n}$ as well as single-particle energies  $\epsilon_{\mathrm{B},n}$ and $\epsilon_{\mathrm{F},n}$ of higher orbitals. In analogy to the purely bosonic system, the critical point of the superfluid to Mott-insulator transition is affected. The phase diagram and the critical lattice depth are shown for a ${^{87}\mathrm{Rb}}$-${^{40}\mathrm{K}}$ mixture in \Fig{MultiOrbitalBoseFermi}. 
In the Bose--Fermi--Hubbard model the transition does not depend on the boson-fermion interaction (dashed lines in \Fig{MultiOrbitalBoseFermi}b),
whereas the generalized  occupation-dependent Hamiltonian predicts a strong dependency on the interspecies scattering length. 
This strong shift of the superfluid to Mott-insulator transition was also observed experimentally \cites{Gunter2006,Ospelkaus2006,Best2009}. \xdl
In Ref.~\cites{Best2009}, the interspecies interaction $a_\mathrm{BF}$ was tuned by a Feshbach resonance, which allows observing the shift of the Mott transition point as function of the interspecies interaction. The transition point shown in \Fig{BoseFermiBest2009} was obtained by measuring the condensate fraction of ${^{87}\mathrm{Rb}}$. For $a_\mathrm{BF}<200a_0$ the experiment finds a shift of the transition which is even stronger than theoretically expected (\Fig{MultiOrbitalBoseFermi}b). However, for $a_\mathrm{BF}<200a_0$ the experiments observes also a strong increase in the particle loss indicating additional processes such as a redistribution of the bosonic atoms. \xe
Note that the experimental lattice ramping procedure can also cause a drop of the bosonic coherence due an adiabatic heating. The latter is caused by different contributions of the atomic species to the total entropy \cites{Cramer2008,Cramer2011}. \xdl
For repulsive interaction, one would expect a phase separation of bosonic and fermionic atoms when the interspecies interactions exceeds the intraspecies interaction of the bosons. Hence, if the interspecies interaction is large enough the bosonic Mott transition is not longer influenced directly by the presence of the fermions. However, the redistribution of bosonic atoms possibly cause higher bosonic filling factors and thereby affects the transition point. \xe




\section{Hubbard models in excited bands} 
\label{orbitalhub}
Up to now, we restricted our considerations to a single band, and took effects of higher bands only in an effective theory into account. 
However, by actively exploiting these higher bands, one may open access to studying orbital physics in optical lattices, with exciting prospects, as reviewed in \cites{Liu11}:  
Multi-orbital physics can lead to unconventional superfluid states \cites{Wirth10,Oelschlaeger2011,SoltanPanahi2012,Oelschlaeger2013}, or additional Mott-insulator phases where atoms localize after undergoing a Tonks--Girardeau-like transition \cites{Alon05}.
Manipulating atoms in higher bands, one may also induce various topological phenomena \cites{Sun2012,Li2013}, and one can go beyond the integer quantum Hall effect that may be obtained in the $s$-orbitals of an optical honeycomb lattice subject to a synthetic gauge field \cites{Kitagawa2010,Goldman10,Alba2011,Hauke2012c}---in the flat $p$-bands of such a lattice, exotic incompressible states analogous to the Laughlin \emph{fractional} quantum-Hall liquid can be created \cites{Wu2007}. 
Several groups have now achieved loading and manipulating ultracold atoms in higher (such as $p$-) bands \cites{Browaeys2005,Kohl2005,Mueller2007,Anderlini07,Wirth10,Oelschlaeger2011,Oelschlaeger2012,Oelschlaeger2013}.
Techniques such as lattice ramping or radio frequency pulses have been used to transfer atoms from the $s$-band to higher bands. There, they can stay in a metastable state for a long time, allowing a detailed study of the effects of orbital degeneracy.

In a broader context, such studies may give important insight into the behavior of strongly-correlated electrons in solid-state samples. 
In many materials, such as transition metal oxides \cites{Tokura2000}, orbital effects play a fundamental role, signing responsible for several important material properties, such as colossal magnetoresistance, ferroelectricity, unconventional superconductivity, or charge ordering.
In many instances, novel quantum phases emerge due to the coupling of the orbital degree of freedom to the charge, spin, or lattice degrees of freedom \cites{Kugel1982,Khaliullin2005}. 
But such coupling not only generates interesting effects, it also complicates the theoretical treatment. 
It is, therefore, desirable to study simpler systems where the orbital degree of freedom is decoupled from all others. 
Here, ultracold atoms provide an ideal tool; loaded into higher bands of optical lattices, they allow to analyse orbital dynamics in a well-controlled environment, including orbital-only models of single-species (spinless) fermions.

\subsection{A three-color $p$-band Hubbard model in optical lattices}

As a first illustrative example for a non-standard Hubbard model in higher bands, we review in this section a three-color model describing spinless fermions in the $p$-band orbitals of an optical lattice close to an optical Feshbach resonance. 
The considered model hosts unconventional phases already in the simple cubic lattice, such as a phase with `axial orbital order' in which $p_z$ and $p_x + i p_y$ (or $p_x - i p_y$) orbitals alternate, thus breaking spatial and time-reversal symmetry \cites{Hauke2011c}.

\begin{figure}
\centering
\includegraphics[width=0.7\columnwidth]{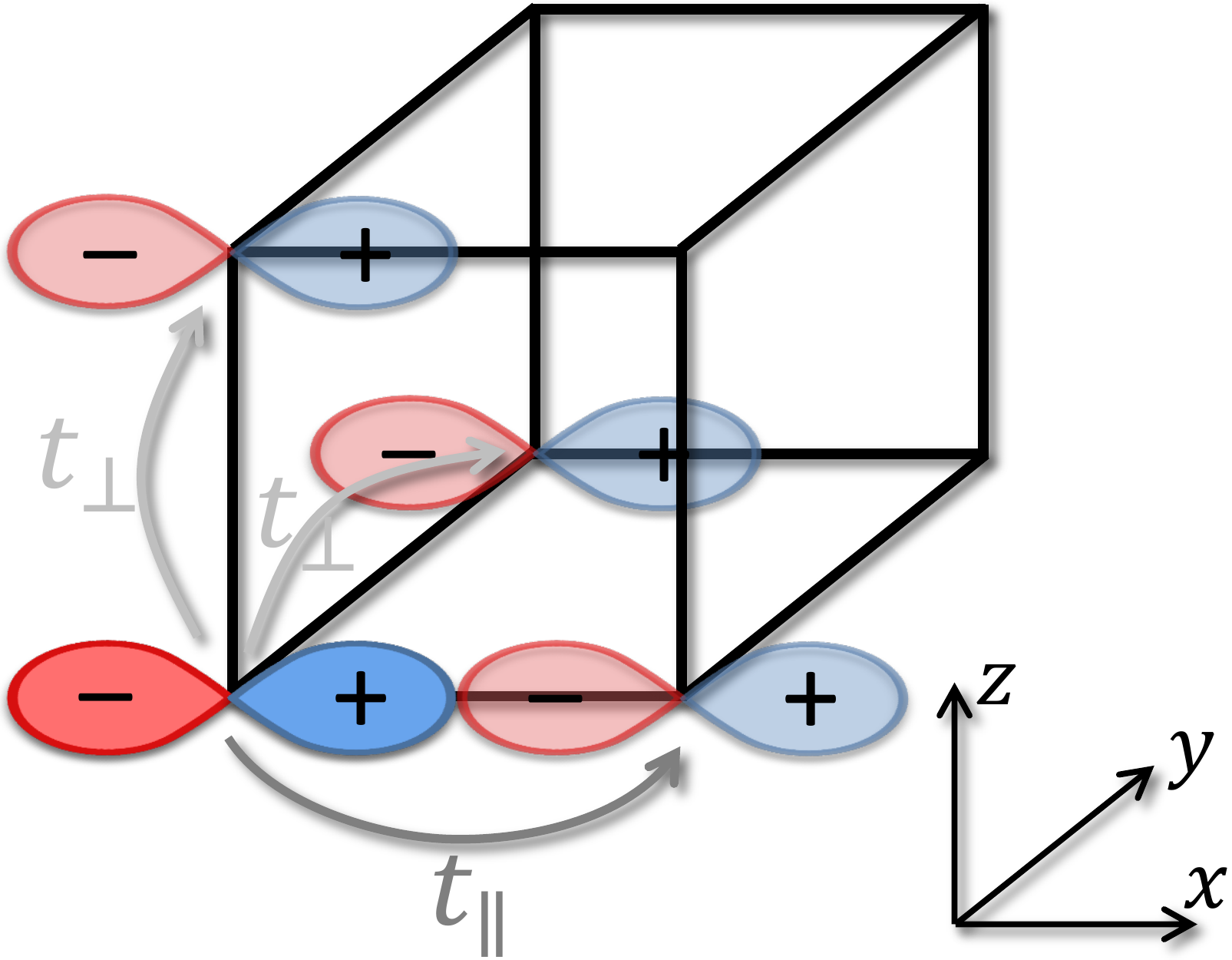}
\caption{
Orbital tunneling, exemplified for the $p_x$ orbital. 
Due to the odd parity of $p$ orbitals [indicated by ``$-$'' (red) and ``$+$'' (blue)], a fermion in a given $p_x$-orbital (solid dumbbell) can tunnel only into neighboring $p_x$-orbitals (semi-transparent dumbbells). 
Since the hopping amplitude is given by the overlap of the anisotropic $p$-Wannier functions, one has typically $\left|t_\parallel\right|\gg \left| t_\perp\right|$, with $t_\parallel=t_{x,x}$ and $t_\perp=t_{x,y}=t_{x,z}$.  
Moreover, the odd parity results in different signs,  $\sgn(t_\perp)=-\sgn(t_\parallel)$. 
\label{fig:tunnelingPOrbitals}
}
\end{figure}

To derive the Hubbard model for ultracold atoms in higher optical-lattice bands, one can proceed similarly to the derivation of the standard Hubbard model of spin-less particles, Eq.~\eqref{BoseHubbard}, explained in Sec.~\ref{cha:standardHub}. 
Generalizing Eq.~\eqref{wannier_field}, we expand the field operators in the Wannier basis of the higher band, 
\begin{equation}
\label{eq:fieldoperatorsToWannierPOrbitals}
\Psio\left(\vect{r}\right)=\sum_{i=1}^N\sum_{\mu=x,y,z} w_i^{\mu}\left(\vect{r}\right) \fo^{\mu}_{{i}}\,.
\end{equation}
Here, the operator $\fo^{\mu}_{i}$ destroys a fermion in the orbital $p_{\mu}$ at site ${i}$. The corresponding Wannier function $w_i^{\mu}\left(\vect{r}\right)$ is a product of $p_{\mu}$ function for $\mu$ and lowest $s$ functions for remaining directions. 
Using this expansion, the dynamics of ultracold atoms in higher optical-lattice bands can be described by the non-standard Hubbard model
\eqa{
\label{eq:Hpband}
\hat{H}&=&-\sum_{{i}=1}^N\sum_{\mu,\nu=x,y,z}
t_{\mu,\nu} (\fo^{\mu\,\dagger}_{i}\fo^{\mu}_{i+\vect{e}_{\nu}} + h.c.) \nonumber\\
& &+\sum_{{i}=1}^N\sum_{\mu,\nu,\mu',\nu'=x,y,z} V_{\mu,\nu,\mu^\prime,\nu^\prime} \fo_{{i}}^{{\mu^\prime}\,\dagger} \fo_{{i}}^{\nu^\prime\,\dagger} \fo_{{i}}^{\mu} \fo_{{i}}^{\nu}\,. 
}
The geometry that we consider here is a simple cubic lattice with spacing set to 1, and with unit vector $\vect{e}_{\nu}$ in direction $\nu=x,y,z$.
The nearest-neighbor tunneling matrix element $t_{\mu,\nu}$ describes the hopping of fermions in orbital $p_{\mu}$ along the direction $\vect{e}_{\nu}$.
As illustrated in Fig.~\ref{fig:tunnelingPOrbitals}, due to the odd parity of $p$-orbital Wannier wave functions, this tunneling does not couple orbitals with different principal axis. 
In conjunction with the anisotropy of the $p$-orbital, the tunneling becomes direction and orbital dependent
\cites{Isacsson05a,Kuklov2006,Liu06}, $t_{\mu,\nu}=t_\parallel \delta_{\mu,\nu} + t_\perp \left(1-\delta_{\mu,\nu}\right)$.
This spatial dependence is responsible for a good part of the rich physics of ultracold atoms in higher bands. 

Additionally, Hamiltonian \eqref{eq:Hpband} contains an on-site inter-orbital interaction term $V_{\mu,\nu,\mu^\prime,\nu^\prime}$. 
Typically, the interaction between fermionic atoms at low temperatures is weak. 
The reason is that the Pauli exclusion principle only allows scattering in high partial-wave channels ($p$, $f$, {\it etc.}), which are suppressed at low temperatures due to the angular momentum barrier. 
To realize strongly-correlated phases, however, strong fermion--fermion interactions are desirable. 
One way to increase the elastic scattering cross section is to employ a Feshbach resonance (FR) 
\cites{Chin10}. 
Typically, the FRs are generated by coupling channels in the electronic ground state through magnetic fields.  
For the case of $p$-waves, however, this method usually leads to significant atom losses through three-body inelastic collisions \cites{Regal2003,Zhang2004,Schunck2005,Guenter2005}.

As discussed in \cites{Goyal2010b,Hauke2011c} optical Feshbach resonances (OFRs) \cites{Theis04,Thalhammer05} should allow to enhance the $p$-wave scattering cross section while avoiding strong losses due to three-body recombination. 
Additionally, the OFR provides for a high degree of control, since, e.g., one can adjust the ratio of interaction strengths among different $p$-orbitals. 
In contrast to previous sections such as Sec.~\ref{GHB}, we consider here a regime where the interactions remain sufficiently small to allow neglecting off-site contributions. 

In \cites{Hauke2011c}, it was shown that in this case Hamiltonian \eqref{eq:Hpband} takes the form
\eqa{
\label{eq:Hpband_V123}
\hat{H} &=& -\sum_{{i}=1}^{N}\sum_{\mu,\nu=x,y,z}
t_{\mu,\nu} \left({{\fo^{\mu\,\dag}}_{i}} \fo^{\mu}_{i+\nu} + \mathrm{h.c.}\right)  \nonumber\\
& &+ \sum_{{i}=1}^{N} \Bigl[ V_1 \noe^{x}_{{i}} \noe^{y}_{{i}}  
+  V_2 \left( \noe^{x}_{{i}} \noe^{z}_{{i}} + \noe^{y}_{{i}} \noe^{z}_{{i}} \right) \\
& &\phantom{+\sum_{{i}=1}^{N}}+ \left( i V_3 {\fo^{x\,\dag}_{i}} \fo^{y}_{{i}} \noe^{z}_{{i}} + \mathrm{h.c.}\right) \Bigr].  \nonumber
}
Here, $\noe^{\mu}_{{i}}={\fo^{\mu\,\dag}}_{{i}}\fo^{\mu}_{{i}}$ is the number operator for fermions in orbital $\mu$ at site $i$. 
Due to the OFR, the relative strengths and signs of $V_{1,2,3}$ can be varied by changing the detuning of the OFR laser or the strength of a Zeeman splitting between internal atomic states.
The terms $V_1$ and $V_2$ denote usual on-site density--density interactions. 
Additionally, the OFR leads to the orbital-changing term $V_3$. 
Physically, it transforms $p_x$ into $p_y$ particles (and \emph{vice versa}). This allows them to explore the entire $xy$ plane, instead of being confined to a one-dimensional line, as is usually the case as long as $t_{\perp}$ can be neglected~\cites{Zhao2008}.

Hamiltonian \eqref{eq:Hpband} generalizes the Hubbard-like models of Refs.\ \cites{Rapp07,Rapp08,Zhao2008,Wu2008a,Miyatake2009,Toth2010}.
For the special case of $V_1=V_2$ and $V_3=0$, Hamiltonian \eqref{eq:Hpband} reduces to the SU(3) symmetric Hubbard model.
One can visualize $p$-band fermions as particles carrying a color index representing the $p_x$, $p_y$, and $p_z$ orbital state. 
Then, Hamiltonian \eqref{eq:Hpband} describes a three-color fermion model with color-dependent interaction $V_{1,2}$, a novel color-changing term $V_3$, and spatially anisotropic and color-dependent tunneling $t_{\mu,\nu}$. 
Since the $V_3$ term explicitly breaks time-reversal symmetry (TRS), we can expect it to lead to novel phases reflecting that intriguing property that lies at the heart of the topological-insulator states \cites{Hasan2010}.

An important limiting case of Hamiltonian \eqref{eq:Hpband} is the one where interactions dominate over tunneling terms, the so called strong-coupling limit. 
In Hubbard models of spinful $s$-band fermions, this limit leads to the emergence of Heisenberg and $t-J$ models, which are relevant for high-$T_c$ superconductivity. 
Different from these situations, in the case studied in reference \cites{Hauke2011c}, three orbital instead of two spin states are involved. 

In the strong-coupling limit of Hamiltonian \eqref{eq:Hpband},
\begin{equation}
\label{eq:Vggt}
\left|t_\parallel\right|\ll V_1 ,\quad \left|t_\parallel\right|\ll V_2-V_3 ,\quad \mathrm{and}\quad \left|t_\parallel\right|\ll V_2+V_3\,,
\end{equation} 
at average $p$-band filling of $1/3$, the low-energy manifold consists of states with one $p$-band particle per site. Since $\left|t_\perp\right|\ll \left|t_\parallel\right|\equiv t$, one can safely neglect perpendicular tunneling $t_\perp$ in this limit \cites{Zhao2008}.

\begin{figure}
\centering
\includegraphics[width=.9\columnwidth]{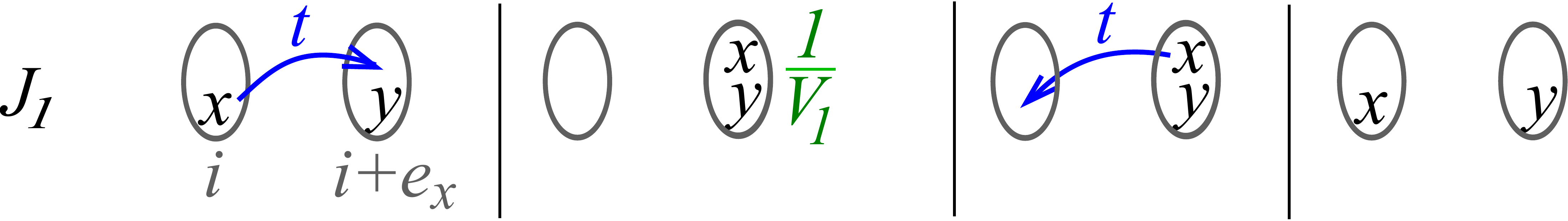}\\
\vspace*{1.3cm}
\includegraphics[width=.9\columnwidth]{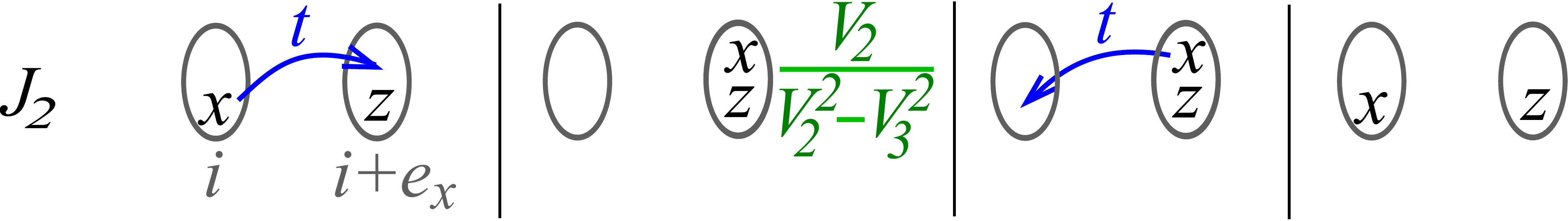}\\
\vspace*{1.3cm}
\includegraphics[width=.9\columnwidth]{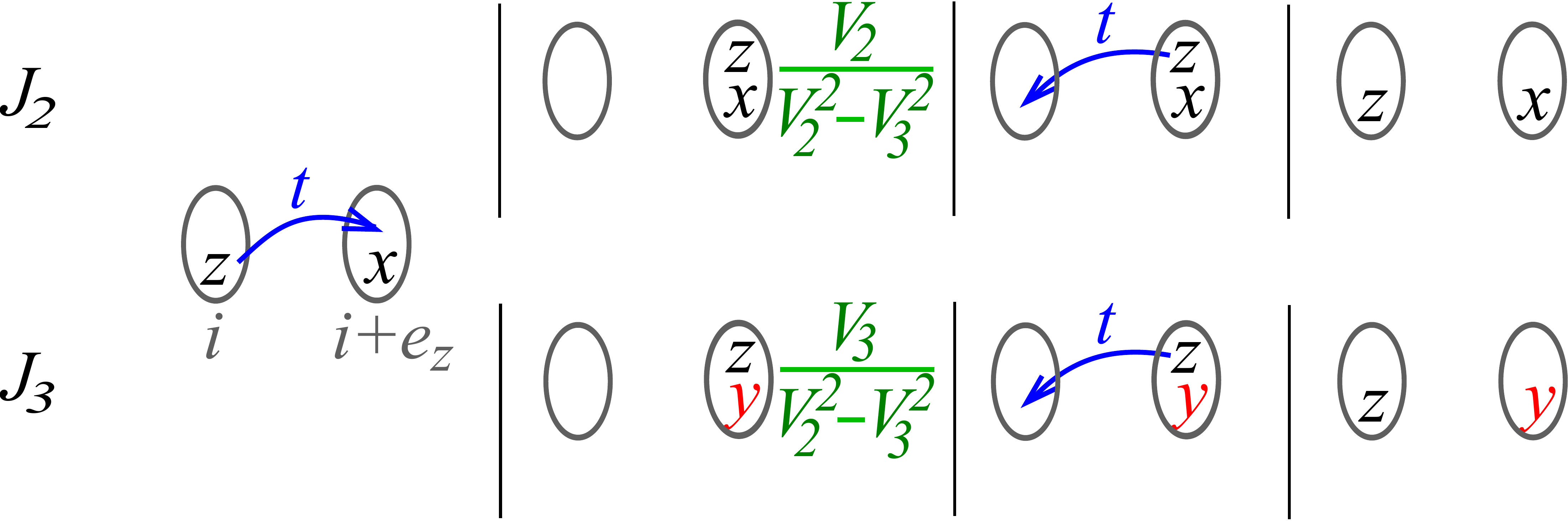}
\caption{
Sketch of the virtual hopping processes at $p$-band filling one-third (one $p$-band particle per site) leading to the effective Hamiltonian \eqref{eq:Heffonethird}.
If neighboring particles are in different orbitals $p_{\mu}$ and $p_{\nu}$  (abbreviated by $\mu$ and $\nu$, respectively), and if they are connected by a bond in $\mu$ or $\nu$ direction, a particle can tunnel with amplitude $t$ (blue) to a neighboring site (leftmost column). There, it experiences on-site interaction (green processes, second column). Due to the anisotropic tunneling, only the same particle can tunnel back (third column). Rightmost column: for the processes $J_{1}$ and $J_{2}$, the final  configuration is the same as the initial one, but in the orbital-changing process $J_3$ an $x$-particle has changed into a $y$-particle (bottom sketch). 
Neglecting $t_\perp$, the sketched processes -- plus the ones obtained by interchanging $x$ and $y$ -- are the only ones that can occur.
\label{fig:virtualHopping}
\from{Hauke2011c}
}
\end{figure}

The low-energy states are coupled via virtual hopping that induces exchange interactions between nearest neighbor orbitals (see Fig.\ \ref{fig:virtualHopping}). 
The resulting physics within the low-energy manifold is captured in an effective Hamiltonian that can be derived from second-order perturbation theory. 
Following this approach and treating the tunneling $t$ in \eqref{eq:Hpband} as a perturbation, one obtains the effective Hamiltonian for the low-energy manifold at $1/3$ filling 
\eqa{
\label{eq:Heffonethird}
\hat{H}_{\mathrm{eff}}&=&-\sum_{{i}} \Big\{ \sum_{\mu=x,y,z} J_{\mu} \noe^{\mu}_{{i}}\left(2- \noe^{\mu}_{{i}+\mu}- \noe^{\mu}_{{i}-\mu}\right) \nonumber\\
& &\quad\quad + \sum_{\mu=x,y}  \left(J_2-J_1\right) \noe^{\mu}_{{i}}\left(\noe^{z}_{{i}+\mu} +\noe^{z}_{{i}-\mu}\right) \\
										 & &\quad\quad 
											 +   J_3\left[ i \fo_{{i}}^{y\,\dagger} \fo^{x}_{{i}} \left(\noe^{z}_{{i}+z}+\noe^{z}_{{i}-z}\right)  +\mathrm{h.c.} \right]\Big\} \,.\nonumber
}
The resulting model is characterized by nearest-neighbor orbital interactions and the `correlated orbital-flipping' term $\sim J_3$. 
To write this model more compactly, we have used 
 $\noe^x_{i}+\noe^y_{i}+\noe^z_{i}=1$, and defined
\begin{eqnarray}
J_1&\equiv& t^2/V_1\,,\\
J_2&\equiv& t^2 V_2 / (V_2^2-V_3^2)\,,\\
J_3&\equiv& t^2 V_3/(V_2^2-V_3^2)\,, 
\end{eqnarray}
as well as 
$J_{x}=J_{y}=J_1$, $J_{z}=J_2$.
For $V_3=0$, $V_1=V_2$, Hamiltonian~\eqref{eq:Heffonethird} reduces 
to terms of the form $J_{\mu} \noe^{\mu}_{{i}}\noe^{\mu}_{i\pm\mu}$, 
a hallmark of the quantum 3-state Potts model.\footnote{Orbital order in a simpler model without OFR, and its relation to the Potts model were discussed  by C.\ Wu in the unpublished version of the work of arXiv:0801.0888v1.}

\begin{figure}
\centering
\begin{minipage}{0.69\columnwidth}
	\begin{flushleft}
		\includegraphics[width=0.9\textwidth]{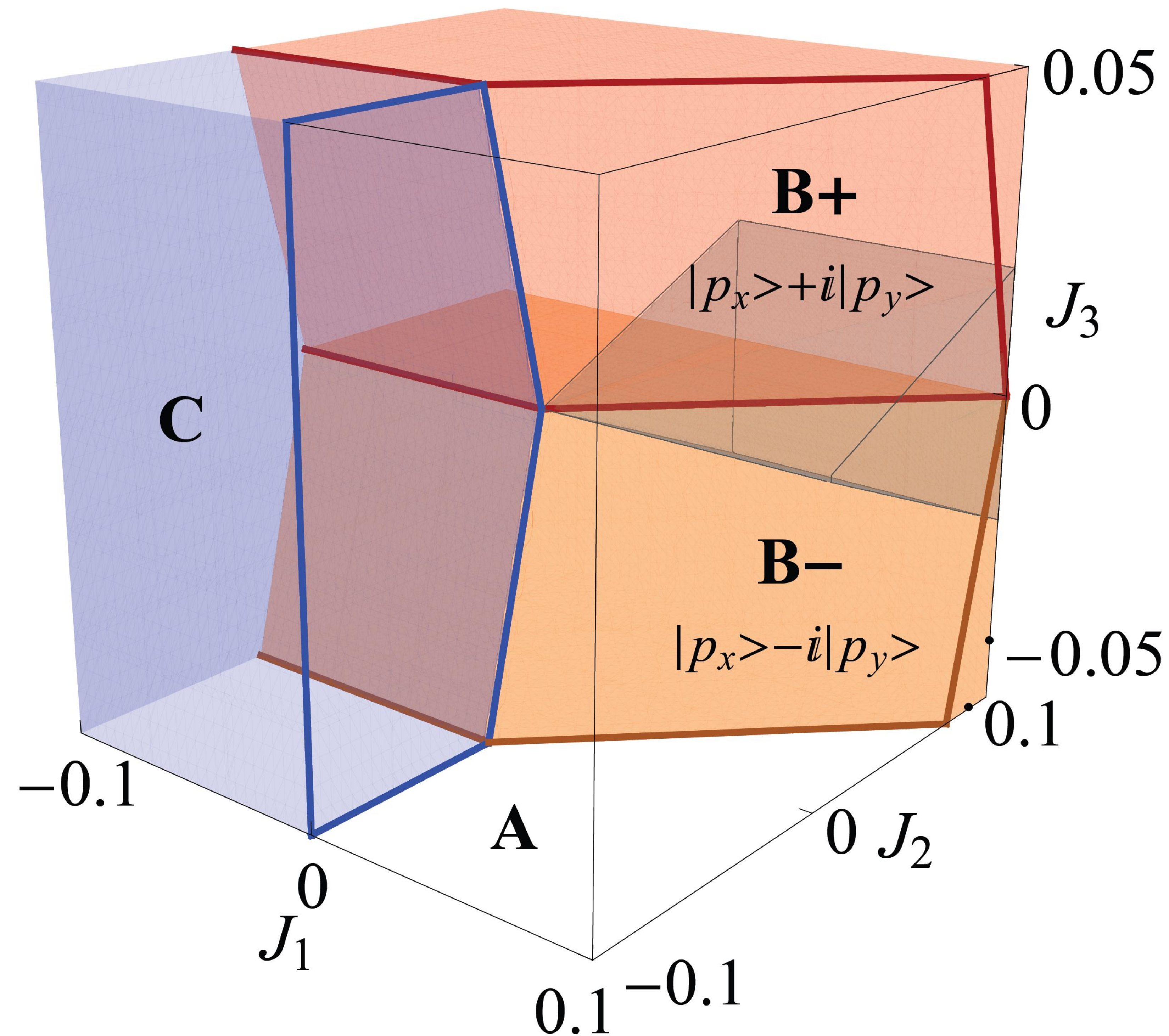}
	\end{flushleft}
\end{minipage} 
\begin{minipage}{0.3\columnwidth}
	\begin{flushright}
		\hspace{-1cm}
		\includegraphics[width=1.1\textwidth]{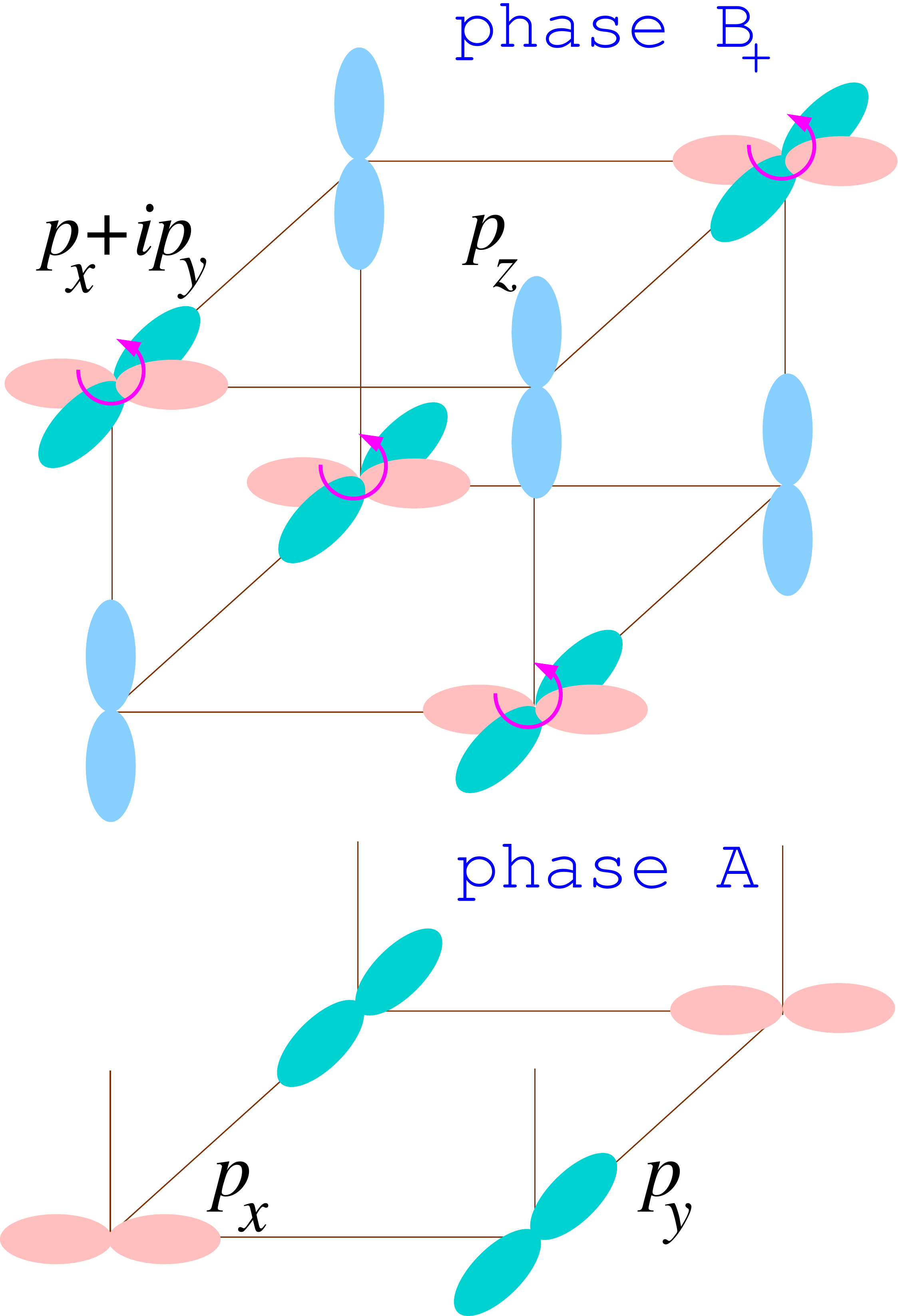}
	\end{flushright}
\end{minipage} 
\caption{ Left: 
Phase diagram of $p$-band Hubbard model,  Eq.~\eqref{eq:Heffonethird}, at $1/3$ filling. 
One finds four phases: phase A with antiferro-orbital order (empty region), phases B$_+$ and B$_-$ with axial orbital order (red/orange, for $J_3\gtrless0$), and 
phase C (blue) with tunneling completely frozen. 
The gray wedge indicates the region satisfying the strong-coupling conditions \eqref{eq:Vggt}. 
Right: sketch of phases A and B$_+$. In phase B$_+$, $\ket{p_z}$ and $\ket{p_x}+ i \ket{p_y}$ orbitals alternate.
Phase B$_-$ can be visualized from this by replacing $\ket{p_x}+ i \ket{p_y}$ with 
$\ket{p_x}- i \ket{p_y}$.
\from{Hauke2011c}
\label{fig:phdonethirdfilling}
}
\end{figure}

For positive couplings $J_{1,2}$, the first term of Hamiltonian \eqref{eq:Heffonethird} favors any configuration where the orbitals at neighboring sites differ, while for negative $J_{1,2}$ it favors configurations where the orbitals at neighboring sites are equal.
The second term favors an alternating pattern between $p_z$- and not-$p_z$-particles if $J_2>J_1$, and an alternating pattern between $p_x$ and $p_y$ if $J_2<J_1$.
The competition between these terms leads to the appearance of three different phases \cites{Hauke2011c}, c.f.\ the phase diagram in Fig.~\ref{fig:phdonethirdfilling}:
\begin{itemize}
\item[(A)] For $J_1>J_2+\left|J_3\right|/2$ and $J_1>0$ , the system is in an \emph{antiferro-orbital} phase: in each $xy$-plane, sites with $p_x$- and $p_y$-orbitals alternate, see Fig.~\ref{fig:phdonethirdfilling}, bottom right (similar to an antiferromagnetic N\'{e}el state in spin systems). 	
	Since $p_x$- and $p_y$-particles do not tunnel in $z$-direction, the $xy$-planes are decoupled. This phase has also been found in the two-dimensional model considered in \cites{Zhao2008}, where $J_3=0$.  
\item[(B)] For $J_1<J_2+\left|J_3\right|/2$ and $J_2>-\left|J_3\right|/2$ , the ground state shows \emph{axial orbital order}. 
	The state is bipartite with $\ket{p_z}$ on one sublattice and $\left(\ket{p_x}\pm i \ket{p_y}\right)/\sqrt{2}$ (for $J_3\gtrless 0$, respectively) on the other sublattice (right panel of Fig.\ \ref{fig:phdonethirdfilling}). 
	The state $\left(\ket{p_x}\pm i \ket{p_y}\right)/\sqrt{2}$ has \emph{finite angular momentum}, whence this novel phase breaks TRS.
At $J_3=0$, any superposition between $\ket{p_x}$ and $\ket{p_y}$ is degenerate, and TRS is restored. 
\item[(C)] For the case $J_1<0$ and $J_2<-\left|J_3\right|/2$, the ground state is highly degenerate, consisting of any configuration where $\alpha\beta$-planes are filled uniformly with $p_{\alpha}$ or $p_{\beta}$, where $\alpha\beta=xy,xz,yz$, thus preventing any tunneling. 
This phase, however, is not physical as it does not fulfill the strong-coupling requirements \eqref{eq:Vggt}.
\end{itemize}

As shown in \cites{Hauke2011c}, the characteristic shape of the $p$-band orbitals provides a direct possibility for distinguishing the different phases experimentally. Due to the non-trivial $p$-orbital Wannier functions, signatures of the distinct phases appear in the momentum distribution that can be observed in standard time-of-flight images. 
Other complex orbital configurations can be obtained even for $J_3=0$ by considering non-cubic lattices such as triangular or Kagome lattices, where frustration effects decrease ordering tendencies \cites{Zhao2008}. These can be detected, for example, in noise--noise correlation functions. 

As demonstrated by the simple example of spinless fermions in a cubic lattice, non-standard orbital Hubbard models allow for the exploration of exotic phenomena, such as time-reversal-symmetry breaking. 
At interfaces of two domains with opposite symmetry breaking, i.e., $p_x+i p_y$ and $p_x-i p_y$, respectively, chiral zero mode fermions may arise, similar to the edge states in spin-Hall insulators \cites{Hasan2010}. 
In the above example, the time-reversal symmetry was broken due to the $J_3$ term that appeared as a consequence of using an optical Feshbach resonance to enhance interactions. 
However, as discussed in the next section, non-standard  orbital Hubbard models offer also the possibility to observe the \emph{spontaneous} formation of topological states \cites{Li2012,Sowinski2013}, adding an exciting new direction to the research on higher bands of optical lattices. 

\subsection{Time-reversal symmetry breaking of $p$-orbital bosons}

In a recent proposal, Li and coworkers addressed the possibility of achieving spontaneous breaking of time-reversal symmetry using $p_x$ and $p_y$ orbitals in a one-dimensional lattice \cites{Li2012}. 
This interesting construction may be realized assuming an optical lattice potential of the form
\begin{equation}\label{poten}
V_{\mathrm{ext}}(\mathbf{r}) = V_{x}\sin^2(\pi x/a_x) + V_y\sin^2(\pi y/a_y) + \frac{m\Omega^2}{2} z^2,
\end{equation}
for a highly nonsymmetric lattice with $V_y\gg V_x$. Assuming $V_x/a_x^2=V_y/a_y^2$, within the harmonic approximation for the lattice sites $p_x$ and $p_y$ orbitals are degenerate  \cites{Li2012}. The asymmetric lattice depths and different lattice constants assure that the tunneling in the $y$ direction is suppressed and that the system consists of a one-dimensional chain of quasi-isotropic sites. In this arrangement, the tunnelings for $p_x$ and $p_y$ orbitals in $x$ direction differ in sign and in magnitude. The $p$-orbital bosons in such a lattice are argued to remain metastable (with a slow decay to $s$ orbitals), similarly to double well experiments \cites{Wirth10}.

The Hubbard-like Hamiltonian obtained using appropriate Wannier functions (the product of Wannier functions in $x$ and $y$ as well as ground state of harmonic oscillator in $z$ direction) reads \cites{Sowinski2013}
\begin{equation} \label{HamSow} 
\hat{\cal H} = \sum_{j} \hat{H}(j)-\sum_{\langle ij\rangle} \left[ t_x \hat{a}_x^\dagger(i) \hat{a}_x(j) + t_y \hat{a}_y^\dagger(i) \hat{a}_y(j)\right].
\end{equation}
Here, the local, on-site Hamiltonian $\hat{H}(j)$ has the form 
\bea \label{HamLoc}
\hat{H}(j) &=&\sum_{\alpha=x,y} \left[E_\alpha \hat{n}_\alpha(j) + \frac{U_{\alpha\alpha}}{2} \hat{n}_\alpha(j)(\hat{n}_\alpha(j)-1)\right] \\
&+&\frac{U_{xy}}{2} \left[4 \hat{n}_x(j)\hat{n}_y(j)+\hat{a}_x^\dagger(j)^2\hat{a}_y(j)^2+\hat{a}_y^\dagger(j)^2\hat{a}_x(j)^2\right]. \nonumber
\eea 
All $U$'s  represent contact interactions between different orbitals, and $E_x$ and $E_y$ are single particle energies. The Hamiltonian commutes with the operator for the total number of particles, $\hat{N}=\hat{N}_x+\hat{N}_y$, where $\hat{N}_\alpha = \sum_i \hat{n}_\alpha(i)$ [this is not valid for $\hat{N}_x$ and $\hat{N}_y$ separately,  due to the last two terms in (\ref{HamLoc}), which  transfer  {\em pairs} of bosons between different orbitals]. Thus the Hamiltonian has a global $Z_2$ symmetry related to the parity of the operator $\hat{N}_y$ (choosing $\hat{N}_x$ leads to the same conclusions), and it commutes with the symmetry operator ${\cal S}=\exp(i\pi\hat{N}_y)$. 

By introducing angular-momentum-like annihilation operators $\hat{a}_\pm(j) = \left[\hat{a}_x(j) \pm i \hat{a}_y(j)\right]/\sqrt{2}$, the local part of the Hamiltonian \eqref{HamSow} can be written in the form
\begin{align} \label{ourlocal}
\hat{H}(j) &= \frac{U}{2} \left[ \hat{n}(j)\left(\hat{n}(j)-\frac{2}{3}\right) - \frac{1}{3}\hat{L}_z^2(j)\right]  \nonumber \\
&+ \delta\left[\left(\hat{n}(j)-1\right)\left(\hat{L}_+(j)+\hat{L}_-(j)\right)\right] \nonumber \\
&+\lambda\left[\frac{1}{4}\hat{L}_z^2(j) -3\left(\hat{L}_+(j)-\hat{L}_-(j)\right)^2-\hat{n}(j)\right],
\end{align}
where $U = (U_{xx}+U_{yy})/2$, $\delta=(U_{xx}-U_{yy})/2$, and $\lambda=U_{xy}-U/3$ with   $\hat{n}(j) = \hat{a}_+^\dagger(j)\hat{a}_+(j)+\hat{a}_-^\dagger(j)\hat{a}_-(j)$, and angular momentum operators $\hat{L}_z(j)=\hat{a}^\dagger_+(j)\hat{a}_+(j) - \hat{a}^\dagger_-(j)\hat{a}_-(j)$, $\hat{L}_\pm(j) = \hat{a}^\dagger_\pm(j)\hat{a}_\mp(j) /2$. 

In the harmonic approximation, when the condition $V_x/a_x^2=V_y/a_y^2$ leading to orbital degeneracy is fulfilled, one has $E_x=E_y$, and considerable simplifications occur. In particular $U_{xx}=U_{yy}=3U_{xy}=U$, independently of 
the lattice depth. Thus, $\delta=\lambda=0$ and  $[\hat{H}(j), \hat{L}_z(j)]=0$ 
(i.e.\ eigenvalues of $\hat{L}_z(j)$ become good quantum numbers). This is no longer true when proper Wannier functions are used. Even for deep optical lattices this leads to important differences between both approaches.
Let us concentrate on the case of site filling of 3/2, as discussed in \cites{Li2012,Sowinski2013}. Consider the staggered angular momentum $\tilde L_z=\sum_j(-1)^j \langle \hat{L}_z(j)\rangle$, the $Z_2$ symmetry  order parameter \cites{Li2012}. In the harmonic approximation \cites{Li2012}, two superfluid phases are observed. For low tunneling, the system shows anti-ferro-orbital (AFO) order with staggered orbital current of $p_x\pm ip_y$ type, which spontaneously breaks time-reversal symmetry. With increasing tunneling strength, a phase transition to a paraorbital (PO) superfluid
is observed, where the staggered angular momentum $\tilde L_z$ vanishes.

\begin{figure}\label{liufig}
\centering\includegraphics[width=0.9\columnwidth]{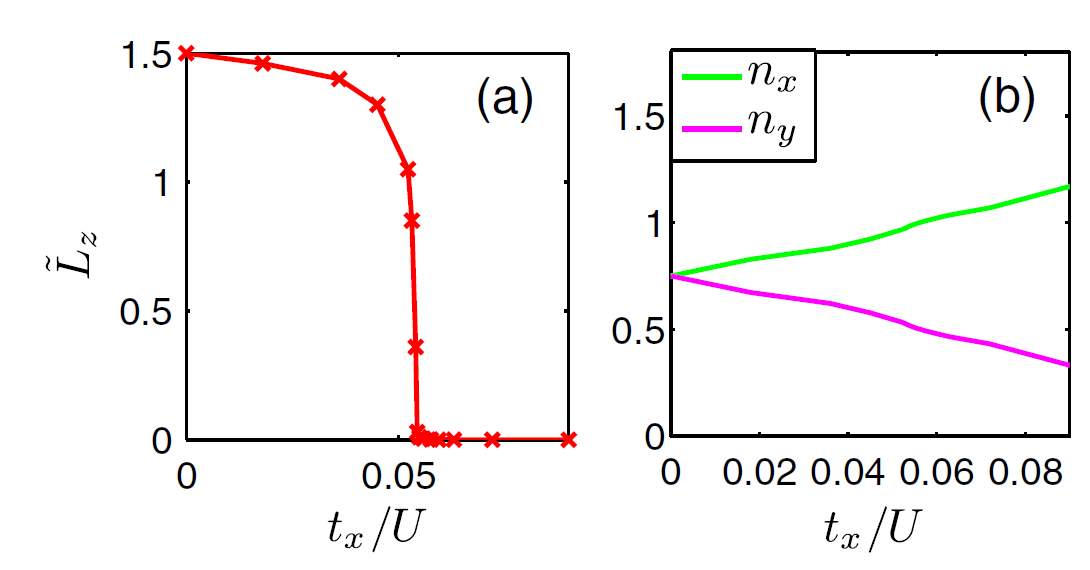}

\caption{ AFO-to-PO phase transition at filling 3/2 in the harmonic approximation. (a) staggered angular momentum ($Z_2$ order parameter)  
as a function of tunneling amplitude, for a fixed ratio of $|t_x/t_y|=9$. Panel (b) shows the filling of $p_x$ and $p_y$ orbitals. Numerical results are obtained using DMRG.
Reprinted figure with permission from X.~Li, Z.~Zhang, and W.~V.~Liu, \href{http://link.aps.org/abstract/PRL/v108/p175302}{Phys. Rev. Lett. {\bf 108}, 175302 (2012)}. Copyright (2012) by the American Physical Society.
}
\end{figure}

\begin{figure}\centering
\includegraphics[width=0.9\columnwidth]{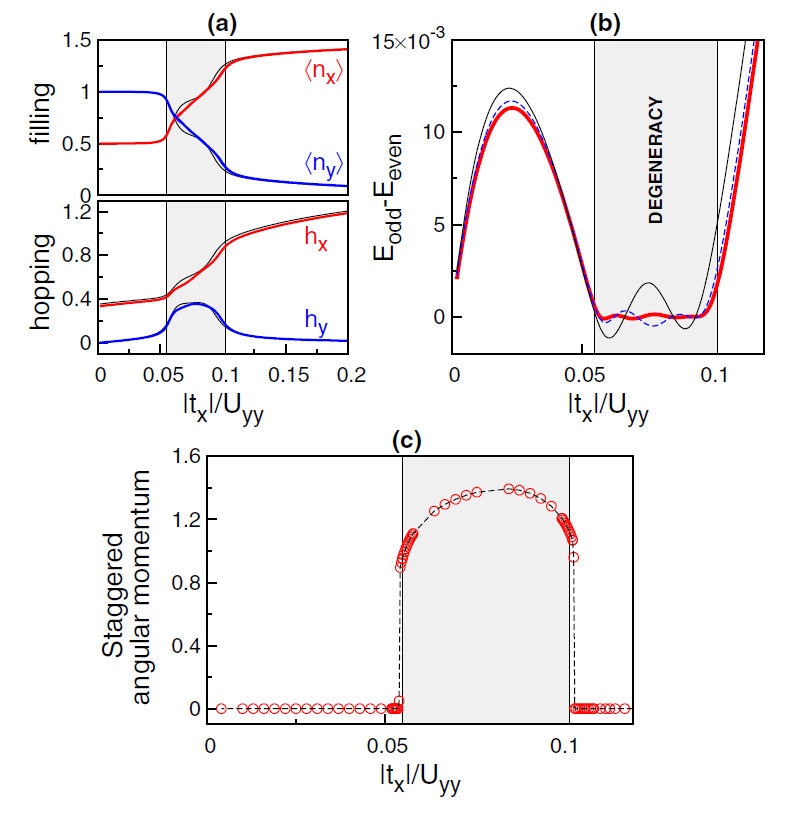}
\caption{(a) Filling and hopping of the $p_x$ (red line) and $p_y$ (blue line) orbitals  $\nu = 3/2$ obtained with ED method on a~lattice with $L=6$ sites. Results agree with corresponding results obtained for $L=4$ (thick black lines) and DMRG calculations (not showed since practically not distinguishable from ED data). (b) The energy difference between the two ground states in even and odd subspaces of the eigenstates of the symmetry operator ${\cal S}$. The energies are obtained with the ED method on the lattice with $L=4,6$, and $8$ sites (thin black, dashed blue, and thick red line respectively). Note that corresponding lines cross the zero energy  $L$ times. (c) Expectation value of the staggered angular momentum  $\hat{\cal L}_z/L$ as a~function of tunneling obtained with DMRG on the lattice with $L=64$ sites. Non vanishing value of $\hat{\cal L}_z$ is present only in the region where ground state is degenerate. In all figures the shaded region denotes the range of tunnelings where the 
ground state of the system 
is degenerate in the thermodynamic limit. Adopted from \cites{Sowinski2013}.\label{figliuour}}
\end{figure}

Interestingly, a quite different picture emerges when ``proper'' Wannier functions are used. Both $\delta$ and $\lambda$ in (\ref{ourlocal}) become different from zero, and, as a result, one has $[\hat{H}(j), \hat{L}_z(j)] \neq 0$, breaking the local axial symmetry. In \cites{Sowinski2013}, the system has been studied via exact diagonalization for small systems of length $L=4,6,$ and $8$ with periodic boundary conditions.  The lowest energy states in two eigen-subspaces of ${\cal S}$ were found independently. Let us call these states $|\mathtt{G}_{\rm even}\rangle$ 
and $|\mathtt{G}_{\rm odd}\rangle$ with corresponding eigenenergies $E_{\rm even}$ and $E_{\rm odd}$ (subscripts even/odd correspond to even/odd number of bosons in orbital $y$). The state with lower energy is the global ground state (GS) of the system. In principle, it may happen that both lowest states have the same energy. In such a case, any superposition $\cos(\theta)|\mathtt{G}_{\rm even}\rangle+\sin(\theta)\mathrm{e}^{i\varphi}|\mathtt{G}_{\rm odd}\rangle$ is a~ground state of the system. In the thermodynamic limit, this $U(1)\times U(1)$ symmetry is spontaneously broken to Ising-like $Z_2$ symmetry and only one of two macroscopic states can be realized \cites{Sowinski2013}.

Exact diagonalization in harmonically approximated system gives for  small tunnelings a degenerate GS,  i.e.\ $|\mathtt{G}_{\rm even}\rangle$ and $|\mathtt{G}_{\rm odd}\rangle$ have the same energy, reproducing the results of \cites{Li2012}. When the anharmonicity of the lattice wells is included, the picture changes (Fig.\ \ref{figliuour}a): for small tunnelings, the GS realizes an insulating state in the $p_y$ orbital with one boson per site, and a fractional superfluid state in the $p_x$ orbital. No significant correlation $\langle \hat{a}_x^\dagger(j)\hat{a}_y(j)\rangle$ is found in this limit. In contrast, for large tunneling all particles occupy the $p_x$ orbital in a superfluid phase, which is manifested by a large hopping correlation $h_x$, defined by 
\begin{equation} \label{hopping-def}
h_\alpha = \frac{1}{L}\sum_j\langle \hat{a}_\alpha^\dagger(j)\hat{a}_\alpha(j+1)\rangle,
\end{equation}
where $\alpha = x, y$.

The most interesting physics takes place for intermediate tunnelings. Figure \ref{figliuour}a, shows  that there exists a particular tunneling value for which both orbitals are equally populated. In the vicinity of this point, the GS is degenerate (Fig.~\ref{figliuour}b). More precisely, the degeneracy occurs exactly for $L$ different values of the tunneling within a certain finite range. The range of tunneling for which $E_{\rm odd}-E_{\rm even}=0$ does not grow with lattice size, but saturates.  This led the authors \cite{Sowinski2012} to claim that in the thermodynamic limit the degeneracy of the ground state is dynamically recovered in a certain well defined range of tunnelings. In this region, whenever the tunneling is changed, one particle is transferred between orbitals to minimize the energy. Since there is no corresponding term in the Hamiltonian, this transfer is directly related to the flip from one eigen-subspace of ${\cal S}$ to the other. 

In the region of recovered degeneracy, both ground states $|\mathtt{G}_{\rm even}\rangle$ and $|\mathtt{G}_{\rm odd}\rangle$ have the same energy. However, in the thermodynamic limit, due to the einselection principle \cites{Zurek}, the macroscopic state that is realized physically should exhibit as low entanglement as possible. Minimizing the von-Neumann entropy of the single-site density matrix, two orthogonal ground states $|\mathtt{G}_\pm\rangle = (|\mathtt{G}_{\rm even}\rangle\pm i|\mathtt{G}_{\rm odd}\rangle)/\sqrt{2}$ with the lowest entropy are found.
Importantly, an independent DMRG calculation revealed that the staggered angular momentum takes nonzero values for the intermediate tunneling region (compare  Fig.~\ref{figliuour}b).

As it turns out, the proper treatment using Wannier functions (and not their harmonic approximation) leads to tunneling-induced restoration of degeneracy and results in time-reversal symmetry breaking \cites{Sowinski2013}. The picture is quite different in the oversimplified harmonic approximation---even for deep lattices.


\section{Hubbard models with dynamical spin} 
\label{orbitaldipole}
\subsection{Mutual interactions of atomic magnets}
Weak dipolar interactions of magnetic moments of atoms, such as Chromium, Erbium, or Disprosium introduce some additional effects that are present only if the spins of the atoms are free. Then (as opposed to frozen spins aligned along the direction of an external magnetic field), the dipole-dipole interactions couple the spin of two particles to their orbital motion. As dipole-dipole interactions conserve the total angular momentum of interacting atoms, they do not conserve spin and orbital components separately. This simple observation leads directly to the Einstein-de Haas effect \cites{Kawaguchi2006, Einstein1915}, which in principle makes it possible to a transfer spin to orbital angular momentum and vice versa. The effect is a macroscopic illustration of the fact that spin contributes to the total angular momentum of a system on the same footing as the orbital angular momentum, and it is the most spectacular manifestation of the spin dynamics driven by the dipolar interactions and coupled to the orbital 
motion. 

In a more general case, when the axial symmetry condition is not met, the total angular momentum is not conserved. Spin-changing dipole-dipole interactions lead to a transfer of atoms from the ground to exited $p$ or/and $d$ states. In a lattice potential, such dipolar interactions with free spin couple the ground and excited bands of the lattice. Therefore, a very interesting class of Bose-Hubard models appears naturally if spin changing collisions are in  play that do not conserve total magnetization. The resulting necessity of taking into account the excited bands, with their relative occupation resulting from the spin-changing dynamics, significantly enriches the Bose-Hubbard physics. 

A number of interesting phases of matter have been predicted theoretically in the context of orbital quantum states in optical lattices. One of the core objectives is the theoretical prediction of conditions under which quantum states with excited Wannier states, in particular those with finite orbital angular momentum, can be realized on demand in the optical lattice. Here, mutual dipolar interactions appear to be a very good candidate for the controlled production of chosen quantum states in higher bands. 

An important feature of dipole-dipole interactions in the optical lattice is their high selectivity---there are very clear selection rules, which allow one to transfer angular momentum between certain, clearly defined spatial quantum states. These selection rules follows directly from the spatial symmetries of the system and energy conservation \cites{Gawryluk2007,Swislocki2011}. The resonant character of spin dynamics was recently observed in \cites{Paz2012}.  In this experiment, the first band excitations correspond to frequencies of $\omega/2\pi \approx 100$ kHz, which corresponds to energies significantly exceeding the dipole-dipole interaction energy  $E_D/\hbar = 0.1$ kHz. The spin dynamics is possible only on the expense of the Zeeman energy if an external magnetic field is applied.  The external magnetic field becomes therefore a very important knob triggering the dynamics and allowing to selectively choose the final band excitation. A theoretical prediction of the resonant values of this external 
magnetic field  in realistic experimental situations is quite difficult, because the spatial shape of the wave function is modified by the presence of  contact interactions between atoms \cites{Pietraszewicz2013}.

The resonant magnetic field is typically on the order of tens or hundreds of micro-Gauss, making the observation of the Einstein-de Haas effect difficult at present. Since dipole-dipole interactions are very weak, the resonances are also very narrow. This means that experimental realization needs high precision. On the other hand, it guarantees that dipolar coupling is highly selective. By choosing an appropriate value of the magnetic field, one can tune the transition of atoms to a particular spatial state. Indeed, controlling dipolar interactions is the crucial point in working with dipolar systems. Such a control has been recently achieved in chromium condensates \cites{Pasquiou2010}. It was shown that the external static magnetic field strongly influences the dipolar relaxation rate---there exists a range of magnetic field intensities where this relaxation rate is strongly reduced, allowing for the accurate determination of $S=6$ scattering length for chromium atoms. In \cites{Pasquiou2011}, a two-
dimensional optical lattice and a static magnetic field are used to control the dipolar relaxation into higher lattice bands. In this work, evidence for the existence of the relaxation threshold with respect to the intensity of the magnetic field is shown. As the authors of \cites{Pasquiou2011} claim, such an experimental setup might lead to the observation of the Einstein-de Haas effect. In the recent experiment of the same group \cites{Paz2012}, the resonant demagnetization of Chromium atoms in a 3D optical lattice was demonstrated. 

In the following, we focus on a model system with on-site axial symmetry. The model discussed will show generic features of Bose-Hubbard systems with dipole-dipole interaction under conditions of free magnetization.

\subsection{The many-body Hamiltonian}
To describe interacting spin-$S$ bosons, it is convenient to introduce the spinor field operator $\hat{\psi}_{m_S}({\bf r})$  annihilating particles in the state $m_S$ ($m_S=-S,\ldots,S$). Then the many-body Hamiltonian can be divided to three parts, 
\begin{equation}
\hat{ H} = \hat{ H}_0 + \hat{ H}_C + \hat{ H}_D.
\end{equation}
The first part is the single-particle Hamiltonian and it has the following form:
\begin{equation}
\label{single_particle}
\hat{ H}_0= \int\! {\rm d}\boldsymbol{r}\sum_m \hat{\psi}_m^{\dagger}(\boldsymbol{r})\left( \frac{\boldsymbol{p}^2}{2m}+V_{\rm ext}(\boldsymbol{r})-\gamma\, \boldsymbol{B}\cdot{\boldsymbol{S}}\right)\hat{\psi}_m(\boldsymbol{r}) \,,
\end{equation}
where, as before,  $m$ is the mass of the atom, $\boldsymbol{S}=(S_x,S_y,S_z)$ is an algebraic vector composed of spin matrices in appropriate representation, and $\gamma$ is the gyromagnetic coefficient. We again take the external potential of the optical lattice $V_{\rm ext}$ in a quasi two-dimensional arrangement \eqref{trap2d}.
The last term in~(\ref{single_particle}) is responsible for the linear Zeeman shift due to a uniform magnetic field $\boldsymbol{B}$. In what follows, we assume that the field is directed along the $z$-axis and it is weak enough to neglect the quadratic Zeeman effect.

The short-range interactions of dipolar atoms are typically described by a pseudo-potential. It can be written in a very general form \cites{Ho1998,Ohmi88,Santos2006}, 
\begin{equation}
\hat{ H}_C = \frac{1}{2}\int \mathrm{d}\boldsymbol{r} \sum_{s=0}^{2S}g_s\hat{\cal P}_s(\boldsymbol{r}),
\end{equation}
where $\hat{\cal P}_s$ are projector operators on different total spins, and $g_s$ is the $s$-wave scattering length for a total spin $s$ and it is given by $g_s=4\pi \hbar^2 a_s/M$. 

The long-range dipolar Hamiltonian can be written as
\begin{equation}
\label{H_D_1}
\hat{ H}_{D}=\frac{\gamma^2}{2} \int\mathrm{d}\boldsymbol{r}\mathrm{d}\boldsymbol{r}'
:\frac{\hat{h}_D(\boldsymbol{r},\boldsymbol{r}')}{|\boldsymbol{r}-\boldsymbol{r}'|^3}:,
\end{equation}
with the Hamiltonian density $\hat{h}_D(\boldsymbol{r},\boldsymbol{r}')$ (in the normal order as indicated by $: .. :$) of the form
\begin{equation}
\hat{h}_D(\boldsymbol{r},\boldsymbol{r}') = \hat{\boldsymbol{F}}(\boldsymbol{r})\cdot\hat{\boldsymbol{F}}(\boldsymbol{r}')-3\left[\hat{\boldsymbol{F}}(\boldsymbol{r})\cdot\boldsymbol{n}\right]\left[\hat{\boldsymbol{F}}(\boldsymbol{r}')\cdot\boldsymbol{n}\right].
\end{equation}
Here, $\hat{\boldsymbol{F}}=(\hat{F}_x,\hat{F}_y,\hat{F}_z)$ is an algebraic vector defined by $\hat{\boldsymbol{F}}(\boldsymbol{r}) = \sum_{ij}\hat{\psi}_i^\dagger(\boldsymbol{r}) \boldsymbol{S}_{ij} \hat{\psi}_j(\boldsymbol{r})$, and $\boldsymbol{n}$ is the unit vector in the direction of $\boldsymbol{r}-\boldsymbol{r}'$. By introducing ladder operators for the spin degree of freedom 
\begin{equation}
\label{F-matrix}
\hat{F}_{\pm}(\boldsymbol{r}) = \hat{F}_x(\boldsymbol{r}) \pm i \hat{F}_y(\boldsymbol{r}),
\end{equation}
one can rewrite the density of the dipolar Hamiltonian as 
\bea
\label{Hdip}
\hat{h}_D(\boldsymbol{r},\boldsymbol{r}')= &&\nonumber\\
\frac{(1-3n_z^2)}{4}[4\hat{F}_z&&(\boldsymbol{r}')\hat{F}_z(\boldsymbol{r})-\hat{F}_{+}(\boldsymbol{r}')\hat{F}_{-}(\boldsymbol{r})
-\hat{F}_{-}(\boldsymbol{r}')\hat{F}_{+}(\boldsymbol{r})] \nonumber \\
- \frac{3}{4} (n_x-in_y)^2&&\hat{F}_{+}(\boldsymbol{r}')\hat{F}_{+}(\boldsymbol{r}) - \frac{3}{4} (n_x+in_y)^2\hat{F}_{-}(\boldsymbol{r}')\hat{F}_{-}(\boldsymbol{r})  \nonumber \\
-\frac{3}{2}\, n_z(n_x-in_y&&) \left(\hat{F}_{+}(\boldsymbol{r}')\hat{F}_z(\boldsymbol{r})+\hat{F}_{z}({\bf  r}')\hat{F}_+(\boldsymbol{r})\right)   \nonumber \\
- \frac{3}{2}\, n_z(n_x+in_y&&)\left(\hat{F}_{-}({\bf r}')\hat{F}_z(\boldsymbol{r})+\hat{F}_{z}({\bf r}')\hat{F}_-(\boldsymbol{r})\right).
\eea
The form of~(\ref{Hdip}) facilitates a physical interpretation of all terms.  The first line represents dipolar interactions that do not lead to a change of the total magnetization of the field: the $z$-components of spin of both interacting atoms remain unchanged, or the $z$-component of one atom decreases by one while the $z$-component of the second atom increases by one. The second line collects terms describing processes where both interacting atoms simultaneously flip the $z$-axis projection of their spin: both by $+1$ or both by $-1$. Notice that the respective terms are multiplied by the phase factor $(n_x\mp in_y)^2$. This corresponds to a change of the projection of the orbital angular momentum of atoms in their center of mass frame by $-2$ or $2$ quanta. The last two lines describe processes in which the spin of one interacting atom is unchanged while the $z$-axis component of the spin of the other atom changes by $\pm 1$. This spin flipping term is multiplied by the phase factor $z(n_x\mp in_y)$, 
which signifies the change of the $z$-projection of relative orbital angular momentum of interacting atoms by $\mp 1$. Evidently, the dipolar interactions conserve the $z$-projection of the total angular momentum of interacting atoms.

\subsection{Two component model system with dynamical spin variable}
The simplest model of the extended Bose-Hubbard system with dynamical spin variable was discussed in \cites{Pietraszewicz2012}. In that model, realistic experimental parameters for spin-3 Chromium atoms confined in the 2D optical lattice were used. A significant simplification of the full many-body physics originates in choosing at each lattice site $(x_i,y_i)$ only two basis wavefunctions $\psi_{a}$ and $\psi_{b}$ of the form
\begin{align}
\psi_a (x,y,z) &= {\cal
W}_0(x){\cal W}_0(y){\cal G}(z), \\
\psi_b (x,y,z) &= \frac{1}{\sqrt{2}}\left[{\cal W}_1(x){\cal W}_0(y) + i{\cal W}_0(x){\cal W}_1(y)\right]{\cal G}(z). \nonumber
\end{align}
The function ${\cal G}(z)=\sqrt[4]{\Omega_z/\pi}\exp(-z^2\Omega_z/2)$ is the ground-state wave function of the 1D harmonic oscillator in $z$ direction, and the functions ${\cal W}_0(x)$ and ${\cal
W}_1(x)$ are the ground and the first excited Wannier states.
In this way, $\psi_a$ and $\psi_b$ form a single-particle basis of the two-component system. This basis
allows to account for the resonant transfer of atoms between
$m_S=3$, $l=0$ and $m_S=2$ and $l=1$ states in the presence of
a magnetic field aligned along the $z$-axis.
The lowest energy state $\psi_a (x,y,z)$ is effectively 
coupled to the excited state with
one quantum of orbital angular momentum $\psi_b (x,y,z)$.
The state is a single-site analogue of 
a harmonic oscillator state $\sim (x+iy) \exp[-(x^2+y^2)/2-z^2 \omega_z/2]$. 
The single-particle energies of the two basis states are denoted by $E_a$ and $E_b$ respectively.

The weakness of the dipolar interactions allows one to select the subspace of the two basis states.
There are several channels of binary dipolar collisions leading to different excited Wannier states.
However, one can choose the desired channel by a proper adjustment of the resonant external magnetic field as shown in \cites{Swislocki2011a}.
The energy difference between
atoms in the ground and in the excited Wannier states
is much larger than the dipolar energy, which is 
the smallest energy scale in the problem (except in the case of vanishing
tunneling), $E_{\rm dip}=10^{-4} E_r \ll E_b-E_a \sim E_r$.
However, at resonant magnetic field $B_0$, $E_a-g\mu_B B_0 = E_b$, the two energies are equal, and
efficient spin transfer between the components appears
on a typical time scale $\hbar/E_{\rm dip} \simeq 10^{-2}$s
($\mu_B$ above is the Bohr magneton, while $g$ is the Lande factor). 
The characteristic width of the resonances is small \cites{Gawryluk2011},
on the order of $E_{\rm dip} \approx g \mu_B B$, i.e.\ $ B \approx 100\mu$G.

Therefore,  a two-component system is realized with the 
$a$-component corresponding to atoms in the $m_S=3$ and $l=0$ state, 
while atoms in the $b$-component have $m_S=2$, $l=1$. The single-site basis states are $|n_a,n_b \rangle$, where $n_{\alpha}$ is a number
of atoms in the $\alpha$-component ($\alpha=a,b$). The Hamiltonian of the system reads
\begin{align}
\hat{ H}_{BH} =&  \sum_i \left[ (E_a-g\mu_B B)\,\hat{a}_i^\dagger \hat{a}_i + E_b\,\hat{b}_i^\dagger \hat{b}_i  +  U_{ab}\,\hat{a}_i^\dagger \hat{b}_i^\dagger \hat{a}_i\hat{b}_i \right. \nonumber \\
& \left. + \frac{U_a}{2}\hat{a}_i^{\dagger 2} {\hat{a}_i}^2  + \frac{U_b}{2}{\hat{b}_i}^{\dagger 2} {\hat{b}_i}^2   + 
D({\hat{b}_i}^{\dagger 2}  {\hat{a}_i}^2 + \hat{a}_i^{\dagger 2} {\hat{b}_i}^2) \right]\nonumber \\
& -\sum_{\langle i,j\rangle}\left[ t_a\, \hat{a}_i^\dagger \hat{a}_j  + t_b\, \hat{b}_i^\dagger \hat{b}_j \right].
\label{BH_Hamiltonian}
\end{align}
The values of parameters  depend on the lattice height $V_0$ and confining frequency $\Omega_z$. 
$U_a, U_b, U_{ab}$ are the contact interaction energies plus
the part of dipolar energy, which has the same form as the corresponding contact term,
$D$ is the on-site dipolar coupling of the two components, while $t_a$
and $t_b$ are tunneling energies (note that $t_a>0$ while $t_b<0$). 
This way, we arrive at a Hamiltonian that is an interesting modification of the
standard Bose-Hubbard model. 

Magnetic dipole-dipole interactions are very weak. Thus, in the above Hamiltonian the  dipole-dipole interactions between atoms at neighboring sites are neglected.
The on-site contact interactions $U_{a}, U_b$, and $U_{ab}$ cannot change the total spin \cites{Kawaguchi2012,Pasquiou2011}, and dipolar two-body interactions are much smaller than the contact ones. Therefore, one can keep only those dipolar terms that lead to spin dynamics.
The structure of the Hamiltonian is general for a two-component system with two spin species coupled by the dipole-dipole interactions of atomic magnetic moments, and can easily be adopted to more realistic situations of not -axially-symmetric and anharmonic lattice potentials. The 
main modification will be in choosing different single-particle basis states.

However, two comments are in order. 

i) The particular choice of the basis states was tailored to account for two-atom spin
flipping processes as selected by a proper adjustment of the magnetic field. Moreover, 
the chosen basis accounts for either two atoms in the $m_S=3$ ground state or two atoms in
the $m_S=2$ and $p_x+ip_y$ orbital state with
one quantum of orbital angular momentum. We neglected coupling 
of the ground-state atoms to the state with one atom in the ground state 
with $m_S=2$ and the second in the $m_S=2$ $d$-band state with two quanta of 
orbital angular momentum. This approximation
is justified if a small energy shift of $p$ and $d$ bands is taken into account while
the on-site potential remains axially symmetric. 

ii) The two-atom orbital $p_x+ip_y$ and $m_S=2$ state is coupled by the contact
interaction to the state with one  atom in $p_x+ip_y$ and $m_S=1$ 
and  the second in $p_x+ip_y$ and $m_S=3$. This coupling can be suppressed
due to the energy conservation by a light shift of the $m_S=2$ state  for example.   

Accounting for both of the above-mentioned  processes would require choosing not two, 
but rather three or four single-particle basis states. This would lead to multi-component Bose-Hubbard systems.

\begin{figure}
\includegraphics[width=90mm,natwidth=610,natheight=642]{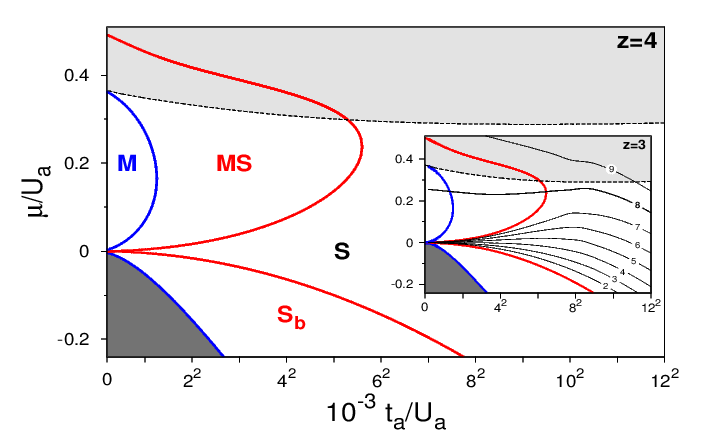}
\caption{Phase diagram for a 2D square lattice at the resonance ($z=4$). The
regions are: $M$ -- Mott insulator with one particle in equal superposition
od $a$ and $b$ states, $MS$ -- superfluid in  $a$ and $b$ components
($b$-dominated) and Mott insulator in the orthogonal superposition,  
$S$ -- superfluid phase of superposition of $a$ and $b$ components,
$S_b$ -- superfluid in the $b$-component.
The inset shows the diagram for $z=3$, together with chemical potential $\mu(N)$
for a given number of particles obtained from exact diagonalization. 
The lines, from bottom to top, correspond to occupation equal to 
$N=2, \ldots, 9$, as indicated. For $\mu> U_b$ (light gray region), the ground state 
of the system is a two-particle state. Therefore, in this regime the phases shown
are thermodynamically unstable. They are stable, however, with respect
to one-particle hopping. \from{Pietraszewicz2012}
}\label{fig:lob2}
\end{figure}

\begin{figure}
\includegraphics[width=85mm,natwidth=610,natheight=642]{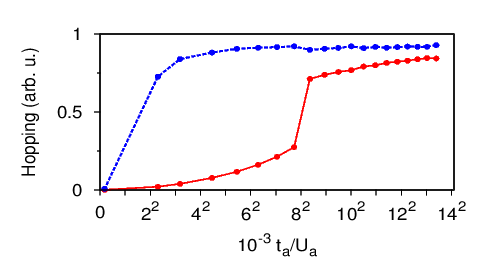}
\caption{Hopping for the lowest energy state
in a $2 \times 4$ plaquette obtained from exact diagonalization. Upper line -- $b$ component, lower line -- $a$ component.
\from{Pietraszewicz2012}}
\label{fig:exact}
\end{figure}

\subsection{Novel ground-state phases}
To get a flavor of physics described by the above model, one can limit the considerations to a small occupation of each lattice site with not more than one
particle per single site on average. We assume a resonant magnetic field with  
 $E_a-g\mu_B B_0=E_b$. This requires adjusting the magnetic field in accordance with the lattice depth, $B_0=B_0(V_0)$. 

The dipolar interactions couple ground and excited Wannier states due to
the tunneling in a higher-order process even for a low density. 
The transfer between $|1,0\rangle$ and  $|0,1\rangle$ states is realized as a sequence of adding an atom to the $a$-component at a given single site $|1,0\rangle \rightarrow
|2,0\rangle$ via tunneling, followed by the dipolar transfer of both $a$-species
atoms to the excited Wannier state $|2,0\rangle \rightarrow
|0,2\rangle$, and finally the tunneling that removes one
$b$-component atom from the site, $|0,2\rangle \rightarrow
|0,1\rangle$. In this way, the  two states are coupled provided that tunneling is nonzero.

Thermodynamically stable phases of the system may be found following the standard mean-field approach of \cites{Fisher1989}. Assuming a translationally invariant ground state
(since the Hamiltonian (\ref{BH_Hamiltonian}) enjoys that symmetry), and introducing
superfluid order parameters for both components, $\phi_{(a)}=\langle
a_i \rangle$ and  $\phi_{(b)}= \langle b_i \rangle$, as well as
the chemical potential $\mu$, the mean-field Hamiltonian of the system 
is a sum of single-site Hamiltonians $\hat{H}_0+\hat{H}_I$, with 
\bea
\label{H0}
\hat{H}_{0} &=& -\mu(\hat{a}^\dagger \hat{a}+\hat{b}^\dagger \hat{b})  + \frac{1}{2}U_a \hat{a}^\dagger \hat{a}^\dagger \hat{a}\hat{a} +\frac{1}{2} U_b \hat{b}^\dagger \hat{b}^\dagger \hat{b} \hat{b} \nonumber \\
&+& U_{ab} \hat{a}^\dagger \hat{b}^\dagger \hat{a}\hat{b} + D(\hat{b}^\dagger \hat{b}^\dagger \hat{a}\hat{a} + \hat{a}^\dagger \hat{a}^\dagger \hat{b}\hat{b}),\\
\hat{H}_{I} &=& -zt_a {\phi}^*_{(a)} \hat{a}  - zt_b {\phi}^*_{(b)}\hat{b} +h.c.
\eea
with site indices omitted and
$z$ being the coordination number (for a 2D square lattice $z=4$). 
The Hamiltonian $\hat{H}_0+\hat{H}_I$ does not conserve the number of particles: it describes 
a single site coupled to a particle reservoir.  
The order parameters $\phi_{(a)}$ and $\phi_{(b)}$ vanish in the MI phase
and hopping of atoms vanishes. Close to  the boundary, on the
SF side, $\phi_{(a)}$ and $\phi_{(b)}$ can be treated as small
parameters in the perturbation theory.

The single-site ground state becomes unstable if the mean fields $\phi_{(a)}$ 
or $\phi_{(b)}$ are different from zero.  The self-consistency condition 
\begin{align}
\phi_{(c)}=\lim_{\beta\rightarrow \infty}\mathrm{Tr}\left[\hat{c}\,\mathrm{e}^{-\beta( H_0+H_I)}\right]/Z({\beta}), 
\label{sfcond}
\end{align}
where $c=a,b$, allows one to  find the mean fields numerically.
In the lowest order, the set of equations (\ref{sfcond})
becomes linear and homogeneous.  Thus, nonzero solutions for $\phi_{(c)}$ are obtained from the necessary condition of a vanishing determinant of (\ref{sfcond}). It yields the lobes shown in Fig.~\ref{fig:lob2}. 

In the limit $\beta \rightarrow\infty$, the partition function reduces to the single contribution of the lowest-energy state, $Z(\beta)=\mathrm{e}^{-\beta E_0}$. 
For $\mu <U_{b}<U_{a}$, the only contribution to Eq.~(\ref{sfcond}) comes from eigenstates of the Hamiltonian  with  zero, one, and two particles.

The single-site ground state is the $|0,0\rangle$ vacuum state (dark gray region in Fig.~\ref{fig:lob2}) for $\mu<0$ and small tunnelings. 
With increasing tunneling (and fixed $\mu$), particles appear in the superfluid vortex $b$-phase (labeled as $S_b$ in Fig.~\ref{fig:lob2}).  
Only at larger tunnelings, some atoms do appear in the $a$-component and both `standard' and $p_x+ip_y$ orbital superfluids coexist ($S$). 

The situation becomes richer for $0<\mu < U_{b}$.
At the resonance, $B=B_0$, the
ground state is degenerate if tunnelings are neglected:  the states $|1,0\rangle$ and $|0,1\rangle$ 
have the same energy, $E_0=-\mu$.
This degeneracy is lifted for non-zero  tunnelings. Additionally, the position of the resonance is shifted then towards smaller magnetic fields. 
The effective Hamiltonian  in the resonant region possesses a single-site ground state that is a superposition of both components, 
$|g\rangle= \alpha_1 |1,0\rangle- \alpha_2 |0,1\rangle$. 
While crossing the resonance, the ground state switches from $|1,0\rangle$ 
to $|0,1\rangle$. Exactly at  resonance $\alpha_1=\alpha_2 =1/\sqrt{2}$. A perturbative analysis
allows one to estimate the width of the resonance $\Delta B$  to be $g\mu_B |\Delta B| \approx 10^{-6} E_r$ 
for $V_0=25 E_r$. For a shallower lattice,
$V_0 = 10 E_r$, the resonant region is broader, $g\mu_B |\Delta B| \approx 10^{-3} E_r$.
Unfortunately, due to its small width,  the resonance  can be hardly accessible particularly for small tunnelings. 
Away from the resonance, the standard phase diagrams for the $a$ or $b$ component emerge.

In Fig.~\ref{fig:lob2}, we show regions of stability of the different possible phases of the system at resonance, i.e.\ when  
$|g\rangle= (|1,0\rangle - |0,1\rangle)/\sqrt{2}$. 
The system is in the Mott insulating phase (M) with one atom per site for small tunnelings. 
Still, every atom is in the superposition of the ground and the vortex Wannier state. At the border of the Mott lobe (blue line), Eq.~(\ref{sfcond}) allows for nonzero solutions for  $\phi_{(a)}$
and $\phi_{(b)}$.  Expressing $\hat{H}_I$ in terms of the composite bosonic operators $\hat{A}^{\dagger}=(\kappa_a \hat{a}^{\dagger} + \kappa_b \hat{b}^{\dagger})$
and $\hat{B}^{\dagger}=(-\kappa_b \hat{a}^{\dagger} + \kappa_a \hat{b}^{\dagger})$, where $\kappa_a^2+\kappa_b^2=1$, 
allows one to diagonalize $\hat{H}_I$, with coefficients $\kappa_i$ depending  on the tunnelings $t_a$ and $t_b$.
These composite operators create an atom in two orthogonal superpositions of $a$ and $b$ states.
At the border of the Mott phase, the mean value of the operator $\hat{B}$ is different from zero, and a non vanishing 
superfluid component,  $\Psi_B = -\kappa_b \phi_{(a)} + \kappa_a  \phi_{(b)}$, 
appears in the MS region. The ratio $(\kappa_b/\kappa_a)^2 \simeq 0.02$ is small  
at the edge of stability of the Mott insulator. Therefore, $\hat{B}^{\dagger} \simeq \hat{b}^{\dagger}$,
i.e.\ the superfluid $\Psi_B$ is dominated by the orbital $b$-component. On the other hand, in the discussed region the 
mean field corresponding to the $\hat{A}^{\dagger} \simeq \hat{a}^{\dagger}$ operator vanishes.
The system is therefore in an equal superposition of the Mott insulating 
and superfluid phases. The Mott phase is 
dominated by the $a$-component and the superfluid phase consists mainly of 
the $b$-particles. Both components, however, contain a small minority of the remaining species.

The system undergoes yet another phase transition for larger tunnelings, as Eq.~(\ref{sfcond}) allows for
another nonzero mean field. At this transition, the departure of the mean value of $\hat{A}$ form zero defines the border of the `bigger'
lobe, and the Mott component of the ground Wannier state becomes unstable.
The additional mean field $\Psi_A = \kappa_a \phi_{(a)} + \kappa_b \phi_{(b)}$ appears in the (S) region.
As before, $(\kappa_b/\kappa_a)^2 \simeq 0.06$ is small. The $a$-species 
dominates the $\Psi_A$ superfluid component. Both $\Psi_A$ and $\Psi_B$ superfluids exist in the (S) region.   

A qualitatively support for the above mean-field findings is obtained by a direct inspection of the 
true many-body ground state, obtained by exact diagonalization of 
the many-body Hamiltonian in a small $2\times 4$ rectangular plaquette
with periodic boundary conditions for total number of particles $N=1, \ldots, 10$.
For such a small system, each site has three neighbors, i.e.\ $z=3$, and the resonance condition is obtained by finding the magnetic field for which both $a$ and $b$ species
are {\it equally populated}. In the inset of Fig.~\ref{fig:lob2}, the exact results are compared with the mean-field results for $z=3$. 
The lines correspond to the constant number of particles per site obtained from the relation
$\mu(N) = \left[E_0(N+1)-E_0(N-1)\right]/2$. In that way, one may
trace the phases the system enters while adiabatically changing the tunneling at fixed particle number. 
The (M) and (MS) phases can be reached with one particle per site only (8 particles in the plaquette).

It is also worthwhile to consider the hopping averages, defined as the mean values of the following hopping operators:
$h_{a} =\sum_{\langle j\rangle}
\langle \hat{a}^\dagger_j \hat{a}_i \rangle$ and $h_b=\sum_{\langle j
\rangle} \langle \hat{b}^\dagger_j \hat{b}_i \rangle$. 
They annihilate a particle at a given site and put it in a neighboring site, and may be thought of as the number conserving analogues of the mean fields $\phi_{(a)}$ and $\phi_{(b)}$, 
which in exact diagonalizations without symmetry-breaking terms always vanish. 
In Fig.~\ref{fig:exact}, the hoppings for the case of one particle per site are shown.
For large tunnelings, both $a$ and $b$ hoppings are large---the components are 
in the superfluid phase. 
When entering the MS phase at $t_a/U_a \simeq 0.064$, the hopping
of the $a$-component rapidly decreases while the hopping of the $b$-phase  remains large---the system
enters an $a$-component-dominated Mott insulator superimposed with a $b$-component dominated superfluid.
At $t_a/U_a \simeq 0.002$, both hoppings tend to zero---the system enters the Mott phase
with equal occupation of both species. These results confirm the findings based on the mean-field approach.

The effective two-state model studied exhibits a number of exotic phases. One might think that the model Hamiltonian crucially depends on the assumed axial symmetry and harmonicity of a single lattice site, which is justified in deep lattices only \cites{Collin2010,Martikainen2011,Pietraszewicz2013}. Including an anharmonic correction requires some modification, but the structure of the system Hamiltonian remains the same in many cases. In an anharmonic and non-axially-symmetry potential the vortex-like final state is no longer an eigenstate of the Hamiltonian. 
Anharmonicity and anisotropy combined with contact interactions lead to a fine structure of two-body energies in the lattice site. The vortex state is split into three two-particle states which can be {\it separately} addressed by an appropriate choice of the magnetic field. Therefore, a two-states structure of the Hamiltonian becomes generic for the systems studied.  
The model discussed here describes the whole class of two-states systems with dipole-dipole interactions and free magnetization, under the resonance condition of equal energies of the two coupled states.



\section{1D and 2D models of the \textit{Salerno type}: the mean-field and
quantum versions}

\subsection{Introduction}

\xdl A natural part of the analysis of Bose--Hubbard models is the consideration 
of their mean-field limit, which corresponds to classical lattice 
models desribed by discrete nonlinear Schr\"{o}dinger (DNLS) equations (see, e.g., recent works
\cites{Carr,Padova} and references therein). In particular,  Ref.\ \cites{Carr} highlights the correspondence between the two descriptions of a system of ultracold bosons in a one-dimensional
optical lattice potential: (1) the discrete nonlinear Schr\"odinger equation, a discrete mean-field theory, and (2) the Bose Hubbard
Hamiltonian, a discrete quantum-field theory. This discussion includes, in particular, formation of solitons. 

 In this vein, the mean-field 
limit of the nonstandard Bose-Hubbard models, whose characteristic feature is a nonlinear 
coupling between adjacent sites of the underlying lattice, is represented by 
classical lattice models featuring a similar nonlinear interaction between 
nearest-neighbor sites. They form a class of systems known as \textit{Salerno models} 
(SMs). In the one-dimensional (1D) form, the SM was first introduced by Mario Salerno in
1992 \cite{Salerno92} as a combination of the integrable \textit{%
Ablowitz-Ladik} (AL) system \cite{Ablowitz76} and nonintegrable
DNLS equation. \xe The former
system is a remarkable mathematical model, but it does not have
straightforward physical implementations, while the DNLS equations
find a large number of realizations, especially in nonlinear optics
and Bose-Einstein condensates (see Sec.~\ref{BEC}). For this reason, the DNLS
equation has been a subject of numerous analytical, numerical, and
experimental studies, many of which were summarized in the book
\cite{Kevrekidis09}. The objective of the present chapter is to
introduce the mean-field (classical) and quantum versions of the SM
in one and two dimensions (in fact, the quantum version is
considered only in 1D), and survey results obtained for localized
models (discrete solitons) in the framework of 1D and 2D
realizations of the mean-field version. An essential peculiarity of
the SM is a nonstandard form of the Poisson bracket in its classical
form, and, accordingly, a specific form of the commutation relations
in its quantum version. These features are, as a matter of fact,
another manifestation of the nonstandard character of Hubbard models
with nonlinear coupling between adjacent sites.

\subsection{One-dimensional Salerno models and discrete solitons}

\subsubsection{The formulation of the model}

It is well known that, while the straightforward discretization of
the 1D nonlinear Schr\"{o}dinger equation is nonintegrable, there is
a special form of the discretization, namely, the AL model, which
keeps the integrability, and admits generic exact solutions for
standing and moving solitons, as well as exact solutions for
collisions between them \cite{Ablowitz76}. Unlike the exceptional
case of the analytically solvable AL model, discrete solitons in
nonintegrable systems are looked for in a numerical form, or
(sometimes) by means of the variational approximation
\cite{Malomed12, Papacharalampous12,VanGorder12}. Nevertheless,
there are some specially devised 1D nonintegrable models in which
\emph{particular} exact soliton solutions can be found, too
\cite{Kevrekidis03,Oxtoby07,Malomed06}.

As the DNLS and AL equations differ in the type of the nonlinear terms
(onsite or intersite ones), and converge to a common continuum limit in the
form of the ordinary integrable nonlinear Schr\"{o}dinger equation, a
combined discrete model may be naturally introduced, mixing the cubic terms
of both types. Known as the SM \cite{Salerno92}, the 1D version of this
combined system is based on the following discrete equation:
\begin{equation}
i\frac{d}{dt}\Phi _{n}=-\left( \Phi _{n+1}+\Phi _{n-1}\right) \left( 1+\mu
\left\vert \Phi _{n}\right\vert ^{2}\right) -2\nu \left\vert \Phi
_{n}\right\vert ^{2}\Phi _{n},  \label{Salerno}
\end{equation}%
where $\Phi _{n}$ is the complex classical field variable at the $n$-th site
of the lattice, while real coefficients $\mu $ and $\nu $ account for the
nonlinearities of the AL and DNLS types, respectively. The celebrated
integrable AL equation proper corresponds to $\nu =0$,
\begin{equation}
i\frac{d}{dt}\Phi _{n}=-\left( \Phi _{n+1}+\Phi _{n-1}\right) \left( 1+\mu
\left\vert \Phi _{n}\right\vert ^{2}\right) .  \label{ALproper}
\end{equation}

In Eq.~(\ref{Salerno}) with $\nu \neq 0$, negative $\nu $ can be made
positive by means of the \textit{staggering transformation}, $\Phi
_{n}\equiv (-1)^{n}\tilde{\Phi}_{n}^{\ast }$ (the asterisk stands for the
complex conjugation), and then one may fix $\nu \equiv +1$, by way of
rescaling, $\tilde{\Phi}_{n}\equiv \tilde{\Phi}_{n}^{\prime }/\sqrt{|\nu |}$%
. Therefore, a natural choice is to fix $\nu \equiv +1$, unless one
wants to consider the AL model per se, with $\nu =0$. On the
contrary to that, the sign of the coefficient $\mu $, which
characterizes the relative strength of the nonlinear AL coupling
between the nearest neighbors, \emph{cannot} be altered. In
particular, the AL model with ($\nu =0$) and $\mu <0$ has no
(bright) soliton solutions.

The SM equation (\ref{Salerno}), as well as its AL counterpart (\ref%
{ALproper}), conserve the total norm, which is different from the
``naive" expression relevant in the case of the DNLS equation,
\begin{equation}
\mathcal{N}_{\mathrm{DNLS}}=\sum_{n}|\Phi _{n}|^{2}.  \label{naive}
\end{equation}%
For both equations (\ref{Salerno}) and (\ref{ALproper}), the conserved norm
is \cite{Ablowitz76, Cai96, Rasmussen97}
\begin{equation}
\mathcal{N}=\frac{1}{\mu }\sum_{n}{\ln }\left( \left\vert 1+\mu |\Phi
_{n}|^{2}\right\vert \right) .  \label{eq:Norm}
\end{equation}%
Note that expression (\ref{eq:Norm}) does not depend on $\nu $. Therefore, it
is identical for the SM\ and AL models, carrying over into the simple
expression (\ref{naive}) in the limit of $\mu \rightarrow 0$.

In addition to the norm, the other dynamical invariant of Eq.~(\ref{Salerno}%
) is its Hamiltonian, which, as well as the norm, has a somewhat
tricky form \cite{Cai96, Rasmussen97} (which has its consequences
for the identification of the
symplectic structure of the SM and its quantization, see below):%
\begin{equation}
\mathcal{H}=-\sum_{n}\left[ \left( \Phi _{n}\Phi _{n+1}^{\ast }+\Phi
_{n+1}\Phi _{n}^{\ast }\right) +\frac{2}{\mu }|\Phi _{n}|^{2}\right] +\frac{2%
}{\mu }\mathcal{N},  \label{eq:SalernoEnergy}
\end{equation}%
where the above normalization, $\nu =+1$, is adopted. In the limiting case of
the DNLS equation $\mu \rightarrow 0$, the expansion of Hamiltonian (\ref%
{eq:SalernoEnergy}) in powers of $\mu $ yields the usual expression
for the DNLS equation:%
\begin{equation}
\mathcal{H}_{\mathrm{DNLS}}=-\sum_{n}\left[ \left( \Phi _{n}\Phi
_{n+1}^{\ast }+\Phi _{n+1}\Phi _{n}^{\ast }\right) +|\Phi _{n}|^{4}\right] .
\end{equation}%
The Hamiltonian of the AL proper can be obtained from the general
expression (\ref{eq:SalernoEnergy}) by taking the limit of $\mu
\rightarrow \infty $, which produces a simple expression \cite{Ablowitz76}:
\begin{equation}
\mathcal{H}_{\mathrm{AL}}=-\sum_{n}\left( \Phi _{n}\Phi _{n+1}^{\ast }+\Phi
_{n+1}\Phi _{n}^{\ast }\right) .  \label{H-AL}
\end{equation}

\subsubsection{Solitons}

The AL equation (\ref{ALproper}) gives rise to exact solutions for
(bright) solitons in the case of the self-focusing nonlinearity, $\mu >0$.
Then, one may set $%
\mu \equiv +1$ by means of obvious rescaling, and the exact soliton
solutions take the following form 
\begin{equation}
\Phi _{n}(t)=\left( \sinh \beta \right) \mathrm{sech}\left[ \beta (n-\xi (t))%
\right] \exp \left[ i\alpha \left( n-\xi (t)\right) -i\varphi (t)\right] ,
\label{ALsoliton}
\end{equation}%
where $\beta $ and $\alpha $ are arbitrary real parameters that determine
the soliton's amplitude, $A\equiv \sinh \beta $, its velocity, $V$, and its
intrinsic frequency, $\Omega $,%
\bea
V&\equiv& \frac{d\xi }{dt}=\frac{2\sinh \beta }{\beta }\sin \alpha ,\nonumber\\
\Omega
&\equiv& \frac{d\varphi }{dt}=-2\left[ \left( \cos \alpha \right) \cosh \beta
+\alpha \sin \alpha \frac{\sinh \beta }{\beta }\right] .  \label{VOmega}
\eea

The SM with $\nu =+1$ (as fixed above) and $\mu >0$, i.e., with \emph{%
noncompeting} onsite and intersite self-focusing nonlinearities, was studied
in a number of works, see Refs.~\cite{Cai96, Rasmussen97, Cai97, Dmitriev03} and references
therein. It has been demonstrated that Eq.~(\ref{Salerno}) gives rise to
static (and, sometimes, moving \cite{Cai97}) solitons at all
positive values of $\mu $. In particular, a nontrivial problem is the
mobility of the discrete solitons in the DNLS limit, which corresponds to $%
\mu =0$ \cite{Ablowitz02, Papacharalampous12}.

\begin{figure}[tbp]
\begin{center}
\includegraphics[width=90mm,natwidth=610,natheight=642]{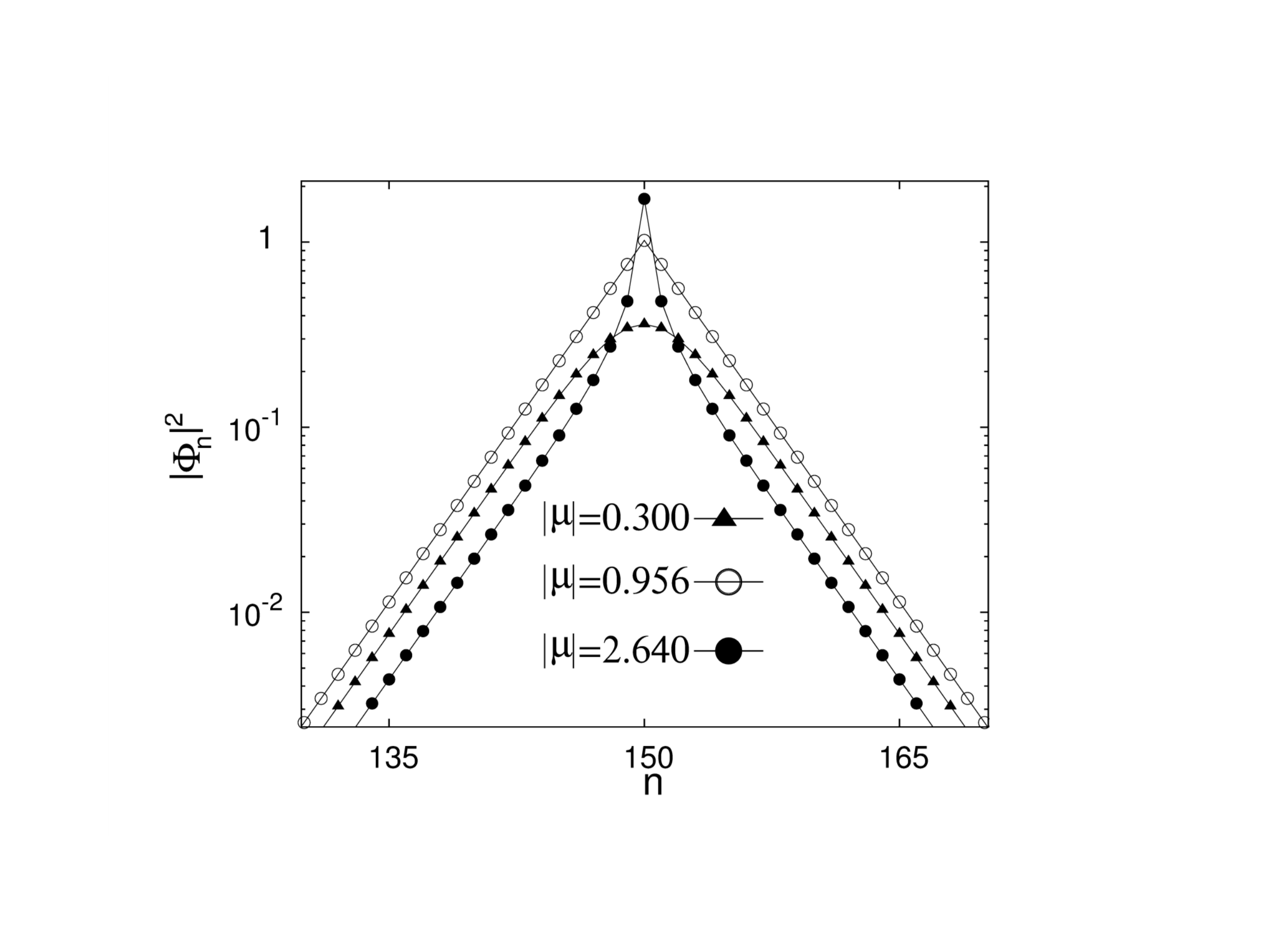}
\end{center}
\caption{Examples of three different types of discrete solitons, shown on
the logarithmic scale at $\protect\omega =-2.091$, in the one-dimensional
Salerno model (\protect\ref{Salerno}) with competing nonlinearities: an ordinary (smooth) soliton at $\protect\mu =-0.3$%
, a peakon at $\protect\mu =-0.956$, and a cuspon at $\protect\mu
=-2.64$. \from{Gomez-Gardenes06}
} \label{fig1}
\end{figure}

The SM based on Eq.~(\ref{Salerno}) with $\mu <0$ features \emph{competing
nonlinearities}, the terms corresponding to $\nu =+1$ and $\mu <0$
representing the self-focusing and defocusing cubic interactions,
respectively. In the 1D setting, the SM with $\mu <0$ was introduced in Ref.~\cite{Gomez-Gardenes06}. In that work, it was demonstrated that this version of the SM
gives rise to families of quiescent discrete solitons, which are looked for
as%
\begin{equation}
\Phi _{n}(t)=e^{-i\omega t}U_{n},  \label{Un}
\end{equation}%
with negative frequency $\omega $ and real amplitudes $U_{n}$
(unlike the complex solutions for moving solitons (\ref{ALsoliton})
in the AL model), of two different types. One family represents
ordinary discrete solitons, which are similar to quiescent solitons
in the SM with $\mu \geq 0$, where $\mu =0$ corresponds to the DNLS
equation, and \textit{cuspons, }that are characterized by a higher
curvature of their profile at the center than in the exponentially
decaying tails, see typical examples in Fig. \ref{fig1}. At the
border between the ordinary solitons and cuspons, a special discrete
soliton appears, in the form of a \textit{peakon}, which is also
shown in Fig.~\ref{fig1}. In the continuum limit of Eq.~(\cite{Gomez-Gardenes06}) with $\mu <0$
(see below), a peakon solution is available in an exact analytical form (\ref%
{peakon}), while cuspons do not exist in that limit. The stability of the
discrete solitons in the SM\ with the competing nonlinearities was also
investigated in Ref.~\cite{Gomez-Gardenes06}, with the conclusion that only a small
subfamily of the ordinary solitons is unstable, while all cuspons, including
the peakon, are stable.

\begin{figure}[tbp]
\begin{center}
\includegraphics[width=90mm,natwidth=610,natheight=642]{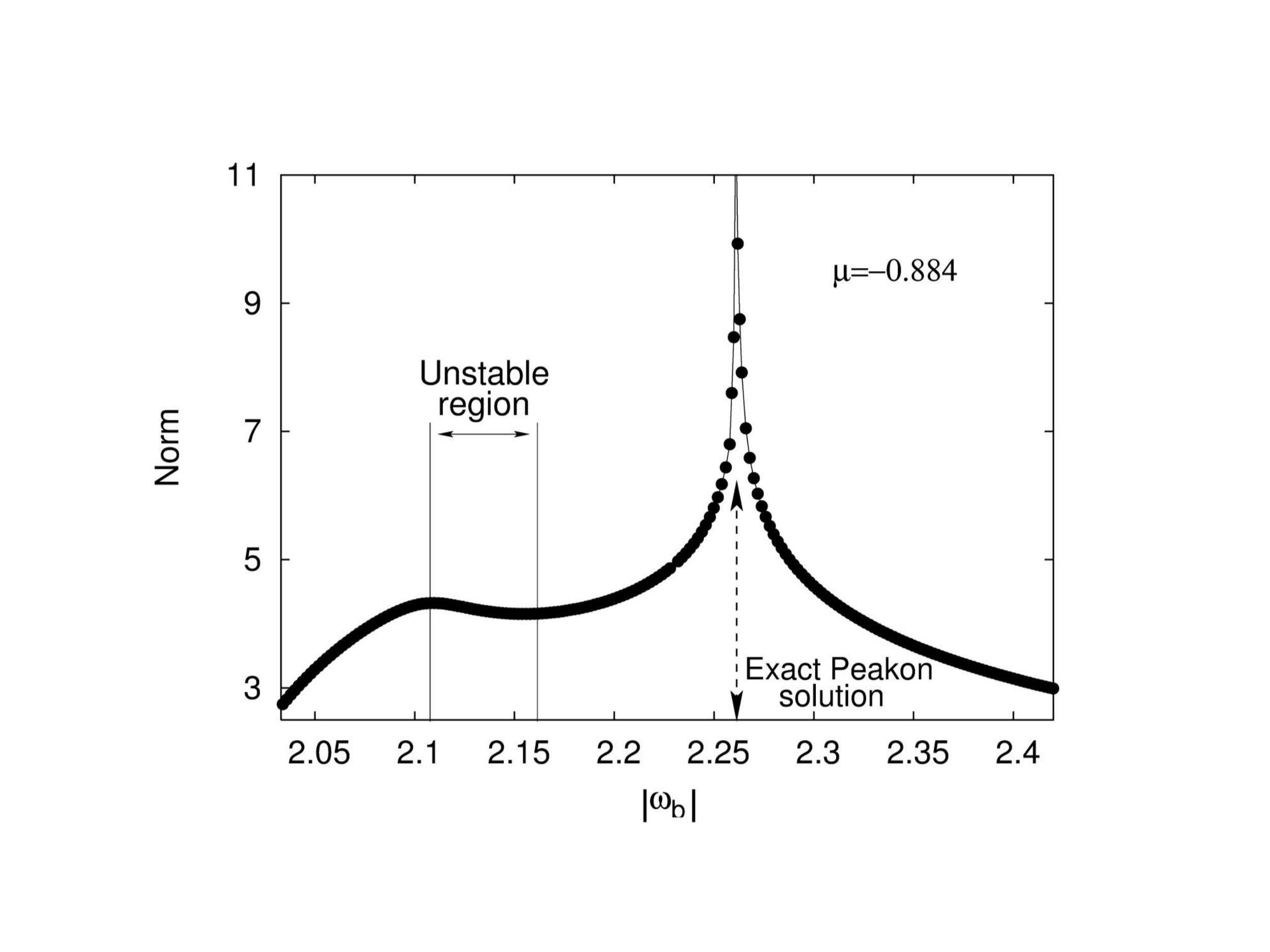}
\end{center}
\caption{The norm of the discrete quiescent solitons, in the Salerno model
with competing nonlinearities, vs.\ the frequency (here the frequency is
denoted $\protect\omega _{b}$, instead of $\protect\omega $), for $\protect%
\mu =-0.884$.
\from{Gomez-Gardenes06}}
\label{fig:2}
\end{figure}

For fixed $\mu =-0.884$, the soliton families are illustrated in Fig.~\ref%
{fig:2}, which shows the norm (\ref{eq:Norm}) as a function of $|\omega |$. The
plot clearly demonstrates that the ordinary solitons and cuspons are
separated by the peakon. Except for the part of the ordinary-soliton family
with the negative slope, $d\mathcal{N}/d(|\omega |)<0$, which is marked in
Fig.~\ref{fig:2}, the solitons are stable. In particular, the peakon and
cuspons are completely stable modes. The instability of the portion of the
ordinary-soliton family with $d\mathcal{N}/d(|\omega |)<0$ agrees with the
prediction of the Vakhitov-Kolokolov (VK) criterion, which gives a necessary
stability condition in the form of $d\mathcal{N}/d\omega ~<0$. The VK
criterion applies to the ordinary solitons, but is irrelevant for the
cuspons.

\subsubsection{Bound states of the discrete solitons and their stability}

\begin{figure}[tbp]
\begin{center}
\includegraphics[width=90mm,natwidth=610,natheight=642]{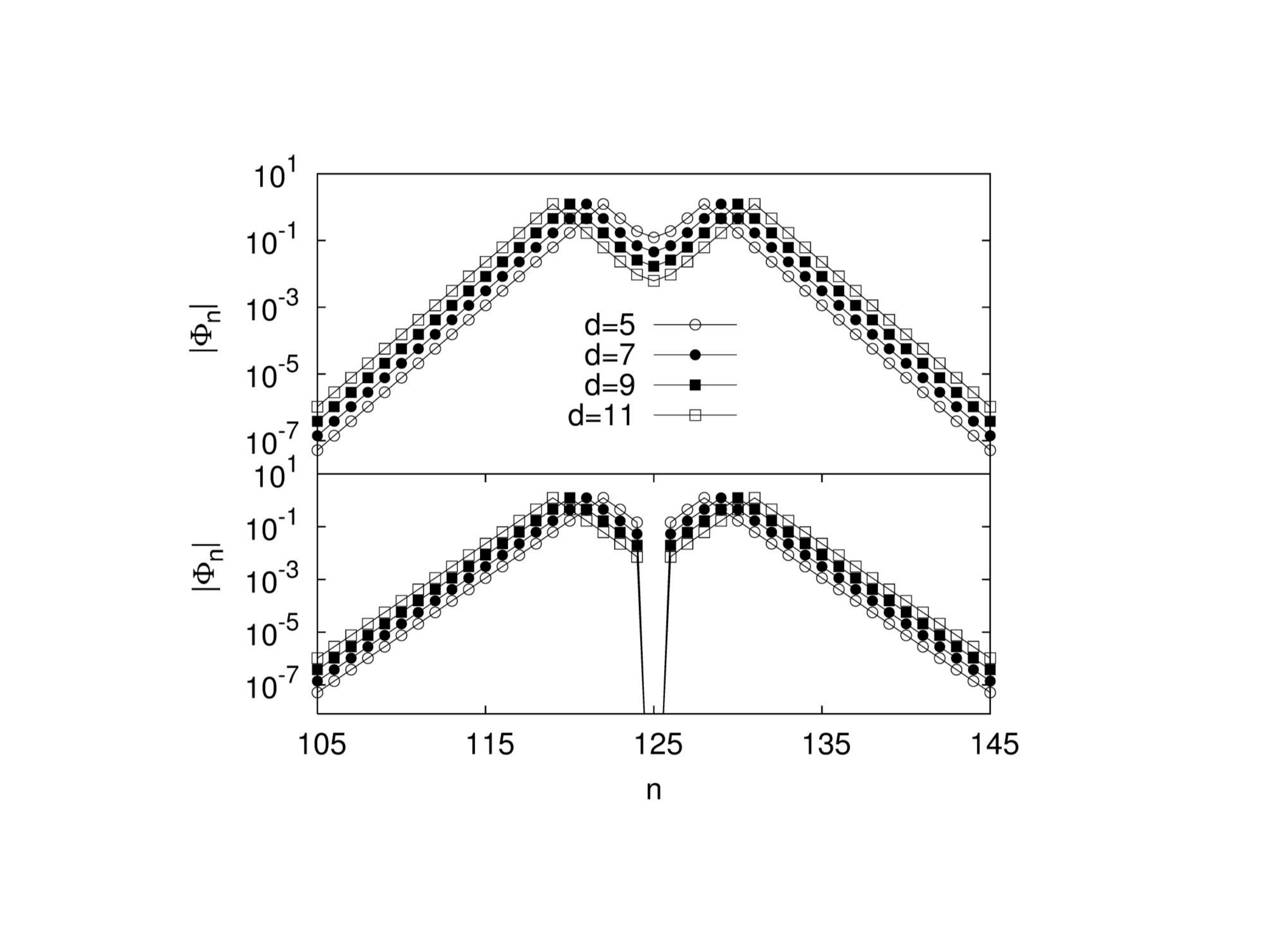}
\end{center}
\caption{Profiles of typical symmetric (top) and antisymmetric (bottom)
bound states of two peakons are shown, on the logarithmic scale, for $\protect\omega =-3.086$ and $\protect\mu ~=-0.645$.
\from{Gomez-Gardenes06}}
\label{fig:3}
\end{figure}

Spatially symmetric (even) and antisymmetric (odd) states of discrete
solitons were also constructed in the framework of Eq.~(\ref{Salerno}), see
examples of bound peakons in Fig.~\ref{fig:3}. It is known that
antisymmetric bound states of discrete solitons in the DNLS\ equation are
stable, while the symmetric ones are not \cite{Kapitula01,Pelinovsky05}. The same is true for
bound states of ordinary discrete solitons in the SM. However, the situation
is \emph{exactly opposite} for the cuspons: their symmetric bound states are
stable, while antisymmetric ones are unstable.

\subsection{The two-dimensional Salerno model and discrete solitons}

The 2D version of the SM was introduced in Ref.~\cite{Bishop06}. It is based on
the following equation, cf.\ Eq.~(\ref{Salerno}),
\begin{eqnarray}
i\frac{d}{dt}\Phi _{n,m} &=&-\left[ \left( \Phi _{n+1,m}+\Phi
_{n-1,m}\right) +C\left( \Phi _{n,m+1}+\Phi _{n,m-1}\right) \right]   \notag
\\
&\times &\left( 1+\mu \left\vert \Phi _{n,m}\right\vert ^{2}\right)
-2\left\vert \Phi _{n,m}\right\vert ^{2}\Phi _{n,m}\;,  \label{2dSalerno}
\end{eqnarray}%
where the same normalization as above, $\nu =+1$, is imposed. In this
notation, $C$ accounts for a possible anisotropy of the 2D lattice [($C=1$ and
$C=0$ correspond, respectively, to the the isotropic 2D lattice and its 1D
counterpart, see Eq.~(\ref{Salerno})]. Accordingly, the variation of $C$
from $0$ to $1$ opens the way for considering the \textit{dimensionality
crossover} from 1D to 2D.

Similar to the 1D version of the SM, Eq.~(\ref{2dSalerno}) conserves the

norm and Hamiltonian, cf.\ Eqs.~(\ref{eq:Norm}) and (\ref{eq:SalernoEnergy}),
\begin{equation}
\mathcal{N}_{\mathrm{2D}}=\frac{1}{\mu }\sum_{m,n}\ln \left( |1+\mu |\Phi
_{n,m}|^{2}|\right) \;, \label{SalernoNorm}
\end{equation}%
\begin{gather}
{\mathcal{H}}_{\mathrm{2D}}=-\sum_{n,m}\left[ \left( \Phi _{n,m}\Phi
_{n+1,m}^{\ast }+\Phi _{n+1,m}\Phi _{n,m}^{\ast }\right) \right.   \notag \\
C\left( \Phi _{n,m}\Phi _{n,m+1}^{\ast }+\Phi _{n,m+1}\Phi _{n,m}^{\ast
}\right)   \notag \\
\left. +\frac{2}{\mu }|\Phi _{n,m}|^{2}\right] +\frac{2}{\mu }\mathcal{N}_{%
\mathrm{2D}}.\;  \label{eq:SalernoHam}
\end{gather}

\begin{figure}[tbp]
\begin{center}
\includegraphics[width=90mm,natwidth=610,natheight=642]{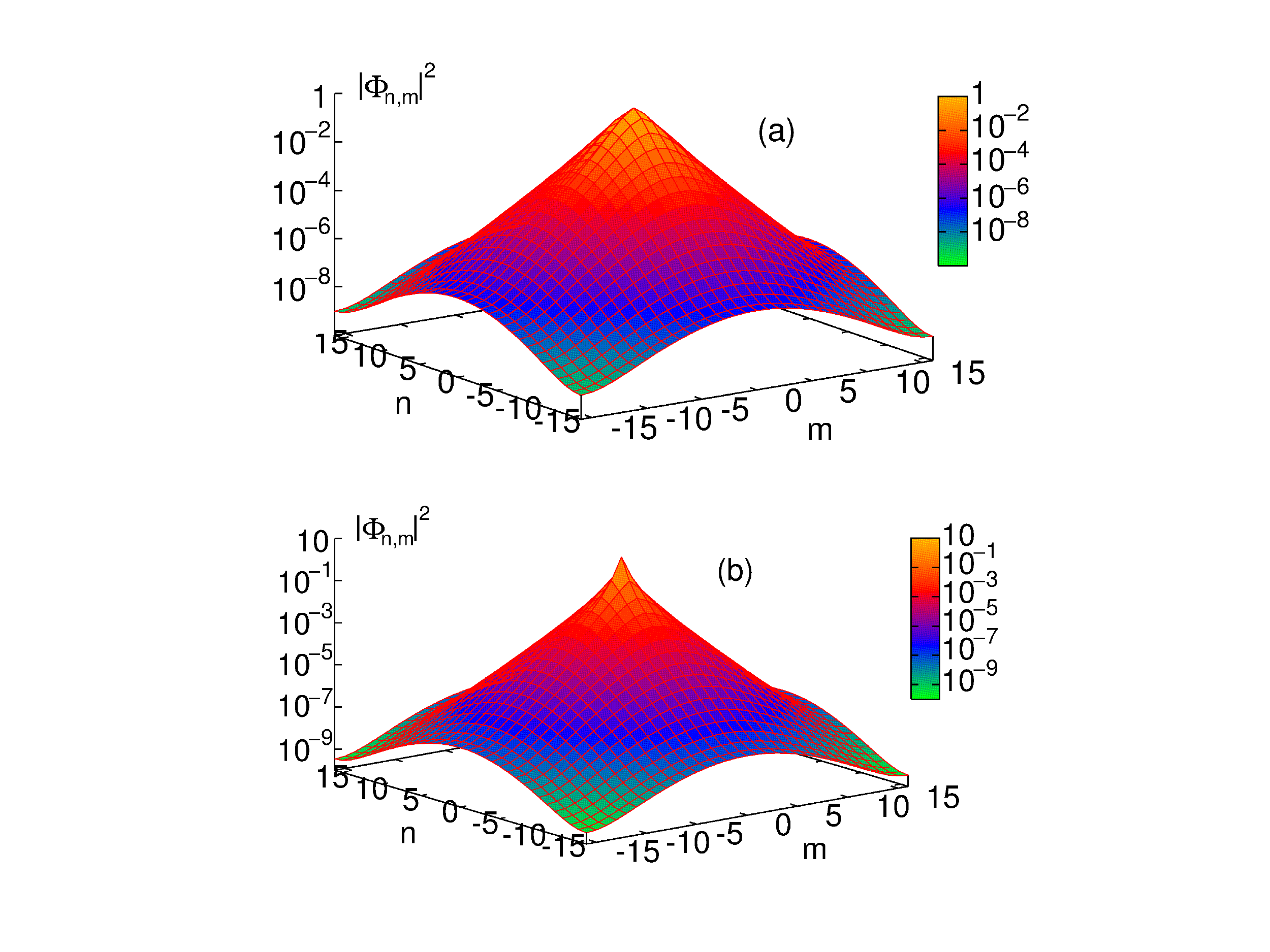}
\end{center}
\caption{ Profiles of discrete solitons in the isotropic ($%
C=1$) 2D Salerno model with competing nonlinearities, found for frequency $%
\protect\omega =-4.22$: (a) a regular
soliton at $\protect\mu =-0.2$; \textbf{(}b\textbf{)} a cuspon at $\protect%
\mu =-0.88$.
\from{Bishop06}}
\label{fig:4}
\end{figure}

Fundamental 2D solitons are looked for in the same form as their 1D
counterparts (\ref{Un}),%
\begin{equation}
\Phi _{mn}(t)=e^{-i\omega t}U_{mn}.  \label{Umn}
\end{equation}%
In the most interesting case of the competing nonlinearities, $\mu
<0$, the general properties of the solitons are similar to those
outlined above in the framework of the 1D version of the SM. There
are ordinary solitons, which have regions of stability and
instability (the stability border depends on $C$), and cuspons,
which are entirely stable. Examples of 2D solitons of both types are
displayed in Fig.~\ref{fig:4}. The families of ordinary solitons and
cuspons are separated by 2D peakons, which are stable, too. Spatially
antisymmetric bound states of the 2D ordinary solitons, and
symmetric bound states of the 2D cuspons are stable, while the bound
states with the opposite parities are unstable, also similar to the
situation in the 1D model.

In addition to the fundamental solitons, the 2D model with the
competing nonlinearities supports solitary vortices, of two
different types, on-site- and off-site-centered ones (alias
``rhombuses" and ``squares"), which have their narrow stability
regions (the stability was investigated in Ref.~\cite{Bishop06} only for
vortex solitons with
topological charge $1$). Examples of the vortices are displayed in Fig.~\ref%
{fig:5}. In the two-dimensional SM with non-competing nonlinearities,
unstable vortices turn into fundamental solitons, losing their vorticity
(obviously, the angular momentum is not conserved in the lattice system).
However, in the SM with the competing nonlinearities, unstable stationary
vortices transform into \textit{vortical breathers}, which are persistent
oscillating localized modes that keep their vorticity.
\begin{figure}[tbp]
\begin{center}
\includegraphics[width=90mm,natwidth=610,natheight=642]{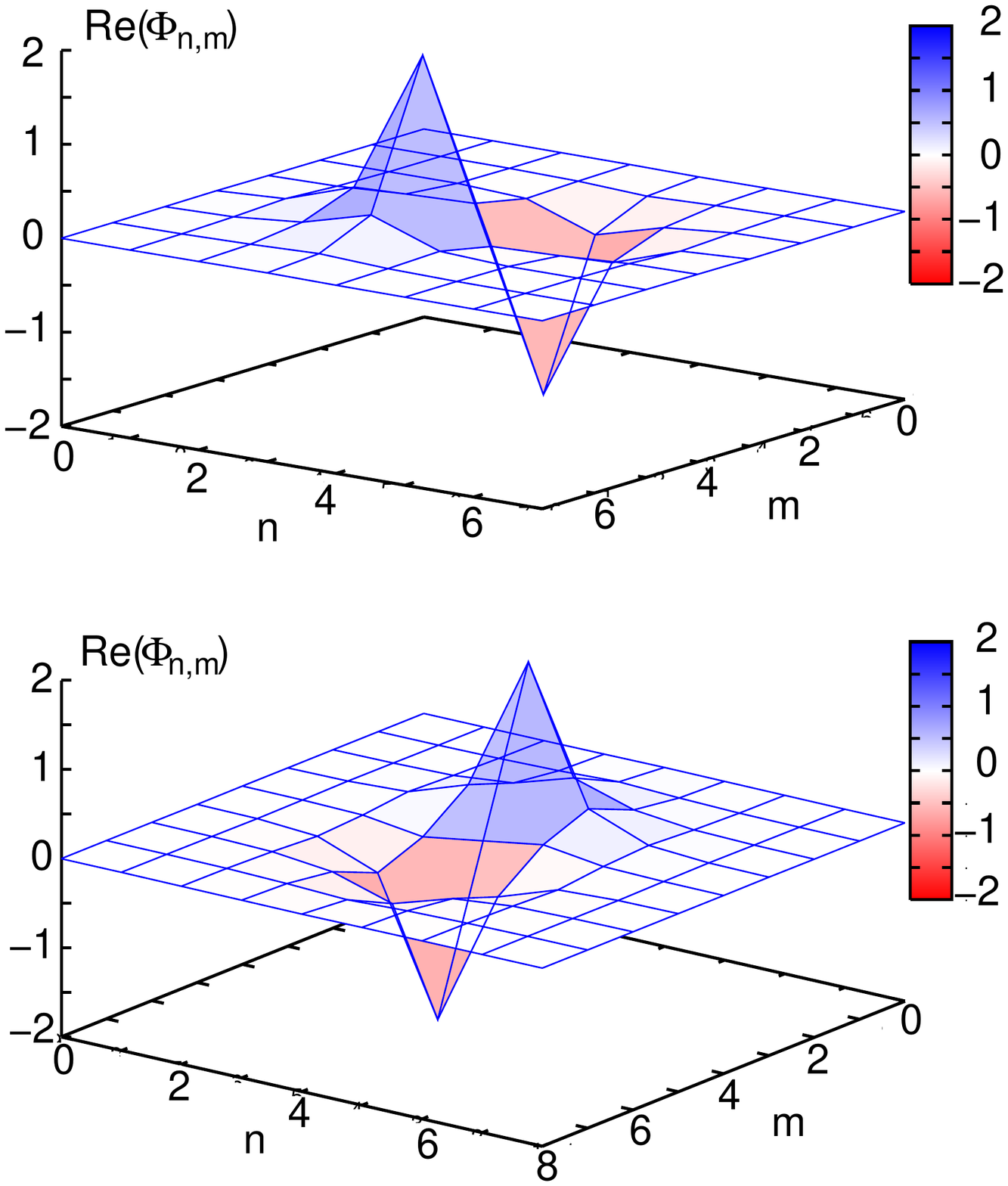}
\end{center}
\caption{ Examples of stable discrete vortices with
topological charge $1$ in the 2D Salerno model with competing
nonlinearities. Profiles of the real
part of the ``vortex-square" and ``vortex-cross" modes are shown in
the top and bottom panels, respectively. Both solutions are found
for $\protect\mu =-0.4$ and $\protect\omega =-7.0$.
\from{Bishop06}} \label{fig:5}
\end{figure}

\subsection{The continuum limit of the 1D and 2D Salerno models}

\subsubsection{One dimension}

The continuum limit of the discrete equation (\ref{Salerno}) deserves
separate consideration. This limit was introduced in Ref.~\cite{Gomez-Gardenes06} by defining $\Phi
(x,t)\equiv e^{2it}\Psi (x,t)$, and using the truncated Taylor expansion, $%
\Psi _{n\pm 1}\approx \Psi \pm \Psi _{x}+(1/2)\Psi _{xx}$, where $\Psi $ is
here treated as a function of the continuous coordinate $x$, which coincides
with $n$ when it takes integer values. Then, the continuum counterpart of
Eq. (\ref{Salerno}) is
\begin{equation}
i\Psi _{t}=-2\left( 1-|\mu |\right) \left\vert \Psi \right\vert ^{2}\Psi
-\left( 1-\left\vert \mu \right\vert \left\vert \Psi \right\vert ^{2}\right)
\Psi _{xx}~,  \label{Psi}
\end{equation}%
where $\nu =+1$ and $\mu <0$ is set as above (i.e., the system with competing nonlinearities is considered). Equation (\ref{Psi})
conserves the norm and
Hamiltonian, which can be derived as the continuum limit of expressions (\ref%
{eq:Norm}) and (\ref{eq:SalernoEnergy}),
\begin{eqnarray}
\mathcal{N}_{\mathrm{cont}} &=&\frac{1}{\mu }\int_{-\infty }^{+\infty }dx~{%
\ln }\left( \left\vert 1-|\mu ||\Psi |^{2}\right\vert \right) ,
\label{Ncont} \\
\mathcal{H}_{\mathrm{cont}} &=&\int_{-\infty }^{+\infty }dx\left[ \left\vert
\Psi _{x}\right\vert ^{2}+2\left( \frac{1}{|\mu |}-1\right) |\Psi |^{2}%
\right]   \notag \\
&&+\frac{2}{\mu }\mathcal{N}_{\mathrm{cont}}.  \label{Hcont}
\end{eqnarray}

Soliton solutions to Eq.~(\ref{Psi}) can be looked for as $\Psi =e^{-i\omega
t}U(x)$, with real function $U$ obeying equation
\begin{equation}
\frac{d^{2}U}{dx^{2}}=-\frac{\omega +2\left( 1-|\mu |\right) U^{2}}{1-|\mu
|U^{2}}U,  \label{U}
\end{equation}%
which may give rise to solitons, provided that $|\mu |<1.$ The absence of
solitons at $|\mu |>1$ implies that (bright) solitons do not exist in the
continuum limit if the continual counterpart of the self-defocusing
intersite nonlinearity is stronger than the onsite self-focusing
nonlinearity. For $|\mu |<1$, the solitons exist in the following frequency
band:%
\begin{equation}
0<-\omega <\left( 1/|\mu |\right) -1.  \label{band}
\end{equation}

Solitons can be found in an explicit form near edges of band (\ref{band}).
At small $|\omega |$, an approximate soliton solution is $U(x)\approx \sqrt{%
|\omega |/\left( 1-|\mu |\right) }\mathrm{sech}\left( \sqrt{2|\omega |}%
x\right) $, while precisely at the opposite edge of the band, at
$-\omega =1/|\mu |-1$, an \emph{exact} solution is available, in the
form of a \textit{peakon }(this time, in the continuum model),
\begin{equation}
U_{\mathrm{peakon}}=\left( 1/\sqrt{|\mu |}\right) \exp \left( -\sqrt{\left(
1/|\mu |\right) -1}|x|\right) .  \label{peakon}
\end{equation}%
The term ``peakon" implies that solution (\ref{peakon}) features a
jump of the derivative at the central point. The norm (\ref{Ncont}) of
the peakon is $\pi ^{2}/[6\sqrt{|\mu |(1-|\mu |)}]$, and its energy
is finite, too.

\subsubsection{Two dimensions}

The continuum limit of Eq.~(\ref{2dSalerno}) may be defined by proceeding
from discrete coordinates $\left( n,m\right) $ to continuous ones, $\left(
x,y\right) \equiv \left( n/\sqrt{\alpha },m/\sqrt{C\alpha }\right) $, and
defining $\Phi _{n,m}\equiv \Psi (x,y)\exp \left( 2(1+C)it\right) $:%
\begin{equation}
i\Psi _{t}+\left( 1+\mu \left\vert \Psi \right\vert ^{2}\right) \left( \Psi
_{xx}+\Psi _{yy}\right) +2\left[ (1+C)\mu +1\right] |\Psi |^{2}\Psi =0,
\label{cont}
\end{equation}%
cf.\ Eq.~(\ref{Psi}). Note that this equation always has the
isotropic form. The dispersive nonlinear term in Eq.~(\ref{cont}),
$\mu \left\vert \Psi \right\vert ^{2}\left( \Psi _{xx}+\Psi
_{yy}\right) $ prevents the collapse, for both positive and
negative $\mu$. Therefore, the \textit{quasi-collapse%
}, which is known in other discrete systems \cite{Laedke94}, is not
expected in the two-dimensional discrete SM either.

\subsection{The Hamiltonian structure of the 1D model, and its quantization}

\subsubsection{The classical version}

The specific form of Hamiltonian (\ref{eq:SalernoEnergy}) of the SM
makes the Poisson brackets in this system different from the
standard form \cite{Cai96, Rasmussen97}. Namely, for a pair of
arbitrary functions of the discrete field variables, $B\left( \Phi
_{n},\Phi _{n}^{\ast }\right) ,$ $C\left( \Phi _{n},\Phi _{n}^{\ast
}\right) $, the Poisson brackets are written as
\begin{equation}
\left\{ B,C\right\} =i\sum_{n}\left( \frac{\partial B}{\partial \Phi _{n}}%
\frac{\partial C}{\partial \Phi _{n}^{\ast }}-\frac{\partial B}{\partial
\Phi _{n}^{\ast }}\frac{\partial C}{\partial \Phi _{n}}\right) \left( 1+\mu
\left\vert \Phi _{n}\right\vert ^{2}\right) ,  \label{Poisson}
\end{equation}%
the last factor being the non-standard feature. In particular, the brackets
of variables $\Phi _{n}$ and $\Phi _{n}^{\ast }$ themselves are
\begin{eqnarray}
\left\{ \Phi _{n},\Phi _{m}^{\ast }\right\}  &=&i\left( 1+\mu \left\vert
\Phi _{n}\right\vert ^{2}\right) \delta _{nm},  \notag \\
\left\{ \Phi _{n},\Phi _{m}\right\}  &=&\left\{ \Phi _{n}^{\ast },\Phi
_{m}^{\ast }\right\} =0.  \label{PhiPhi}
\end{eqnarray}

Instead of dynamical variables $\Phi _{n}$, one can attempt to define another set,%
\begin{equation}
\chi _{n}\equiv f\left( \left\vert \Phi _{n}\right\vert ^{2}\right) \Phi
_{n},  \label{chi}
\end{equation}%
so that they will obey the usual commutation relations,
\begin{equation}
\left\{ \chi _{n},\chi _{m}^{\ast }\right\} =i\delta _{mn},~\left\{ \chi
_{n},\chi _{m}\right\} =0,  \label{standard}
\end{equation}%
instead of ``exotic" ones (\ref{PhiPhi}). Substituting
ansatz (\ref{chi}) into Eq.~(\ref{standard}), and making use of definition (%
\ref{Poisson}), one arrives at the following equation for the function
$f(x)$, which secures that the Poisson brackets for the new
variables indeed take the standard form of Eq.~(\ref{standard}):
\begin{equation}
2xf\frac{df}{dx}+f^{2}=\frac{1}{1+\mu x}.  \label{ODE}
\end{equation}%
A solution of Eq.~(\ref{ODE}) is $f(x)=\sqrt{\left\vert \ln \left( 1+\mu
x\right) \right\vert /\left( \mu x\right) }$. Thus, the new set of canonical
variables (\ref{chi}) is%
\begin{equation}
\chi _{n}=\sqrt{\frac{\left\vert \ln \left( 1+\mu \left\vert \Phi
_{n}\right\vert ^{2}\right) \right\vert }{\mu \left\vert \Phi
_{n}\right\vert ^{2}}}\Phi _{n}~.  \label{new}
\end{equation}

The definition (\ref{new}) may be inverted, to express $\Phi _{n}$ in terms of $%
\chi _{n}$,%
\begin{eqnarray}
\left\vert \Phi _{n}\right\vert ^{2} &=&\mu ^{-1}\left[ \exp \left( \mu
\left\vert \chi _{n}\right\vert ^{2}\right) -1\right] ,  \notag \\
\Phi _{n} &=&\sqrt{\frac{\exp \left( \mu \left\vert \chi _{n}\right\vert
^{2}\right) -1}{\mu \left\vert \chi _{n}\right\vert ^{2}}}\chi _{n}~.
\label{inverse}
\end{eqnarray}%
Making use of Eqs.~(\ref{inverse}), the norm (\ref{eq:Norm}) and Hamiltonian (%
\ref{eq:SalernoEnergy}) can be written in terms of the new canonical
variables as%
\begin{equation}
\mathcal{N}=\sum_{n}\left\vert \chi _{n}\right\vert ^{2}~,  \label{Nnew}
\end{equation}%
\begin{eqnarray}
\mathcal{H} &=&\sum_{n}\left\{ -\left(\mathcal{A}_{n,n+1} \left( \chi
_{n}\chi _{n+1}^{\ast }+\chi _{n+1}\chi _{n}^{\ast }\right) \right) \right.
\notag \\
&&\left. -\frac{2}{\mu ^{2}}\left[ \exp \left( \mu \left\vert \chi
_{n}\right\vert ^{2}\right) -1\right] \right\} +\frac{2}{\mu }\mathcal{N}\;,
\label{Hnew}
\end{eqnarray}%
with the shorthand notation
\begin{equation}
\mathcal{A}_{n,n+1}= \sqrt{\frac{\left[ \exp \left( \mu
\left\vert \chi _{n}\right\vert ^{2}\right) -1\right] \left[ \exp \left( \mu
\left\vert \chi _{n+1}\right\vert ^{2}\right) -1\right] }{\mu ^{2}\left\vert
\chi _{n}\right\vert ^{2}\left\vert \chi _{n+1}\right\vert ^{2}}}.
\end{equation}
Finally, Eq.~(\ref{Salerno}) (with $\nu \equiv 1$), if rewritten in terms of
variables $\chi _{n}$, may be represented in the standard Hamiltonian form,
with the usual Poisson brackets,%
\begin{equation}
i\frac{d\chi _{n}}{dt}=\frac{\partial \mathcal{H}}{\partial \chi _{n}^{\ast }%
}~,  \label{normal}
\end{equation}%
where Hamiltonian $\mathcal{H}$ is taken as per Eq. (\ref{Hnew}).

\subsubsection{The quantum version of the Salerno Model}

The SM was actually introduced from the very beginning in its quantum form
\cite{Salerno92}. As usual, the quantization of the classical model is
performed by replacing the canonically conjugate variables, $\Phi _{n}$ and $%
\Phi _{n}^{\ast }$, by the creation and annihilation operators,
\begin{equation}
\Phi _{n}\rightarrow \hat{\Phi}_{n},\Phi _{n}^{\ast }\rightarrow \hat{\Phi}%
_{n}^{\dag }.  \label{Q}
\end{equation}
This correspondence replaces the classical deformed Poisson algebra (\ref%
{PhiPhi}) by the following deformed Heisenberg algebra:%
\begin{eqnarray}
\left[ \hat{\Phi}_{n},\hat{\Phi}_{m}^{\dag }\right] &=&\hbar \left( 1+\mu
\hat{\Phi}_{n}^{\dag }\hat{\Phi}_{n}\right) \delta _{nm},  \notag \\
\left[ \hat{\Phi}_{n},\hat{\Phi}_{m}\right] &=&\left[ \hat{\Phi}_{n}^{\dag },%
\hat{\Phi}_{m}^{\dag }\right] =0\text{.}  \label{comm}
\end{eqnarray}%
These operators act on the standard Fock states as%
\begin{eqnarray}
\hat{\Phi}_{n}^{\dag }\left\vert N_{n}\right\rangle &=&\sqrt{\mu ^{-1}\left[
\left( 1+\hbar \mu \right) ^{N+1}-1\right] }\left\vert N_{n}+1\right\rangle ,
\notag \\
\hat{\Phi}_{n}^{\dag }\left\vert N_{n}\right\rangle &=&\sqrt{\mu ^{-1}\left[
\left( 1+\hbar \mu \right) ^{N+1}-1\right] }\left\vert N_{n}-1\right\rangle .
\label{Fock}
\end{eqnarray}

Further, the operator of the total number of particles is constructed as the
quantum counterpart of the classical expression (\ref{eq:Norm}) for the
total norm,%
\begin{equation}
\hat{\mathcal{N}}=\frac{1}{\ln \left( 1+\hbar \mu \right) }%
\sum_{n}\ln \left( 1+\mu \hat{\Phi}_{n}^{\dag }\hat{\Phi}_{n}\right) .
\label{N}
\end{equation}%
It acts on the global Fock's state as the proper number operator, $\hat{\mathcal{N}}\left\vert N\right\rangle =N\left\vert
N\right\rangle $ \cite{Salerno92}.

The quantum Hamiltonian can be derived directly from its classical
counterpart (\ref{eq:SalernoEnergy}),%
\begin{equation}
\hat{\mathcal{H}}=-\sum_{n}\left[ \hat{\Phi}_{n}^{\dag
}\left( \hat{\Phi}_{n-1}+\hat{\Phi}_{n+1}\right) +\frac{2}{\mu }\hat{\Phi}%
_{n}^{\dag }\hat{\Phi}_{n}\right] +\frac{2}{\mu }\hat
{\mathcal{N}}.
\end{equation}

This Hamiltonian and  commutation relations (\ref{comm}) lead to the
Heisenberg's equation of motion, $id\hat{\Phi}_{n}/dt=\left[ \hat{\Phi}_{n},%
\hat{\mathcal{H}}\right] $, which can be derived in a
straightforward way from the classical SM equation, (\ref{Salerno}),
replacing the classical variables by their quantum counterparts as per Eqs. (%
\ref{Q}, yielding
\begin{equation}
i\frac{d\hat{\Phi}_{n}}{dt}=-\left( 1+\mu \hat{\Phi}_{n}^{\dag }\hat{\Phi}%
_{n}\right) \left( \hat{\Phi}_{n-1}+\hat{\Phi}_{n+1}\right) -2\hat{\Phi}%
_{n}^{\dag }\hat{\Phi}_{n}^{2}~.  \label{Heis}
\end{equation}

The transformation of the classical canonical variables as per Eq.~(\ref%
{inverse}), which ``rectifies" the deformed Poisson brackets
(\ref{PhiPhi}) into their standard form (\ref{standard}), suggests
one to perform a similar canonical transformation in the quantum SM,
which is
possible indeed. The transformation is carried out as follows:%
\begin{eqnarray}
\hat{\Phi}_{n}^{\dag } &=&\hat{F}_{n}\hat{\chi}_{n}^{\dagger },~\hat{\Phi}%
_{n}=\chi _{n}\hat{F}_{n},  \notag \\
\hat{F}_{n} &\equiv &\sqrt{\frac{\left( 1+\hbar \mu \right) ^{\chi
_{n}^{\dagger }\chi _{n}}-1}{\mu \chi _{n}^{\dagger }\chi _{n}}}~.  \label{F}
\end{eqnarray}%
The operators $\chi _{n}^{\dagger }$ and $\chi _{n}$, unlike the original ones, $%
\hat{\Phi}_{n}^{\dag }$ and $\hat{\Phi}_{n}$, obey the usual commutation
relations
\begin{equation}
\left[ \hat{\chi}_{n},\hat{\chi}_{m}\right] =\left[ \hat{\chi}_{n}^{\dagger
},\hat{\chi}_{m}^{\dagger }\right] =0,\left[ \hat{\chi}_{n},\hat{\chi}%
_{m}^{\dagger }\right] =\delta _{nm},
\end{equation}
cf.\ Eqs.~(\ref{comm}), and they act on the Fock states in the usual way:%
\begin{eqnarray}
\hat{\chi}_{n}^{\dag }\left\vert N_{n}\right\rangle  &=&\left(
N_{n}+1\right) \left\vert N_{n}+1\right\rangle , \\
\hat{\chi}_{n}\left\vert N_{n}\right\rangle  &=&N_{n}\left\vert
N_{n}-1\right\rangle ,
\end{eqnarray}%
see Eqs.~(\ref{Fock}). Further, the operator (\ref{N}) of the total number of
particles also takes the usual form in terms of these $\hat{\chi}%
_{n}^{\dagger }$ and $\hat{\chi} _{n}$, $\hat{\mathcal{N}}%
=\sum_{n}\hat{\chi}_{n}^{\dagger }\chi _{n}$, cf.\ Eq.~(\ref{N}), while the
Hamiltonian, expressed in terms of $\hat{\chi}_{n}^{\dagger }$ and $\hat{\chi} _{n}
$, is a counterpart of its classical form (\ref{Hnew}):%
\begin{gather}
\hat{\mathcal{H}}~=-\sum_{n}\left\{ \hat{F}_{n}\hat{\chi}%
_{n}^{\dag }\left( \hat{\chi}_{n+1}\hat{F}_{n+1}+\hat{\chi}_{n-1}\hat{F}%
_{n-1}\right) \right.  \\
\left. +\frac{2}{\mu ^{2}}\left[ \left( 1+\hbar \mu \right) ^{\hat{\chi}%
_{n}^{\dag }\hat{\chi}_{n}}-1\right] \right\} +\frac{2}{\mu }\hat{\mathcal{N}},
\end{gather}%
where the operators\ $\hat{F}_{n}$ are defined in Eq.~(\ref{F}).

Finally, it is relevant to mention that quantum counterparts of the
classical solitons, which were reported in the classical
(mean-field) versions of the SM with non-competing and competing
nonlinearities in Refs.
\cite{Cai96,Rasmussen97,Cai97,Dmitriev03,Gomez-Gardenes06}, have not
been constructed yet.

\section{Conclusions}

We have shown on a number of selected cases that the standard Hubbard model for fermions or its bosonic counterpart the Bose-Hubbard model, even supplemented with nearest-neighbor  interactions (extended models) are often insufficient to quantitatively describe the physics of ultracold atoms in optical lattices. While already Hubbard was aware of additional terms contributing to tunneling in a nonlinear, density-dependent (interaction-based) fashion, it was  Hirsch and coworkers  who stressed the importance of these terms (called also bond-charge interactions) in the condensed-matter context. Somehow only in the last few years the ultracold-atom community became aware of that fact, starting with the then puzzling observations of a shift of the Mott-superfluid border for Bose-Fermi mixtures. The density-induced tunneling effects become especially important for long-range (e.g.\ dipolar) interactions, although they may significantly contribute also for  contact interactions, provided these are sufficiently 
strong. 

With increasing interaction strength the  higher bands become important, which one can easily understand, since the Wannier functions are originally constructed for a periodic, interaction-free, single-particle potential. For sufficiently strong interactions, however, different Bloch bands become coupled. One may be tempted to try and treat the problem by multiband expansions---an approach that is doomed to fail due to the strongly increased complexity. Moreover, the tunneling between highly-excited, extended Wannier states cannot be restricted to nearby sites only---and the advantage of a tight-binding approximation is, somehow, lost. 

For moderate contact interactions, an effective approach is possible, described in detail   in \Sec{sec:MO}. A possible prediagonalization of the on-site many-body Hamiltonian forms a convenient many-body ``dressed'' basis. After expressing the tunnelings in that basis, one is led to an effective single-band Hamiltonian with population-dependent coefficients, thus obtaining effective three-body, four-body, etc.\ terms. The importance of these terms has been already verified in Bose-Fermi mixtures as well as in collapse and revival experiments. Clearly, however, for sufficiently strong interactions one expects problems with that approach, and the general solution is not yet known. 

For longer-range dipolar interactions, the problems are even more severe. Due to the nature of the dipolar interactions, the integrals (\ref{intint}) increase in value for higher Bloch bands. As soon as the interaction couples to higher bands the multiorbital approach presented above for contact interactions does not converge. Presently there is no known solution to this problem. One possible way of attacking it is to resign from the Wannier localized basis for higher excited bands, and to work directly with the Bloch
functions \cites{Dutta14}. Yet there exist another potential problems for realistic polar molecules---namely the high density of ro-vibrational molecular states, which may lead   to a formation of long-lived molecular complexes as described in Ref.~\cites{Mayle13}.
That effect will lead a to loss of molecules and potentially may limit the density of molecules in an optical lattice.

All of these effects may complicate the treatment of ultracold atoms in optical lattices, but they also generate a manifold of novel quantum phenomena not present in the standard Hubbard model. 
Despite the recent progress, there is still much to be learnt about interacting ultra-cold atoms and molecules in optical lattice potentials. There are a lot of questions arising beyond the  standard Hubbard model.
\section{Acknowledgments}
All of us would like to acknowledge long-term collaboration
with ICFO. We thank Ravi Chhajlany and Przemek Grzybowski for discussions and critical reading of the manuscript.
D.-S. L. thanks Ole J\"urgensen and Klaus Sengstock for stimulating discussions.
This work has been supported by Polish
National Science Centre within project No. DEC-2012/04/A/ST2/00088 (O.D. and J. Z.) , DEC-
2011/01/D/ST2/02019 (T.S.) and DEC-2012/04/A/ST2/00090 (M.G.).
P.H. and M.L. acknowledge support from  EU IP SIQS. P.H. was also supported by
SFB FoQuS (FWF Project No. F4006-N16) and
ERC synergy grant UQUAM.
M.L.  acknowledges financial support from Spanish
Government Grants TOQATA (FIS2008-01236) and FOQUS,  EU STREP EQuaM, and ERC Advanced
Grants QUAGATUA and OSYRIS. D.-S. L. acknowledges funding by the Deutsche Forschungsgemeinschaft (grants SFB 925 and GRK 1355).
\section{Appendix}
\label{Appendix}
\xdl
\subsection{Multi-orbital dressing of off-site processes}
As briefly described in \Sec{MODressing} and \Sec{MOHamiltonians}, the transformation to a dressed band which incorporates higher orbital contributions allows treating multi-orbital Hamiltonians effectively with single-band methods.  \xe
In the following, it is shown how a two-site operator, such as the tunneling, can be represented and computed within the dressed-band approach following Refs.~\cites{Luhmann2012,Jurgensen2012,Bissbort2012}. First, we turn to the representation of operators in the dressed band using the ground state $\Psi(n)$ of the $n$ particle on-site problem \eqref{eq:Hsite}.
Within a single-orbital treatment any tight-binding two-site operator can be decomposed in the form
\begin{equation}
\label{eq:SO-operator}
 \hat O_\mathrm{SO} = A\ \hat O_\mathrm{L}\, \hat O_\R,
\end{equation}
with an amplitude $A$ and operators $O_i$ consisting of creation/annihilation operators $\hat b_i^\dagger\,/\,\hat b_i$ on the left (L) or right (R) site, e.g.\ the single-particle tunneling $-J \hat b^\dagger_\mathrm{L} \hat b_\R$. 
The multi-orbitally dressed band (indicated by a tilde) is constructed with creation and annihilation operators that fulfill the usual relations  
\begin{equation}\begin{split}
\tilde b_i \ket{\Psi(n)}_i &= \sqrt{n} \ket{\Psi(n-1)}_i, \\
\tilde b_i^\dagger \ket{\Psi(n)}_i &= \sqrt{n+1} \ket{\Psi(n+1)}_i. 
 \end{split}\end{equation}
Note that by construction the states $\Psi(n)$ are still orthogonal in respect to the particle number $n$ and therefore 
the particle number operator in Wannier and dressed basis are equivalent, $\tilde n_i= \tilde b_i^\dagger \tilde b_i = \hat n_i$.
Formally, by replacing in  $\hat O_\mathrm{L}$ and $\hat O_\R$ the operators $\hat b_i^\dagger\,/\,\hat b_i$ with their dressed counter-parts $\tilde b_i^\dagger\,/\,\tilde b_i$, the operator 
\begin{equation}
\tilde O =  \tilde O_\mathrm{L}\ \tilde O_\R\ \tilde A_{\hat n_\R,\hat n_\mathrm{L}}
\end{equation}
of the dressed band  is constructed. Here, the indices of $\tilde A$ have operator form, which expresses that  $\tilde A$ is projected to the respective
occupation-number dependent amplitude $\tilde A_{\nR,\nL}$ (cf.\ \Eq{eq:DressedU}).

While the definitions for the interaction-dressed band are given above, the actual problem is to compute the dressed-band amplitudes $\tilde A_{\nL,\nR}$ that effectively include all orbital processes. In general, a multi-orbital two-site operator can be decomposed into 
\begin{equation}
\label{eq:MO-operator}
 \hat O_\mathrm{MO}  = \sum_{\{\alpha\},\{ \beta\}} A^{\{\alpha\},\{ \beta\}} \hat O_\mathrm{L}^{\{\alpha\}} \hat O_\R^{\{\beta\}},
\end{equation}
where the summation is over all possible sets of orbitals $\{\alpha\}=\{\alpha_1,\alpha_2,... \}$ and $\{\beta\}=\{\beta_1,\beta_2,... \}$, $A^{\{\alpha\},\{ \beta\}}$ is the amplitude, and $\hat O_i^{\{\alpha\}}$ consists of creation and annihilation operators $\hbd{$i$}{\alpha_k}$ and $\hb{$i$}{\alpha_k}$ at site $i$ in the orbital $\alpha_k$. For the multi-orbital tunneling
\begin{equation}
	 \J_\text{MO}=- \sum_\alpha \J^{\a}\ \hbd{L}{\alpha} \hb{R}{\alpha},
\end{equation} 
the operators on the left and the right site, $\hat O_\mathrm{L}^{\alpha} = \hbd{L}{\alpha}$ and $\hat O_\R^{\beta} = \hb{R}{\beta}$, depend only on a single orbital $\{\alpha\}=\alpha$ and $\{\beta\}=\beta$, with an orbital-conserving amplitude $A^{\alpha,\beta}=-\J^{\a} \delta_{\alpha,\beta}$.  
The effective amplitude $\tilde A_{\nR,\nL}$ is obtained from the matrix element $\bra{\Psi_\text{F}} \hat O_\mathrm{MO}  \ket{\Psi_\text{I}}$, where $\Psi_\text{I}(\nL,\nR)$ denotes the initial and $\Psi_\text{F}=\Psi(\nL',\nR')$ the final state of the process. The occupation-dependent amplitude includes the summation over all multi-orbital processes. Since the states are product states of the individual lattice sites $\ket{\Psi(\nL)}\ket{\Psi(\nR)}$,  the effective amplitude $\tilde A$ decomposes into individual site contributions, 
\begin{equation}\begin{split}
 \tilde A_{\nR,\nL} \!= \!\frac{1}{N} \!\sum_{\{\alpha\},\{ \beta\}} \!\! A^{\{\alpha\},\{ \beta\}} &\bra{\Psi(\nL')} \hat O_\mathrm{L}^{\{\alpha\}} \ket{\Psi(\nL)} \\ 
\times & \bra{\Psi(\nR')} \hat O_\R^{\{\beta\}} \ket{\Psi(\nR)}.
\end{split}\end{equation}
The prefactor $N=\bra{\Psi_\text{F}} \tilde O_\mathrm{L} \tilde O_\R \ket{\Psi_\text{I}}$ is needed for the correct normalization, e.g.\ $N=\sqrt{\nL(\nR+1)}$ for the tunneling process. Since $\hat O_i^{\{\alpha\}}$ acts on the single-site multi-orbital Wannier basis, $ \tilde A_{\nR,\nL}$ can be evaluated using the on-site coefficients of the many-particle state \eqref{eq:ManyParticleGS} discussed in the previous section.

\bibliography{references_v02}

\end{document}